\documentclass[aps,showpacs,twocolumn,amsmath,amssymb,nofootinbib]{revtex4-1}
\usepackage{graphicx,epsfig}
\pdfoutput=1
\usepackage{amsfonts}
\usepackage{amsmath}
\usepackage{amssymb}
\usepackage{mathrsfs}
\usepackage{multirow}
\usepackage{color}
\usepackage{graphicx}  
\usepackage{subfig}
\usepackage{hyperref}

\def\Journal#1#2#3#4{{#1} {\bf #2}, #3 (#4)}

\def\EPJC{{Eur. Phys. J.} C}

\def\LRR{Liv. Rev. Relat.}

\def\NPB{{Nucl. Phys.} B}
\def\PLB{{Phys. Lett.}  B}
\def\PLA{{Phys. Lett.}  A}
\def\PRL{Phys. Rev. Lett.}
\def\PRD{{Phys. Rev.} D}
\def\PR{{Phys. Rev.}}

\def\APJ{{Astrophys. J.}}
\def\APJL{{Astrophys. J. Lett.}}
\def\AA{{Astron. Astrophys.}}
\def\MNRAS{{Mon. Not. R. Astr. Soc.}}
\def\NAT{{Nature}}
\def\SCI{{Science}}
\def\CQG{{Class. Quant. Grav.}}
\def\CMP{{Commun. Math. Phys.}}
\def\JCAP{{JCAP}}
\def\JMP{{J. Math. Phys.}}

\def\GRG{{Gen. Relativ. Gravit.}}
\def\IJMPA{{Int. J. Mod. Phys.} A}
\def\IJMPD{{Int. J. Mod. Phys.} D}

\def\Sc{Schwarzschild }

\newcommand{\be}{\begin{equation}}
\newcommand{\ee}{\end{equation}}
\newcommand{\bea}{\begin{eqnarray}}
\newcommand{\eea}{\end{eqnarray}}

\begin{document}

\title{High precision numerical sequences of rotating hairy black holes}

\date{\today}


\author{Gustavo Garc\'ia$^1$}
\email{gustavo.garcia@correo.nucleares.unam.mx} 
\author{Eric Gourgoulhon$^2$}
\email{eric.gourgoulhon@obspm.fr}
\author{Philippe Grandcl\'ement$^2$}
\email{Philippe.Grandclement@obspm.fr}
\author{Marcelo Salgado$^1$}
\email{marcelo@nucleares.unam.mx} \affiliation{$^1$Instituto de Ciencias
Nucleares, Universidad Nacional Aut\'onoma de M\'exico,\\
A.P. 70-543, CDMX 04510, M\'exico\\
  $^2$Laboratoire Univers et Th\'eories, Observatoire de Paris, Universit\'e PSL, Universit\'e Paris Cit\'e, UMR 8102 CNRS, F-92190 Meudon, France}

\begin{abstract}
  We analyze numerically the existence of regular stationary rotating hairy black holes within the
  framework of general relativity, which are the
  result of solving the Einstein-Klein-Gordon system for a complex-valued scalar field under suitable
  boundary (regularity and asymptotically flat) conditions. To that aim we solve the
  corresponding system of elliptic partial differential equations using spectral methods which are 
  specially suited for such a numerical task.
  In order to obtain such system of equations we employ a parametrization for the metric that
  corresponds to quasi-isotropic coordinates (QIC) that have been used in the past for analyzing
  different kinds of stationary rotating relativistic systems. Our findings are in agreement with those reported
  originally by Herdeiro \& Radu \cite{Herdeiro2014,Herdeiro2015}.
  The method is submitted to several analytic and numerical tests, which include the recovery
  of the Kerr solution in QIC and the cloud solutions in the Kerr background.
  We report different global quantities that allow us to determine the contribution of the boson hair to the
  spacetime, as well as relevant quantities at the horizon, like the surface gravity. The latter indicates to what
  extent the hairy solutions approach the extremal limit, noting that for this kind of solutions the
  ratio of the angular momentum per squared mass $J_\infty/M^2_{\rm ADM}$ can be larger than unity due to the contribution of the scalar hair,
  a situation which differs from the Kerr metric where this parameter is bounded according to $0\leq |J/M^2| \leq 1$
  with the upper bound corresponding to the extremal case.
 \end{abstract}

 \maketitle

\section{Introduction}
\label{Introduction}
Black holes (BH) are one of the most fascinating predictions of general theory of relativity (GR).
A BH region within an asymptotically flat (AF) spacetime is a no escape region where the event horizon, ${\cal H^+}$,
a null surface, separates this region from the ``observable universe'', called the domain of outer communication (DOC).

Uniqueness theorems establish that the only regular, AF and stationary BH in GR coupled to
Maxwell theory, are contained within the Kerr-Newman family of solutions which are characterized by only three
parameters, the mass, electric charge and angular momentum \cite{uniqueness}. In particular, in the absence of rotation, this family reduces
to the Reissner-Nordstrom BH, and for a vanishing electric charge the only possible solution is the \Sc solution, which has only
the mass as a single parameter.

The no-hair conjecture, first proposed by Carter in 1968 \cite{Carter1968}, stipulates that the only stationary AFBH are
precisely described in terms of those three parameters. This conjecture has been reinforced by no-hair theorems
which establish the circumstances under which some non-trivial fields cannot coexist outside a BH \cite{nohairtheorems,Sudarsky1995,Pena1997}. Similar theorems are extended as to include non-minimally coupled scalar-fields to gravity \cite{NMCfields}. Several of these ``no-go theorems'', are mainly restricted to static, spherically symmetric and asymptotically flat  scenarios, leading to the conclusion that the only possible BH corresponds to the \Sc solution.

From the observational point of view, there is strong, albeit indirect evidence, that BH's do exist in the universe. For instance, the gravitational-wave (GW) observatories LIGO-VIRGO together
with massive numerical simulations and a detailed statistical analysis indicate that the source of several
GW signals detected in recent years come from the collision of two BH that presumably are of Kerr or \Sc type \cite{VIRGOLIGO}.
Moreover, the dynamics of stars near the center of our galaxy (Sgr A*) also shows that the central object is undoubtedly
a supermassive BH \cite{Ghezetal,Genzeletal}. Finally, the Event Horizon Telescope (EHT) has produced radio images from the matter in the
neighborhood of the center of galaxy M87 and our own galaxy that are consistent with the hypothesis of an accretion luminous disk around
a Kerr BH \cite{EHT}, which is in addition, one of the simple mechanisms to explain the emission of relativistic jets in this
kind of environments \cite{EHTjets}.

This overwhelming observational evidence about the existence of BH's in the universe leaves, however, some room
for speculating about the specific kind of BH that have been observed, notably, BH's other than Kerr or \Sc$\!$. At this respect, we mention that despite the elaboration of several no-hair theorems, hairy BH solutions have been found in different kind of field theories coupled to GR \cite{hairyBH,Nucamendi2003} (for a review see \cite{Volkov1999,CQGfocushairyBH}), showing that more general BH's require more parameters and thus, indicating that the no-hair conjecture might be wrong. Nevertheless, most, if not all, of those kind of hairy BH solutions have proved to be unstable under linear perturbations (cf. Refs. \cite{hairyunstable,Nucamendi2003}), and correspond to static and spherically symmetric situations.
Thus, it is possible that if the stability condition is included the no-hair conjecture may hold.

In 2014 a remarkable result that puts in jeopardy this conjecture was obtained by Herdeiro \& Radu~
\cite{Herdeiro2014,Herdeiro2015} (hereafter referred to as HR).
Namely, that a regular, stationary and rotating AFBH solution in GR can coexist with a complex-valued boson field
(hereafter hairy solutions). This is
a surprising result in view that a specific no-hair theorem for this kind of matter was proved in the
past, although assuming a static and spherically symmetric spacetime \cite{Pena1997}. The hairy solutions
found by HR under different kind of values for the boundary conditions at the horizon are numerical, and the corresponding partial differential equations (PDE) involved  in the problem are not easy to solve, even numerically. The authors also showed that
non-trivial configurations for the same kind of fields can be also found when the background is fixed, notably,
a Kerr spacetime background. These solutions were termed {\it clouds}. The numerical analysis of these clouds was then extended
as to include electric charge, and non-trivial charged scalar clouds were also found within a fixed Kerr-Newman background \cite{Benone2014}.

A further study by some of us pinpointed the reason of why the no-hair theorems for static situations could not be
extended to stationary and rotating scenarios \cite{Garcia2019,Garcia2020}. This analysis provides a further hint to understand in a simple and heuristic fashion the existence of those cloud solutions.

It is still premature to assess the impact of those hairy solutions from the observational point of view, as the current
data from the observations alluded above have several uncertainties, which in the future might be accommodated
as to include or rule out this kind of hairy black holes. Theoretically, these hairy solutions might be unstable but they still cannot be dismissed since a sound stability analysis is still lacking. Only a preliminary analysis of this sort has been performed recently \cite{Ganchev2018,Degollado2018},
but it only includes a very small region of the parameter space of solutions. Other hairy solutions in alternative gravity theories have also been found \cite{hairyalter,CQGfocushairyBH}. Thus, observations can also validate, rule out or constrain those solutions as well, or the
alternative theories themselves \cite{obshairy}.

Since the hairy solutions found by HR might have an impact in the interpretation of the forthcoming observations regarding BH's,
it is important to provide an independent study of this kind of solutions.

The goal of this paper is to provide that analysis using an approach that differs significantly from the
HR method. Namely, a)  we use a parametrization of the metric in terms of quasi-isotropic coordinates (QIC), that
leads to a rather compact system of elliptic PDE for the metric potentials; b) we use a spectral method decomposition
that has proved to be well suited to solve those kind of PDE with great numerical precision\footnote{At the time this work was completed we became aware about Ref.~\cite{Fernandes2022} where the authors analyze different kind of BH's using spectral methods and perform similar sort of tests that appear in our Appendix A.}. Spectral methods have been used systematically in the past by the Meudon group and collaborators \cite{Grandclement2009} to
analyze numerically a large variety of scenarios involving similar kind of relativistic rotating systems, like
neutron stars, boson stars, galileon BH's, among others. Thus, these numerical methods are reliable and relatively easy to
implement thanks to the spectral-method library KADATH \cite{Grandclement2010}.

The article is organized as follows. In Section~\ref{sec:formalism} we provide the formalism and equations.
Section~\ref{sec: BC} discusses the boundary conditions, Section~\ref{sec:globalq} introduces some formulae to compute global
quantities associated with the BH solutions, like the Komar mass, angular momentum, and the surface gravity.
Section~\ref{sec:numint} provides the numerical analysis and Section~\ref{sec:compare} compares our results
with previous studies reported recently by  HR. Finally, in Section~\ref{sec:conclusion} we present the conclusions and 
outlook for further studies along this line of research. Several appendices are included in order to complete the different
sections and to make the article more self-contained.


\section{Formalism and field equations}
\label{sec:formalism}
We consider the action functional
\bea
S[g_{ab},\Psi] &= &\int\left\lbrace\frac{R}{16\pi} - \left[\frac{1}{2}(\nabla_c \Psi^*)(\nabla^c \Psi) + \frac{1}{2}\mu^2\Psi^*\Psi\right]\right\rbrace\
\nonumber \\ && \times \sqrt{-g}d^4x\;,
\eea
associated with GR and a matter contribution from a complex-valued scalar field $\Psi$, submitted to a potential that includes only a mass term. Variation of the metric and the scalar field leads
respectively to the Einstein's field equations
endowed with an energy-momentum tensor (EMT) that describes the distribution of the scalar-field, and a Klein-Gordon
equation for the field:
\bea
\label{EFE}
&& G_{ab} = R_{ab}- \frac{1}{2} g_{ab} R= 8\pi T_{ab} \,,\\
\label{EMT}
&& T_{ab} = \nabla_{(a} \Psi^* \nabla_{b)} \Psi - g_{ab}\Big[\frac{1}{2}(\nabla_c \Psi^*) (\nabla^c \Psi) +
  \frac{1}{2}\mu^2\Psi^*\Psi \Big]\nonumber \\\\
&& \Box \Psi = \mu^2\Psi \,,
\label{KG}
\eea  
where $\Box= g^{cd}\nabla_c \nabla_d$, and as remarked above, we assume a potential describing a free but massive scalar-field $\Psi$
(hereafter boson field) with mass $\mu$. We use geometrized units where $G=c=1$.

Our task is to find solutions of the above system of field equations under several symmetries: we focus on stationary, axisymmetric, asymptotically flat (AF) spacetimes without ``meridional currents", i.e., 
under the circularity condition. Therefore, we assume the existence of two Killing fields, $\xi^a$ and $\eta^a$ which are, respectively, timelike and
spacelike, in the asymptotic regions, and are associated with the above symmetries. Due to the above assumptions, these Killing fields commute and therefore can be taken as a part of the coordinate basis. We use the time $t$ and the spatial angle $\varphi$ as coordinates adapted to such vector fields,
\be
\label{KF}
\xi^a  = \left(\frac{\partial}{\partial t}\right)^a \,\,\,,\,\,\,\eta^a = \left(\frac{\partial}{\partial\varphi}\right)^a \,.
\ee
The circularity condition translates into $\xi^a R_a^{\,\,[b} \xi^c \eta^{d]}=\eta^a R_a^{\,\,[b} \xi^c \eta^{d]}=0$ \cite{Wald1984},
which upon the use of the Einstein field Eq.~(\ref{EFE}), this condition imposes the following restrictions on the
EMT $\xi^a T_a^{\,\,[b} \xi^c \eta^{d]}= \eta^a T_a^{\,\,[b} \xi^c \eta^{d]}=0$. In addition, the AF allows for the
following conditions to be verified $\xi_{[a}\eta_{b}\nabla_c \xi_{d]}=0$, $\xi_{[a}\eta_{b}\nabla_c \eta_{d]}=0$
on the (rotation) axis of symmetry \cite{Wald1984}. These assumptions imply that the planes orthogonal to
$\xi_a$ and $\eta_a$ are integrable (cf. Refs. \cite{Papapetrou1966,Carter1969,Carter1973}).
    
Thus far, we have local coordinates $(t, x^1,x^2,\varphi)$, where the coordinates $x^1,x^2$ are associated with
the two coordinate basis vectors $\left(\frac{\partial}{\partial x^1}\right)^a$, $\left(\frac{\partial}{\partial x^2}\right)^a$
respectively, which due to the previous hypothesis, are orthogonal
to $\xi^a$ and $\eta^a$, and thus, the metric components $g_{\mu\nu}$ with respect to the basis vectors
satisfy $g_{t1}= g_{t2}= g_{\varphi 1}= g_{\varphi 2}=0$, and the remaining no null components are functions of $x^1,x^2$
solely. Furthermore,
for the two coordinates $x^1,x^2$ we adopt the so-called Quasi-Isotropic gauge (QI)  so that
these coordinates are of spherical type $r,\theta$ and the metric reads\footnote{The reader is urged not to confuse
  the QI coordinates with the usual Boyer-Lindquist radial coordinates of Kerr geometry (cf. Appendix~\ref{app:KerrQI}).}
\bea
 \label{metric}
 ds^2 &=& -N^2dt^2 + A^2\left(dr^2 + r^2d\theta^2\right) \nonumber \\
 && + B^2r^2\sin^2\theta\left(d\varphi + \beta^{\varphi}dt\right)^2, 
\eea
where the metric potentials $N,A,B,\beta^\varphi$ are functions of the coordinates $r,\theta$, solely.
In particular, in the limit of spherical symmetry and staticity, the shift-vector component
$\beta^{\varphi}$ vanishes, and
$N,A,B$ become independent of the coordinate $\theta$. Moreover, in that scenario $A(r)=B(r)$, and in vacuum such parametrization leads to the \Sc solution in isotropic coordinates (cf. Appendix~\ref{app:KerrQI}).

The reason behind the use of QI coordinates (QIC) is because the Einstein equations lead to a system of elliptic
PDE for the metric potentials (or combinations of them) that can be solved numerically with great accuracy using
spectral methods. In a previous work configurations of black holes with scalar hair could also be computed by means of spectral methods
 using another choice of coordinates: the spatial harmonic ones \cite{Grandclement2022}.

Since the QI gauge for spacetimes with the above symmetries have been extensively used in the past \cite{Gourgoulhon2011,Gourgoulhon2022},
notably in the study of rotating boson stars \cite{Grandclement2014}, we proceed directly to write the specific Einstein equations for
the metric potentials, which have been derived under the 3+1 formalism of GR (for a review see \cite{Gourgoulhon2012,Alcubierre2008,Baumgarte2010}) and 
arranged in a more convenient fashion to take into account the spectral solver KADATH \cite{Grandclement2010},
and the fact that at the inner boundary, which represents the BH horizon ${\cal H}^+$, the
lapse function $N$ vanishes (cf. Sec.~\ref{sec: BC}):
\bea
\label{eq:S11}
&& \Delta_3N + \frac{\partial N\partial B}{B} -
\frac{B^2r^2\sin^2\theta}{2}\frac{\partial\beta^{\varphi}\partial\beta^{\varphi}}{N} \nonumber \\
&& = 4\pi A^2N\left(E + S\right)\;, \\ \nonumber \\
\label{eq:S22}
&&\Delta_3\left(\beta^{\varphi} r\sin\theta\right) - \frac{\beta^{\varphi}}{r\sin\theta} - r\sin\theta\frac{\partial\beta^{\varphi}\partial N}{N} \nonumber \\
&& + 3r\sin\theta\left(\partial\beta^{\varphi}\partial\ln B\right)= 16\pi\frac{NA^2p_{\varphi}}{B^2r\sin\theta}\;,\\ \nonumber \\
\label{eq:S33}
&& N\Delta_2\left[\left(B - 1\right)r\sin\theta\right] + \left[\left(B - 1\right)r\sin\theta\right]\Delta_2N \nonumber \\
\nonumber \\
&+& 2\partial N\partial\left[\left(B - 1\right)r\sin\theta\right] +
\Delta_2\left[\left(N - 1\right)r\sin\theta\right] \nonumber \\ \nonumber \\
&& = 8\pi NA^2Br\sin\theta\left(S - S_{\varphi}^{\varphi}\right)\;,\\ \nonumber \\
\label{eq:S44}
&& N\Delta_2A + A\Delta_2N - \frac{N}{A}\partial A\partial A - \frac{3}{4}AB^2r^2\sin^2\theta\frac{\partial\beta^{\varphi}\partial\beta^{\varphi}}{N} \nonumber \\
&& = 8\pi A^{3}NS^{\varphi}_{\varphi}\;.
\eea

In this set of equations the following notation has been adopted
\begin{equation}
\Delta_3 \equiv \frac{\partial^2}{\partial r^2} + \frac{2}{r}\frac{\partial}{\partial r} + \frac{1}{r^2}\frac{\partial^2}{\partial\theta^2} + \frac{1}{r^2\tan\theta}\frac{\partial}{\partial\theta}, 
\end{equation}
\begin{equation}
\Delta_2 \equiv \frac{\partial^2}{\partial r^2} + \frac{1}{r}\frac{\partial}{\partial r} + \frac{1}{r^2}\frac{\partial^2}{\partial\theta^2},
\end{equation}
and
\begin{equation}
\partial u\partial v \equiv \frac{\partial u}{\partial r}\frac{\partial v}{\partial r} + \frac{1}{r^2}\frac{\partial u}{\partial\theta}\frac{\partial v}{\partial\theta}.
\end{equation}

As concerns the scalar field $\Psi$, we assume the following harmonic form on the time and angular dependence\footnote{In order to match the notation of Ref.~\cite{Grandclement2014} one has to use $m=k$. Moreover, the integer $m$ of this paper is not to be confused with the scalar field mass of that reference.
Finally, to compare with~\cite{Garcia2019,Garcia2020} 
one has to transform 
$\omega\rightarrow -\omega$, $m\rightarrow-m$, i.e., to change a global sign in the harmonic dependence of $\Psi$.}:
\begin{equation}
\label{Psians}
\Psi(t,r,\theta,\varphi)=  \phi(r,\theta)e^{i(\omega t - m \varphi)}\;,
\end{equation}
where $\phi(r,\theta)$ is real valued and $m$ is a non-zero integer. The harmonic dependence 
of the boson field is such that the EMT respects the symmetries of the underlying spacetime.
The Klein-Gordon Eq.~(\ref{KG}) then reads,
\begin{eqnarray}
  \label{KGG}
\nonumber && \Delta_{3}\phi + \frac{A^2}{N^2}\left(\omega + \beta^{\varphi}m\right)^2\phi - \frac{m^2\phi}{r^2\sin^2\theta}\left(\frac{A^2}{B^2}\right) \\   
&& + \:  \frac{1}{N}\partial\phi\partial N + \partial\phi\partial\ln B = A^2\mu^2\phi.
\end{eqnarray}

The system of elliptic PDE provided by Eqs. (\ref{eq:S11})--(\ref{eq:S44}) is equivalent 
but different from the system used previously by HR \cite{Herdeiro2015} under a different coordinate gauge.
The source terms that appear at the r.h.s of Eqs. (\ref{eq:S11})--(\ref{eq:S44}) are provided by the 3+1 decomposition
of the EMT (\ref{EMT})
\cite{Gourgoulhon2012,Alcubierre2008,Baumgarte2010,Grandclement2014}~\footnote{We stress a difference (possibly due to a typo --a misplaced right bracket--) in a similar expression 
in Eq.(A14) of \cite{Grandclement2014}.}:
\begin{eqnarray}
\label{traceE}
&& E = \left[\frac{\left(\omega + m\beta^{\varphi}\right)^2}{N^2} +
  \frac{m^2}{B^2r^2\sin^2\theta}\right]\frac{\phi^2}{2} \nonumber \\
&& + \frac{1}{2A^2}\left[ \left(\frac{\partial\phi}{\partial r}\right)^2 
+ \frac{1}{r^2}\left(\frac{\partial\phi}{\partial \theta}\right)^2\right] + \frac{\mu^2\phi^2}{2}\;,\\
\label{Momentum}
&& p_\varphi = \frac{m}{N}\left(\omega + m\beta^{\varphi}\right)\phi^2 \;,\\
\label{S_phiphi}
&& S_{\varphi}^{\varphi} = \left[\frac{\left(\omega + m\beta^{\varphi}\right)^2}{N^2} +
  \frac{m^2}{B^2r^2\sin^2\theta}\right]\frac{\phi^2}{2} \nonumber \\
&&  - \frac{1}{2A^2}\left[ \left(\frac{\partial\phi}{\partial r}\right)^2 
   + \frac{1}{r^2}\left(\frac{\partial\phi}{\partial \theta}\right)^2\right] 
    - \frac{1}{2}\mu^2\phi^2 \\
 \label{traceS}
 && S = \left[3\frac{\left(\omega + m\beta^{\varphi}\right)^2}{N^2} -
  \frac{m^2}{B^2r^2\sin^2\theta}\right]\frac{\phi^2}{2} \nonumber \\
&& - \frac{1}{2A^2}\left[ \left(\frac{\partial\phi}{\partial r}\right)^2 
   + \frac{1}{r^2}\left(\frac{\partial\phi}{\partial \theta}\right)^2\right] - \frac{3}{2}\mu^2\phi^2 \;.
\end{eqnarray}
The above components of the EMT in 3+1 form are compatible with the circularity condition, namely,
the components $T_{t\theta}$, $T_{tr}$, $T_{\varphi\theta}$, $T_{\varphi r}$ vanish identically. Incidentally,
the component $T_{r \theta}$, which does not vanish, is not necessary to solve the above system of equations.
We performed some basic tests to check the consistency of this system, namely, we verified that
in vacuum they satisfy the exact Kerr solution in QIC (cf. Appendix~\ref{app:KerrQI}), as opposed to the Kerr solution in Boyer-Lindquist (BL) coordinates. At this point it is important to emphasize that BL coordinates have been used to find solutions (with the Kerr background fixed)
of cloud configurations for the boson field $\Psi$
\cite{Herdeiro2014,Herdeiro2015,Garcia2019,Garcia2020,Hod}, which in turn allowed the use of separation of variables for the radial and angular dependence $\theta$ of the boson field
leading to Teukolsky-type of equations~\cite{Teukolsky1972}.


\subsection{Boundary (regularity) conditions}
\label{sec: BC}

In order to solve the system of elliptic PDE's, appropriate regularity conditions are implemented.
We consider an inner boundary corresponding to the BH horizon
${\cal H}^+$, that we assume to have a spherical topology, and thus, it is located at $r=r_H$.
Therefore, we impose regularity conditions at $r=r_H$ for each of the metric potentials
$N$, $A$, $B$ and $\beta^{\varphi}$. According to Hawking's rigidity theorem~\cite{Hawking},
when physically reasonable conditions are satisfied (weak energy condition and matter obeying hyperbolic equations--in dynamic situations),
the event horizon of a stationary BH coincide with a Killing horizon. These conditions are verified in the current
scenario. This theorem is compatible with the circularity condition~\cite{Carter1969,Carter1973}.
In the present case the  null geodesic generators of the horizon have as tangent the {\it helical} Killing field
\be
\label{Killingchi}
\chi^a = \xi^a + \Omega_H \eta^a \,,
\ee
in the region where its norm vanishes, like in the Kerr BH, where
\be
\Omega_H = - \frac{\xi_a \xi^a}{\xi_a\eta^a}\Big|_{{\cal H}^+}= - \frac{\eta_a \xi^a}{\eta_a\eta^a}\Big|_{{\cal H}^+} = {\rm const}.\,,
\ee
is the {\it angular velocity} of the horizon~\cite{Heusler1996}. In terms of the 3+1 variables
\be
\Omega_H = - \beta^\varphi (r_H,\theta) = {\rm const} \,.
\ee
This is the condition imposed on the shift vector at $r_H$. 

As mentioned above, the BH horizon is located at the place where $\chi^a \chi_a|_{{\cal H}^+}=0$. In terms of the 3+1 variables this corresponds at the place where 
the lapse function vanishes $N(r_H,\theta)=0$. Additionally, the {\it ergosphere} is defined as the region where the Killing field $\xi^a$ becomes 
spacelike. The ergosurface corresponds to the spacetime points where $\xi^a \xi_a=0$, this is where the metric component
$g_{tt}=0$. Outside the ergosphere the Killing field $\xi^a$ is {\it timelike}, where it is possible to find {\it static observers}, 
i.e. observers with 4-velocity $u^a_S$ parallel to $\xi^a$. However, inside the ergosphere, those observers no longer exist,
and must rotate with a component in the direction of the {\it rotational}
Killing field $\eta^a$ [see Eq.~(\ref{KF})]. It is precisely within the ergosphere where the Penrose process of {\it energy extraction} can take place.

As concern the metric potentials $A$ and $B$, we assume that the event horizon ${\cal H}^+$ coincides with an
{\it apparent horizon}, and so these potentials have to satisfy the following condition (see Appendix~\ref{app:appHor})
\be
  \label{Hstationary2}
\left[\frac{\partial}{\partial r}\left(\frac{1}{A}\right) + \frac{1}{A}\left(\frac{2}{A}\frac{\partial A}{\partial r} + \frac{1}{B}\frac{\partial B}{\partial r} + \frac{2}{r}\right)\right]_{r_H} =0\;,
\ee
In particular, we checked that the condition (\ref{Hstationary2}) is verified exactly for the
Kerr metric in QIC. It may seem strange to use just one boundary condition for the two quantities $A$ and $B$. However, by looking at Eqs.  (\ref{eq:S33}) and (\ref{eq:S44}) one can see that they are degenerate, in the sense that the factor in front of the highest derivative (i.e. the $\Delta_2$ operators) vanishes at the inner boundary. Such degenerate equations cannot be solved with any boundary conditions. In the case at hand, it appears that Eq. (\ref{eq:S33}), on the horizon, becomes a regularity condition and is treated as such by the solver. Equation (\ref{Hstationary2}) is treated as being the boundary condition associated with Eq. (\ref{eq:S44}).

As regards the boson field, we consider the so-called {\it synchronicity} or {\it no flux} condition across the horizon to ensure the existence of ``bound states'' associated with a stationary boson field \cite{Herdeiro2014,Herdeiro2015},

\be
\chi^a \nabla_a 
\Psi\Big|_{{\cal H}^+} = 0 \,.
\ee

In view of Eq.~(\ref{Psians}), this condition translates into
\begin{equation}
\label{fluxcond1}
(\omega- m\Omega_H)\Psi_H=0 \;.
\end{equation}
So for any finite $\Psi_H \neq 0$ we obtain 
\begin{equation}
\label{fluxcond2}
\omega= m\Omega_H \;.
\end{equation}
This condition stems also from demanding regularity in Eq.~(\ref{KGG}) at ${\cal H}^+$ as one can appreciate from the first term within brackets, which must vanish as the horizon is approached
and where $N_H=0$.

The Klein-Gordon equation (\ref{KGG}) also requires some regularity condition to be fulfilled, due to the division by $N$. Close to the horizon the solver works with $N$ times Eq. (\ref{KGG}) to remove possible divergences. Doing so, the equation becomes degenerate (as for Eqs. (\ref{eq:S33}) and (\ref{eq:S44})) and so it is solved without any boundary condition. This procedure allows to find the right value of the scalar field, consistent with regularity, without the need to enforce that regularity explicitly.

Asymptotic flatness (AF) is to be imposed on the spacetime, and thus, $\beta^\varphi\rightarrow 0$,
$N\rightarrow 1$, $A\rightarrow 1$, $B\rightarrow 1$, as $r\rightarrow \infty$. Thanks to a compactification of space 
those conditions are enforced at exact infinity.

Now, when the background spacetime is fixed, one needs to solve
only for the scalar field Eq.(\ref{KGG}). With  a Kerr background, the solution for $\Psi$ lead to the
cloud solutions mentioned before \cite{Herdeiro2014,Herdeiro2015,Garcia2019}. In this context, Eq. (\ref{KGG}) is linear with respect to the scalar field and it admits 
non-zero solutions only for discrete values of the parameters ($a$, $M$, $\Omega_H$, $\omega$ and $\mu$). So this is an eigenvalue problem. The numerical method thus needs to
prevent the code from converging to the trivial solution $\phi(r,\theta)=0$ and to find the right values of the parameters. A method for dealing with this class of problems (i.e. eigenvalue problems arising from a linear approximation) has already been used several times (see \cite{Grandclement2011, Fodor2014} and especially \cite{Grandclement2022} in the context of clouds). Using the same technique, clouds are constructed using a Kerr background in QI coordinates. They correspond to the limit of the full system when $\phi \rightarrow 0$.  In this work, clouds are used as an initial guess to construct black holes with scalar hair, thus 
leading to the solution of the full non-linear problem
once we take into account the backreaction of the scalar-field into the spacetime. By changing the parameters away from those of the cloud solutions, sequences of hairy BH can be obtained. For instance, in Fig. \ref{fig:KomarMassAngular}, the cloud solution corresponds to the rightmost point (with $\Omega_H/\mu \approx 0.98$) and the sequence is constructed by slowly varying $\Omega_H$. We elaborate more about this point in
Section~\ref{sec:numint}.

As remarked before, the first test for the spectral code consisted in recovering numerically the
exterior Kerr solution by setting the sources to zero, i.e., assuming a vacuum scenario with a vanishing scalar field.
This tests are included in the Appendix \ref{app:KerrQI} where we compare the numerical solution
with the exact Kerr solution in QIC coordinates. The relative numerical errors are typically $\sim 10^{-14}-10^{-16}$ or lower for 17 coefficients in the radial spectral expansion.

We perform a second test where we fix the Kerr background and recover the same kind of cloud solutions
provided in Ref.\cite{Garcia2019} for the subextremal scenarios, but in QIC. 


\section{Global quantities, surface gravity and BH thermodynamics}
\label{sec:globalq}

\subsection{Boson number, mass and angular momentum}
\label{sec:massangtot}

There are basically three global quantities that emerge as a consequence of the AF conditions which are measured at
spatial infinity (as opposed to global quantities defined at ${\cal H}^+$).
The first one is related to the internal symmetry associated with the boson field, namely,
the invariance of the model with respect to a global phase transformation $\Psi\rightarrow \Psi'= e^{i\alpha} \Psi$,
where $\alpha=const.$ This symmetry leads to the local conservation of the 
{\it boson current} $\nabla_{a}j^{a} = 0$, where\footnote{Some differences on sign in the current depend on conventions. Here 
we agree with \cite{Herdeiro2014}, but we use a different variable $\Psi$ that introduces a factor $1/2$ (cf. the EMT). For all the numerical calculations we take units where $\hbar=1$.}

\begin{equation}\label{current}
j^{a} = \frac{i}{2\hbar}\left(\Psi^{*}\nabla^{a}\Psi - \Psi\nabla^{a}\Psi^*\right)  \;.
\end{equation}
Furthermore, the conserved current implies that the {\it total particle number} of the boson field is also conserved. 
This is given in terms of a volume integral on a spatial hypersurface $\Sigma_t$ of the component of the flux $j^a$
that is normal to the hypersurfaces $\Sigma_t$:
\begin{equation}\label{numberB}
\mathcal{Q} \equiv \int_{\Sigma_t} -n_{a}j^{a}\sqrt{\gamma}d^3x  \;,
\end{equation}
where $\gamma = {\rm det}(\gamma_{ij}) = A^{4}B^2r^{4}\sin^{2}\theta$, is the determinant of the three-dimensional metric, and 
$d^3x= dr d\theta d\varphi$. In this case the timelike normal is given by
\bea
\label{normaln}
n^a &=& \frac{1}{N}\left(\xi^a -\beta^\varphi \eta^a\right)\;,\\
n_a &=& -N\nabla_a t= - N (dt)_a \;,
\eea
which is interpreted physically as the 4-velocity of observers that have {\it zero angular momentum} (ZAMO's)
$\ell \equiv -n_a \eta^a\equiv 0$. 

Using the harmonic form for the scalar field (\ref{Psians}), the current (\ref{current}) reduces to
\be
\label{current1}
j^{a} = \hbar^{-1}\phi^{2}\nabla^{a}\left(m\varphi - \omega t\right)  \;, 
\ee
and the total boson number is 
\be
\label{numberB1}
\mathcal{Q} \equiv \frac{2\pi }{\hbar}\int_{r_H}^\infty \int_{0}^{\pi}\frac{1}{N}\left(\omega + m\beta^{\varphi}\right)\phi^{2} A^{2}Br^2\sin\theta
dr d\theta \;.
\ee
Where we used the fact that all the quantities involved in the integral are independent of the angle $\varphi$ due to the axial symmetry.

On the other hand, there are two global quantities at spatial infinity related to the spacetime symmetries. The first one
is the {\it Komar mass}, which is associated with the Killing vector $\xi^a = (\partial/\partial t)^a$~\cite{Wald1984}:
\be
  \label{Kmass1}
M_K \equiv  -\frac{1}{8\pi}\oint_{\mathcal{S}}\nabla^{a}\xi^{b}dS_{ab}  \;.
\ee
Since in the current scenario the scalar field is present outside the BH, this integral can be split into two contributions~\cite{Gourgoulhon2012},
one at the inner boundary corresponding to the intersection ${\cal H}_t \equiv \Sigma_t \cap {\cal H}^+$,
between the spatial hypersurfaces $\Sigma_t$ defined by the parameter $t$ associated with the timelike Killing field $\xi^a$
[cf. Eq.~(\ref{KF})], and the event horizon ${\cal H}^+$. The second contribution corresponds to a volume integral within a 2-sphere
$\mathcal{S}_\infty \subset \Sigma_t$ that extends to spatial infinity. In both cases $dS_{ab}$ is the area-element 2-form normal to both boundaries. 
Thus, Eq.~(\ref{Kmass1}) reads 
\be
\label{Kmass2}
M_K = M^{{\mathcal H}_t} + M^{\Sigma_t} \;.
\ee
The mass $M^{{\mathcal H}_t}$ can be written using the 3+1 variables \cite{Gourgoulhon2012}\footnote{In Ref.~\cite{Gourgoulhon2012} the normal $s^a$ is taken pointing inward the horizon ${\cal H}_t$, i.e., $s^a\rightarrow -s^a$, and therefore there is a global sign difference in Eqs.~(\ref{KmassH}) 
and (\ref{KJH}) as compared with the same equations of Ref.~\cite{Gourgoulhon2012}.}:
\bea
M^{{\mathcal H}_t} &=&-\frac{1}{8\pi}\oint_{{\mathcal H}_t}\nabla^{a}\xi^{b}dS_{ab}^{{\mathcal H}_t} \nonumber \\
\label{KmassH}
&=& \frac{1}{4\pi}\oint_{{\mathcal H}_t} \left(D_a N - K_{ab} \beta^b\right)s^a\sqrt{q}d^2x \;,
\eea
where, $D_a N =\gamma_{a}^{\;b}\nabla_b N$, $\gamma_{a}^{\;b}=\delta_{a}^{\,b} + n_a n^b$, $K_{ab}= -\gamma_{a}^{\;c} \gamma_{b}^{\;d}\nabla_c n_d$,
$s^a=  \sqrt{g^{rr}} \left(\partial/\partial r\right)^a
$ is the {\it outward} normal to ${\mathcal H}_t$, $q = A^{2}B^2r^{4}\sin^{2}\theta$, is the determinant of
the two-dimensional metric induced on $r={\rm const.}$ surfaces, and $d^2x= d\theta d\varphi$. More specifically,
$s^a D_a N=  \sqrt{g^{rr}}\partial_r N=\partial_r N/A $ and $K_{ab} \beta^b s^a=
 K_{r\varphi} \beta^{\varphi}/A$.
Therefore\footnote{In principle, this integral can be evaluated at $r\rightarrow \infty$ as well as at $r=r_H$ when
  matter is absent.} 
\bea
\!\!\!\!\!\!\!\!\!\! && M^{{\mathcal H}_t} = \frac{r_H^2}{2}\int_0^\pi \left(\partial_r N - K_{r\varphi} \beta^{\varphi} \right)B\sin\theta d\theta \Big{|}_{r_H}\nonumber \\
\label{KmassH2}
\!\!\!\!\!\!\!\!\!\! && = \frac{r_H^2}{2}\int_0^\pi \left(\partial_r N - B^2r^2\sin^2\theta \frac{\partial_r\beta^{\varphi}}{2N} \beta^{\varphi}\right)B\sin\theta d\theta \Big{|}_{r_H}
\eea
where we used $K_{r\varphi}= B^2r^2\sin^2\theta \frac{\partial_r\beta^{\varphi}}{2N}$.

The volume integral at spatial infinity associated with the contribution $M^{\Sigma_t}$ can be expressed in terms of the
EMT associated with the matter~\cite{Wald1984,Gourgoulhon2012}, in this case associated with the EMT of the boson field (\ref{EMT}):
\be
\label{Kmassinfty}
M^{\Sigma_t} = 2\int_{\Sigma_t}\left(T_{ab}n^{a}\xi^{b} - \frac{1}{2}Tn_{a}\xi^{a}\right)\sqrt{\gamma}d^3x \;,
\ee
this mass can be obtained as well from the conserved current
$\nabla_a \left(-T^{ab} \xi_b\right)=0$, which is associated with the stationary condition.

For the problem at hand $T_{ab}n^{a}\xi^{b} = N^{-1}\left( T_{tt} - T_{t\varphi}\beta^{\varphi}\right)$.
The components of the EMT involved in the previous expressions and its trace $T$ appearing in Eq.~(\ref{Kmassinfty})
can be obtained directly from (\ref{EMT}), which yield
\bea
\label{Kmass3}
M^{\Sigma_t} &=& 2\pi \int_{r_H}^\infty \int_{0}^{\pi}\left[\frac{2\omega}{N}\left(\omega + m\beta^{\varphi}\right)\phi^{2} - N\mu^2\phi^2\right]
\nonumber \\
&& \times A^{2}Br^2\sin\theta dr d\theta \;.
\eea
For instance, in vacuum, $M^{\Sigma_t}\equiv 0$, and for \Sc and Kerr spacetime in QIC $M^{{\mathcal H}_t} \equiv M$ (cf. Appendix~\ref{app:KerrQI}).

On the other hand, since the spacetime considered in our analysis is AF, one can obtain the total energy of the system from the mass ADM, which is given by a surface integral at infinity in the following form \cite{Gourgoulhon2012,Baumgarte2010}
\be
\label{ADMmass}
M_{\rm ADM} \equiv \frac{1}{16\pi}\oint_{r\rightarrow\infty}\left[\bar{D}^{b}\gamma_{ab} - \bar{D}_a\left(f^{cd}\gamma_{cd}\right)\right]s^a\sqrt{q}d^2x\;,
\ee
where $f^{cd}$ is the flat 3-metric, $\gamma_{ab}$ is the 3-metric and $\bar{D}_a$ denotes the covariant derivative associated with the metric $f^{cd}$. For the physical system that we study, it is found that
\bea
\label{ADMmass2}
\nonumber && M_{\rm ADM} = \frac{1}{8\pi}\int_{0}^{2\pi}\int_{0}^{\pi}\left[\bar{D}^{j}\gamma_{rj} - \bar{D}_r\left(f^{kl}\gamma_{kl}\right)\right]\\ && \times Br^2\sin\theta d\theta d\varphi\Big{|}_{r\rightarrow \infty} \nonumber \\
&=& -\frac{1}{8}\int_{0}^{\pi}\left[\frac{\partial}{\partial r} (A^2 + B^2) + \frac{B^2 - A^2}{r}\right]Br^2\sin\theta d\theta
\Big{|}_{r\rightarrow \infty}\;.
\eea
When we consider the Kerr spacetime in QIC, as $r \rightarrow \infty$ the leading terms of the metric potentials  are as follows (cf. Appendix~\ref{app:KerrQI}): $A^2 \sim 1$, $B^2 \sim 1$, $\frac{\partial A^2}{\partial r} \sim -\frac{2M}{r^2}$ and $\frac{\partial B^2}{\partial r} \sim -\frac{2M}{r^2}$, so from Eq.(\ref{ADMmass2}), it is found that $M_{\rm ADM} = M$.
Due to the AF assumption, the Komar mass $M_{\rm K}$ given by Eq.~(\ref{Kmass2}) coincides with the ADM mass (\ref{ADMmass}).

The second global quantity defined at spatial infinity which is associated with the axial Killing vector $\eta^a = (\partial/\partial \varphi)^a$, is
the {\it Komar angular momentum}. This quantity is given by an integral similar to the Komar mass (\ref{Kmass1}) but replacing $\xi^a = (\partial/\partial t)^a$ by 
$\eta^a = (\partial/\partial \varphi)^a$ and introducing a factor `$-\frac{1}{2}$'~\cite{Wald1984}:
\be
\label{JKomar}
J_{\rm K} \equiv \frac{1}{16\pi}\oint_{\mathcal{S}_\infty} \nabla^{a}\eta^{b}dS_{ab} \;.
\ee

Similarly, this expression can be split into two contributions~\cite{Gourgoulhon2012}
\be
\label{JKomar2}
J_{\rm K} = J^{{\mathcal H}_t} + J^{\Sigma_t} \;,
\ee
where $J^{{\mathcal H}_t}$ is given by the following expression in terms of 3+1 variables~\cite{Gourgoulhon2012}:
\be
\label{KJH}
J^{{\mathcal H}_t} = \frac{1}{8\pi}\oint_{{\mathcal H}_t} K_{ab} s^a \eta^{b} \sqrt{q}d^2x \;.
\ee
More explicitly,
\bea
\label{KJH2}
J^{{\mathcal H}_t} &=& \frac{r_H^2}{4}\int_0^\pi K_{r\varphi} B\sin\theta d\theta \Big{|}_{r_H}\nonumber \\
&=& \frac{r_H^4}{8}\int_0^\pi \frac{\partial_r\beta^{\varphi}}{N} B^3\sin^3\theta d\theta \Big{|}_{r_H}\;.
\eea

The contribution $J^{\Sigma_t}$ can be written in terms of a volume integral
from the conserved current $\nabla_a \left(T^{ab} \eta_b\right)=0$:
\be
\label{JKomarS}
J^{\Sigma_t} = -\int_{\Sigma_t}T_{ab}n^{a}\eta^{b}\sqrt{\gamma}d^3x \;.
\ee
In this case $T_{ab}n^{a}\eta^{b} = N^{-1}\left(T_{t\varphi} - \beta^{\varphi}T_{\varphi\varphi}\right)$,
and from (\ref{EMT}), one obtains
\be
\label{JKomarS2}
J^{\Sigma_t} = 2 \pi m\int_{r_H}^\infty \int_{0}^{\pi} \frac{1}{N}\left(\omega + m\beta^{\varphi}\right)\phi^2 A^{2}Br^2\sin\theta dr d\theta \;.
\ee
From (\ref{JKomarS2}) and (\ref{numberB1}) note that
\be
\label{JmQ}
J^{\Sigma_t}= m \mathcal{Q}\hbar \;,
\ee
that is, $J^{\Sigma_t}$ is an integer multiple of the boson number $\mathcal{Q}\hbar$, as it happens also in the rotating boson-star scenario~\cite{Schunck1999,Grandclement2014}.

An equivalent expression of the Komar angular momentum is
(see e.g. Eq.~(8.77) in Ref.~\cite{Gourgoulhon2012})
\be
\label{JADM}
J_\infty
\equiv \frac{1}{8\pi}\oint_{\mathcal{S}_\infty}K_{ab}\eta^as^bdS\;,
\ee
where, as before, $\eta^a = (\partial/\partial\varphi)^a$ and $s^a = \sqrt{g^{rr}} (\partial/\partial r)^a$. For our current analysis, this becomes
\bea
\label{JADM2}
J_\infty &=& \frac{1}{8\pi}\int_{0}^{2\pi}\int_{0}^{\pi}K_{r\varphi}\eta^{\varphi}s^r r^2 \sin \theta d\theta d\varphi \Big{|}_{r\rightarrow \infty}\nonumber \\
&=& \frac{1}{8}\int_{0}^{\pi}\frac{\partial_r\beta^{\varphi}}{N} B^3r^4\sin^3\theta d\theta
\Big{|}_{r\rightarrow \infty}\;.
\eea
For instance, in Kerr spacetime in QIC, as $r \rightarrow \infty$ the leading terms  of the metric potentials  are as follows (cf. Appendix~\ref{app:KerrQI}): $N \sim 1$, $A \sim 1$, $B \sim 1$ and $\beta^\varphi \sim -2J/r^3$, in agreement with Eq.(\ref{JADM2}).

While $M_{\rm K}\equiv M_{\rm ADM}$ and $J_{\rm K} = J_\infty$
are identities that hold exactly, from 
the numerical point of view this is not necessarily the case as 
each quantity is computed differently. At the end of the next section we show that both pairs of quantities converge to the same value as the number of coefficients in the spectral decomposition increase. This is an important convergence test that allow us to determine the precision of the spectral code that has been used to solve the Einstein-Klein-Gordon (EKG) system.

\subsection{Surface gravity}
\label{sec:surgrav}
Following Ref.~\cite{Wald1984} the surface gravity $\kappa$ of a stationary BH is given in
terms of the Killing field (\ref{Killingchi}) by
\be
\label{SurGra1}
\kappa^{2} = -\frac{1}{2}\left(\nabla^{a}\chi^{b}\right)\left(\nabla_{a}\chi_{b}\right)\big|_{{\cal H}^+} \;. 
\ee
This expression leads to the following formulae for the surface gravity associated with a stationary, axisymmetric and circular spacetime containing a BH with metric (\ref{metric}) 
(see Appendix~\ref{app:surgrav} for the details):
\bea
\label{kappa}
\kappa &= & \left\{\frac{1}{A}\left[\left(\partial_rN\right)^2 + \frac{1}{r^2}\left(\partial_{\theta}N\right)^2\right]^{1/2}\right\}_{r_H}
\nonumber \\
&=& \left\{\frac{1}{2AN}\left[\left(\partial_rN^2\right)^2 + \frac{1}{r^2}\left(\partial_{\theta}N^2\right)^2\right]^{1/2}\right\}_{r_H}
\,.
\eea
From (\ref{kappa}) one can check that the surface gravity for \Sc and Kerr spacetimes using QIC
(cf. Appendix~\ref{app:KerrQI}) leads to the usual expressions \cite{Wald1984} in terms of the mass $M$ and
the angular momentum $a=J/M$ (cf. Appendix~\ref{app:surgrav}). 

\subsection{Mass and BH thermodynamics formula}
\label{sec:massthermo}
According to Ref.~\cite{Wald1984},  stationary, rotating, axisymmetric and AFBH's like the ones we are analyzing here satisfy 
the following Smarr relation (using the notation of this section)
\be
\label{Massthermo0}
M_K= M^{\Sigma_t} +\frac{1}{4\pi}\kappa {\cal A} + 2\Omega_H J^{{\cal H}_t} \,.
\ee
where ${\cal A}$ is the {\it area} of the horizon, which in terms of the QIC is 
given by 
\bea
{\cal A} &=& r_H^2 \int_{0}^{2\pi} \int_{0}^{\pi} A B \sin\theta d\theta d\varphi \,\Big|_{r_H} \nonumber \\
&=& 2\pi r_H^2 \int_{0}^{\pi} A B \sin\theta d\theta 
\,\Big|_{r_H}\,.
\eea
Using Eq.(\ref{Kmass2}) in 
Eq.(\ref{Massthermo0}), we can write the following formula at the horizon
\be
\label{Massthermo}
M^{{\cal H}_t}= \frac{1}{4\pi}\kappa {\cal A} + 2\Omega_H J^{{\cal H}_t} \,.
\ee
Formula (\ref{Massthermo0}), obtained first by Bardeen, Carter and Hawking~\cite{Bardeen1973}, was instrumental to derive the {\it first law} of BH mechanics. In particular, in the absence of matter outside the BH, the formula reduces to the one provided by Smarr~\cite{Smarr1973} which takes the same form as (\ref{Massthermo}).

For instance, in the case of Schwarzschild and Kerr BH's (cf. Appendix \ref{app:KerrQI}) Eq.(\ref{Massthermo}) 
simply reduces to $M^{{\cal H}_t}=M$. 

The temperature of the BH horizon is defined 
as $T_H= \kappa/(2\pi)$, whereas 
the entropy $S_H= {\cal A}/4$.


\section{Numerical Analysis}
\label{sec:numint}

To solve the system of elliptic partial differential equations (\ref{eq:S11})-(\ref{eq:S44}) together with the Klein-Gordon equation (\ref{KGG}), we have used the KADATH library, which, as mentioned above, implements the spectral methods to solve this type of systems. Furthermore, to solve the EKG system, we have included the regularity conditions for the metric potentials $N$, $A$, $B$ and $\beta^{\varphi}$, and for the amplitude of the scalar field $\phi(r, \theta)$ that were described in the Subsection~\ref{sec: BC}. In this work we focus only on solutions corresponding to $n=0$ (i.e. where $\phi$ has no nodes) and $m=1$.

\begin{figure}[h!]
 \centering
    \includegraphics[width=0.45\textwidth]{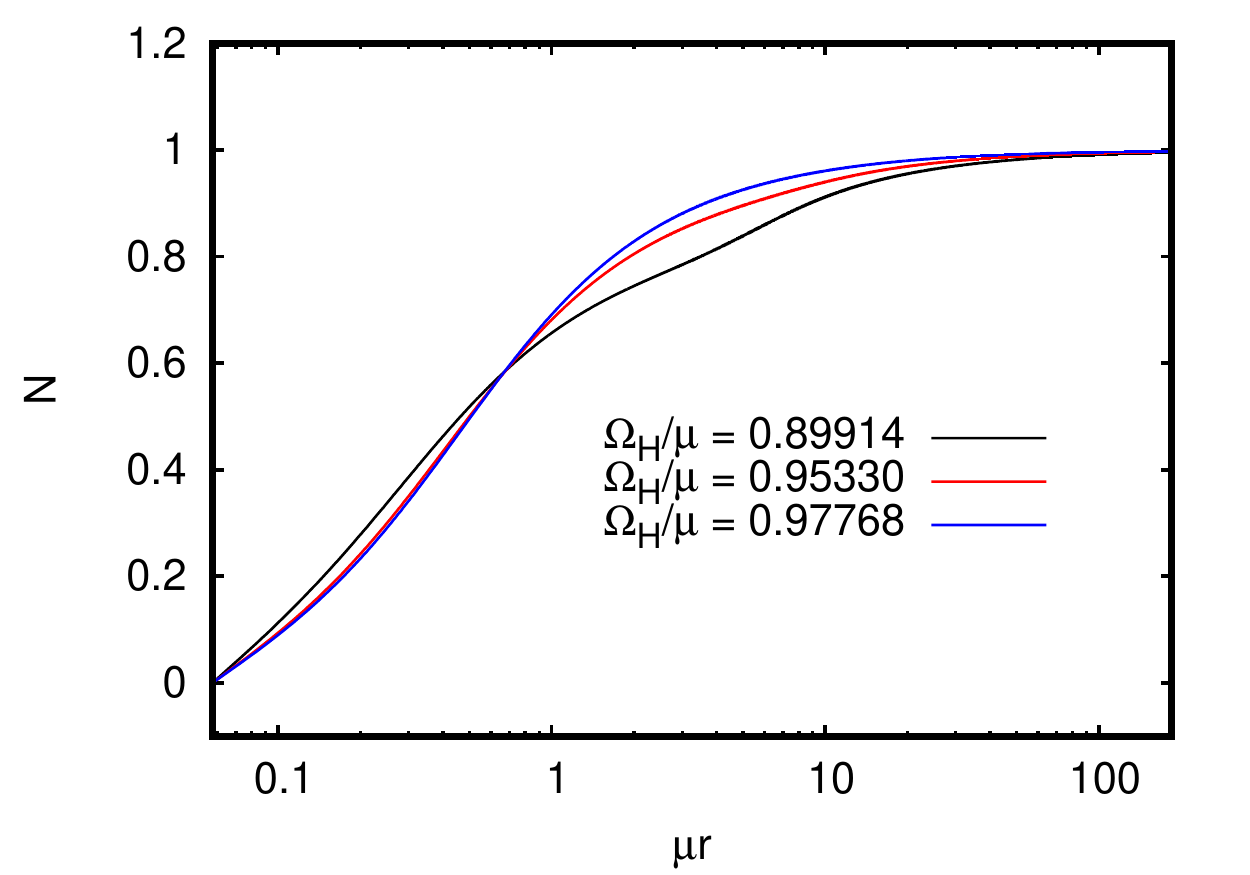}
    \includegraphics[width=0.45\textwidth]{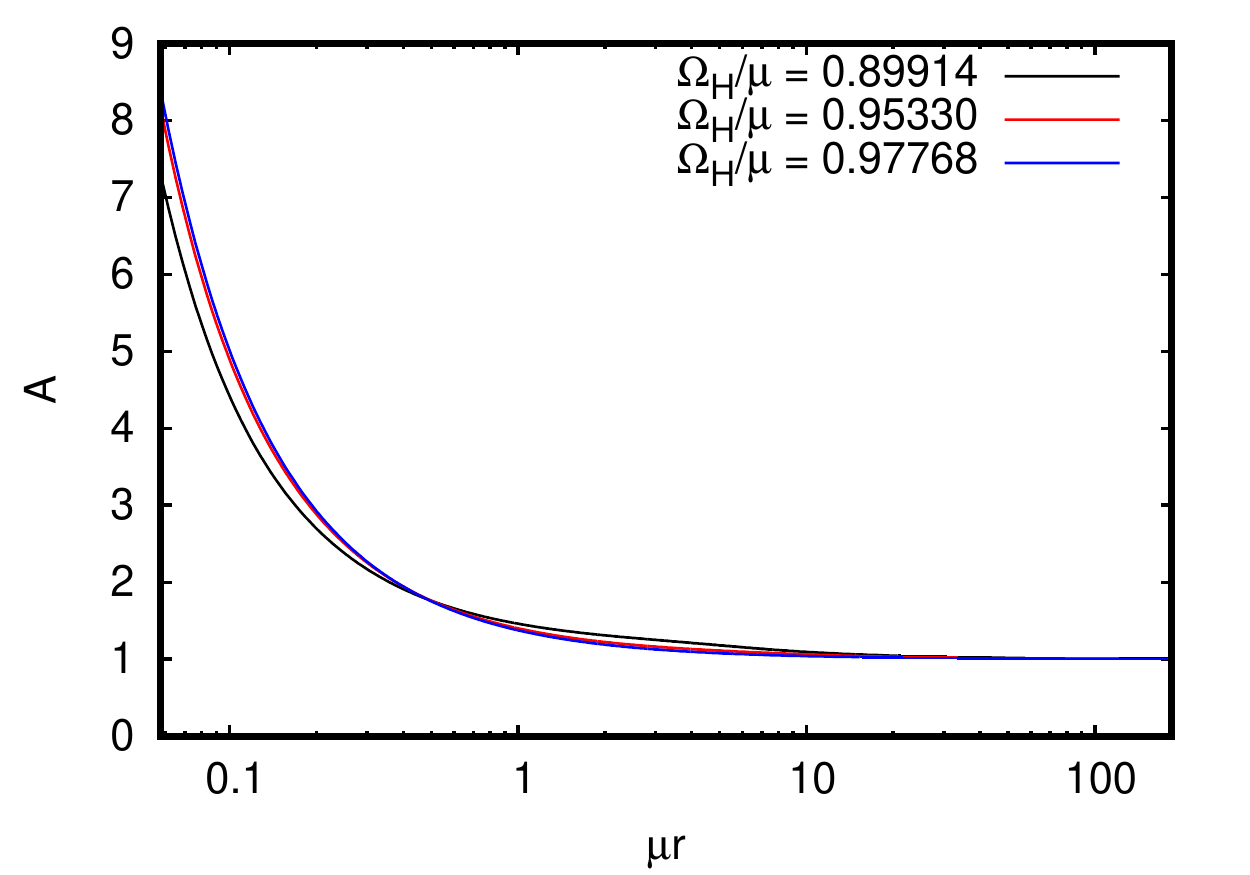}  
 \caption{Solutions for the lapse $N$ and the metric function $A$ (evaluated at the equator $\theta=\pi/2$) associated with the elliptic system Eqs.(\ref{eq:S11})--(\ref{eq:S44}),  with $\mu r_H = 0.057648$ and three different values of $\Omega_H$.}
 \label{fig:CompNA}
\end{figure}

\begin{figure}[h!]
 \centering
    \includegraphics[width=0.45\textwidth]{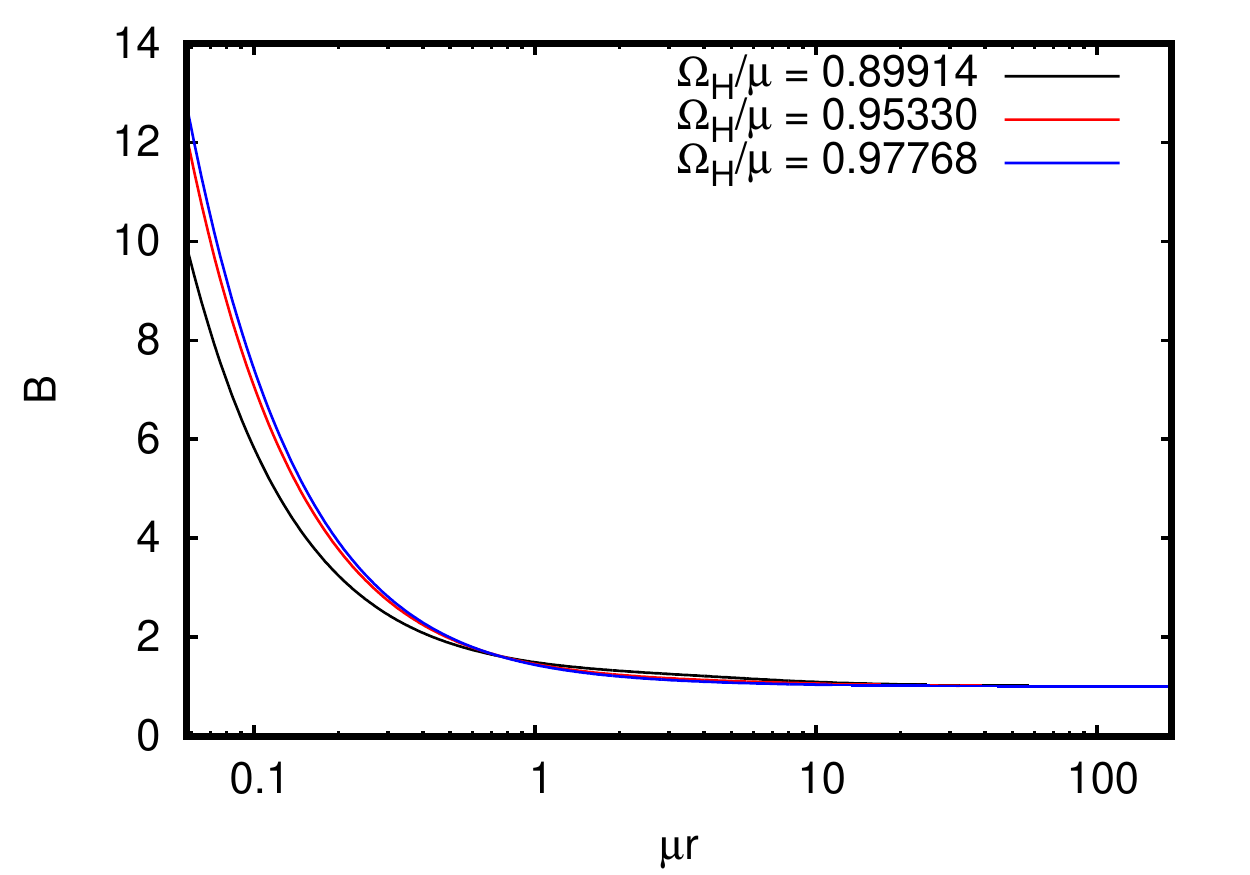}
    \includegraphics[width=0.45\textwidth]{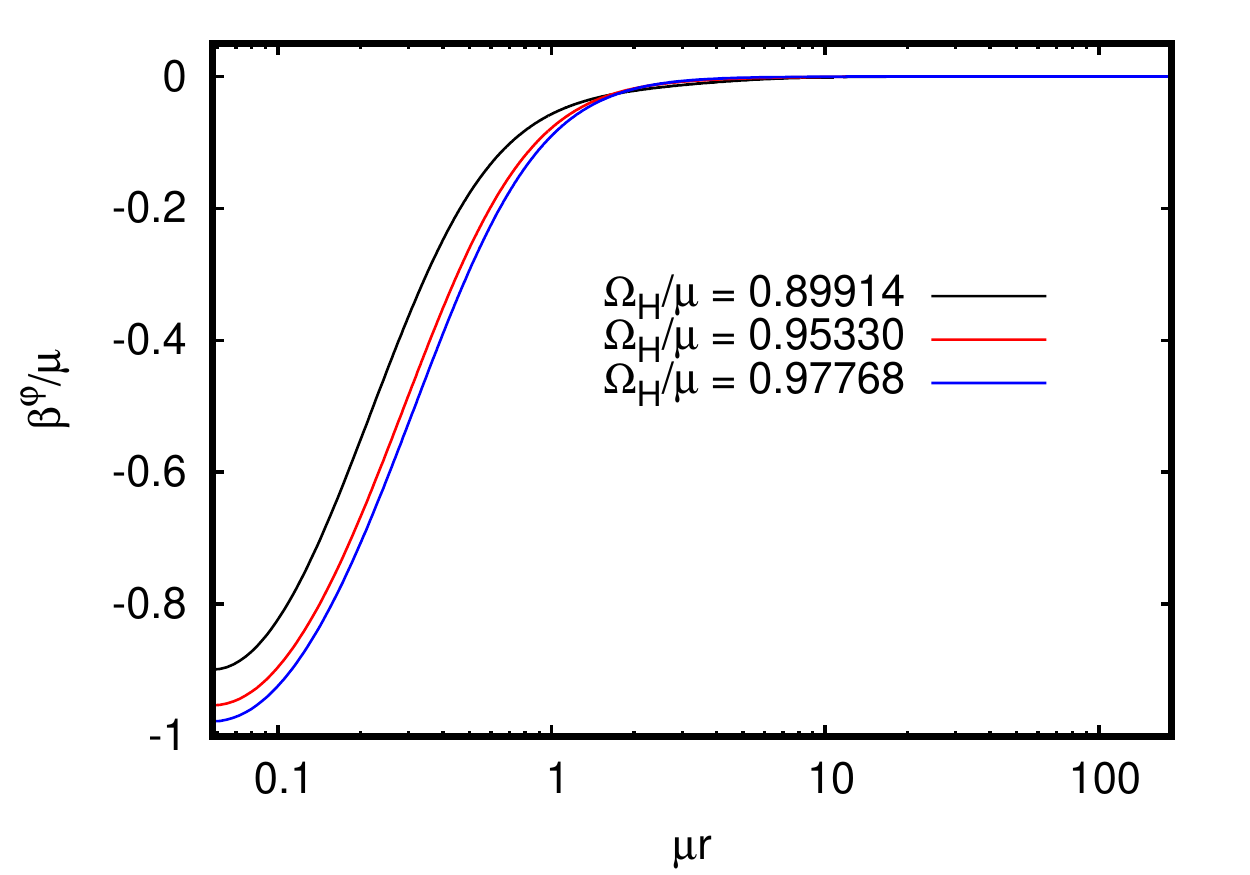} 
 \caption{Solutions for the function $B$ and the shift $\beta^\varphi$ (evaluated at the equator $\theta=\pi/2$) associated with the  elliptic system Eqs. (\ref{eq:S11})--(\ref{eq:S44}) with $\mu r_H = 0.057648$ and three different values of $\Omega_H$.}
 \label{fig:CompBNp}
\end{figure}

Figure~\ref{fig:CompNA} shows the metric potentials $N$ and $A$, while Figure~\ref{fig:CompBNp} depicts $B$ and $\beta^\varphi$, both for black hole solutions with scalar hair by fixing the coordinate at the horizon as $r_H = 0.057648/\mu$ and taking three different values of $\Omega_H$.

\begin{figure}[h!]
\begin{center}
\includegraphics[width=0.45\textwidth]{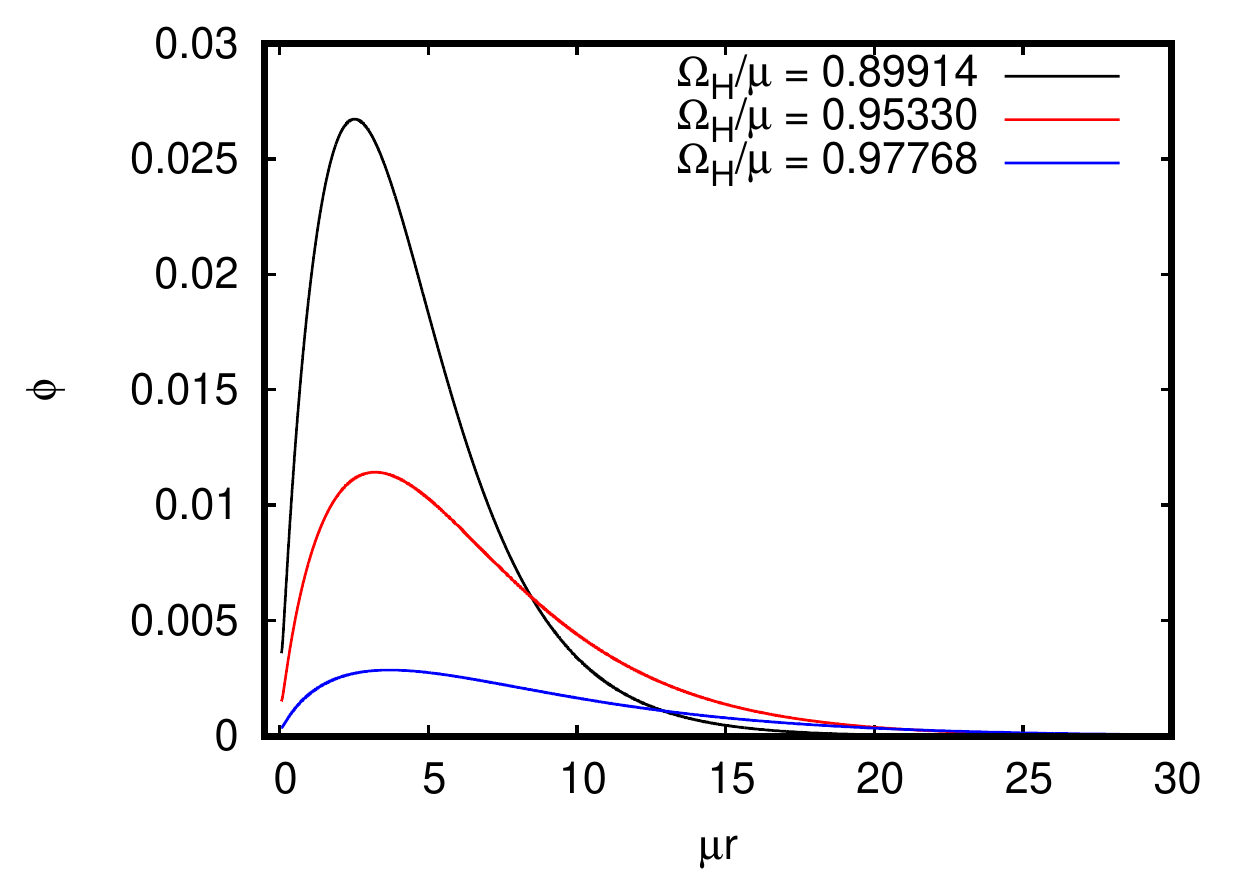}
\caption{Scalar field solutions $\phi(r, \pi/2)$ for a black hole with $\mu r_H = 0.057648$ and three different values of $\Omega_{H}$.}\label{fig:ScalarField1}
\end{center}
\end{figure}

Figure~\ref{fig:ScalarField1} shows the corresponding solutions for the scalar field amplitude $\phi(r, \theta)$ at the equatorial plane ($\theta = \pi/2$). The values of $\phi$ together with the corresponding metric potentials shown in Figures~\ref{fig:CompNA} and \ref{fig:CompBNp} are examples 
of solutions of the full EKG system, and as such, they represent black hole solutions endowed with scalar hair.

We stress that these solutions are regular in the DOC, notably, at the event horizon. Moreover, we appreciate that asymptotically the scalar field vanishes and the metric functions match the Minkowski values ($N\rightarrow 1$, 
$A\rightarrow 1$, $B\rightarrow 1$, and $\beta^\varphi \rightarrow 0$).

\begin{figure}[h!]
\begin{center}
\includegraphics[width=0.45\textwidth]{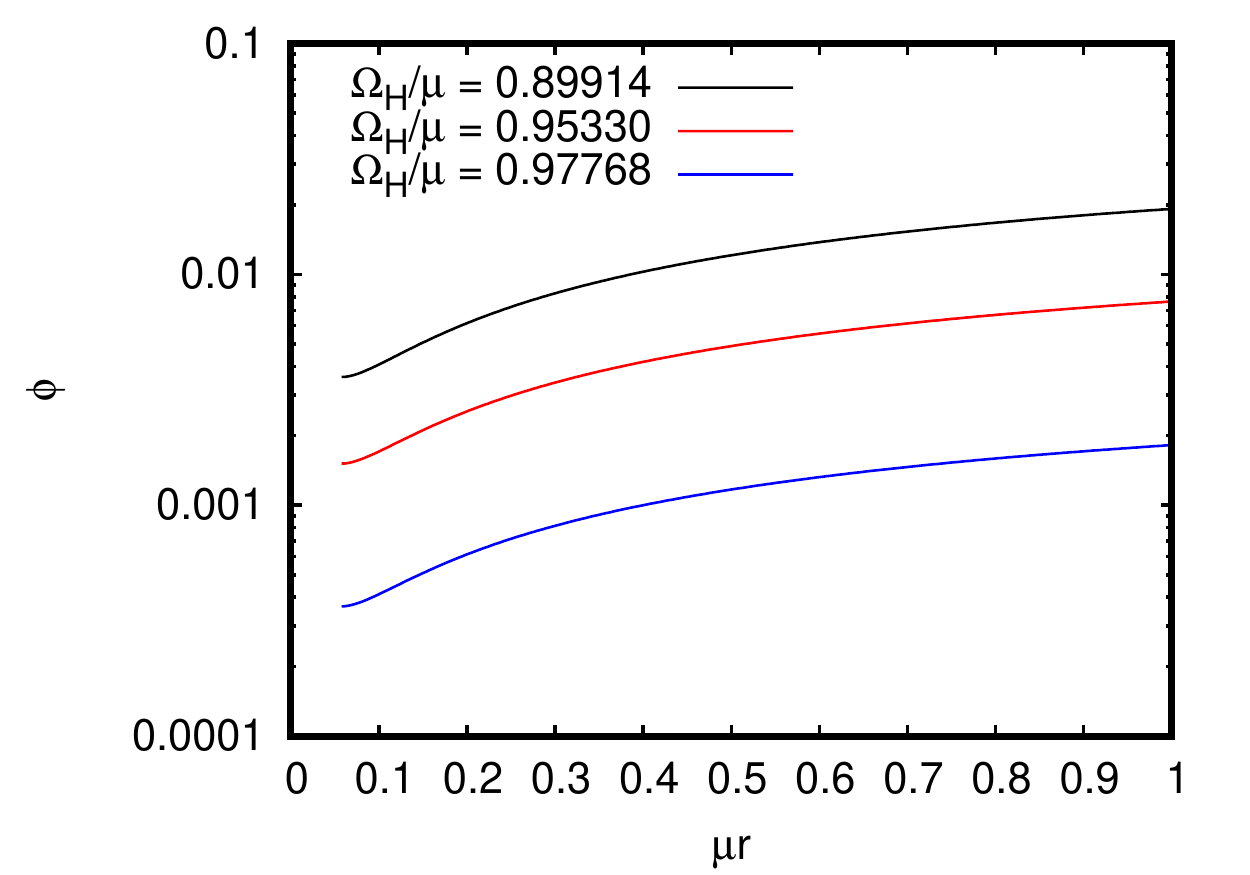}
\caption{Scalar field solutions $\phi(r, \pi/2)$ associated 
with Figure~\ref{fig:ScalarField1}, but depicted close to the horizon located at $\mu r_H =0.057648$.}\label{fig:ScalarField3}
\end{center}
\end{figure}

\begin{figure}[h!]
\begin{center}
\includegraphics[width=0.45\textwidth]{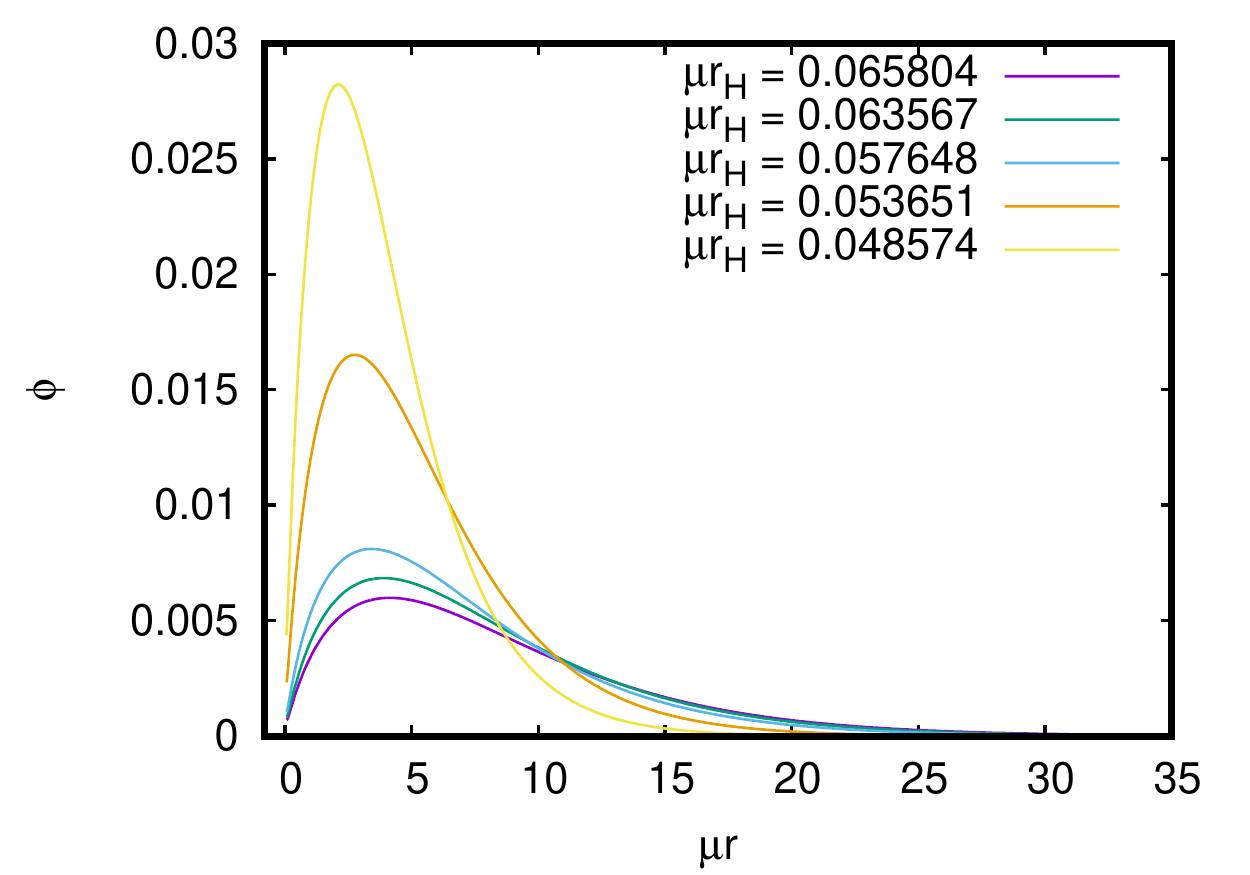}
\caption{Scalar field solution $\phi(r, \pi/2)$ for different values of $r_H$ and $\Omega_{H}$ (the latter are not shown).}\label{fig:ScalarField2}
\end{center}
\end{figure}

Figure~\ref{fig:ScalarField3} shows  the scalar field near the horizon ($r_H\approx 0.058/\mu$)
in order to appreciate closely the regularity conditions there.
Figure~\ref{fig:ScalarField2} displays a family of scalar field solutions $\phi$ at the equatorial plane taking different values 
for $r_H$ and $\Omega_H$.

\begin{figure}[h!]
 \centering
    \includegraphics[width=0.45\textwidth]{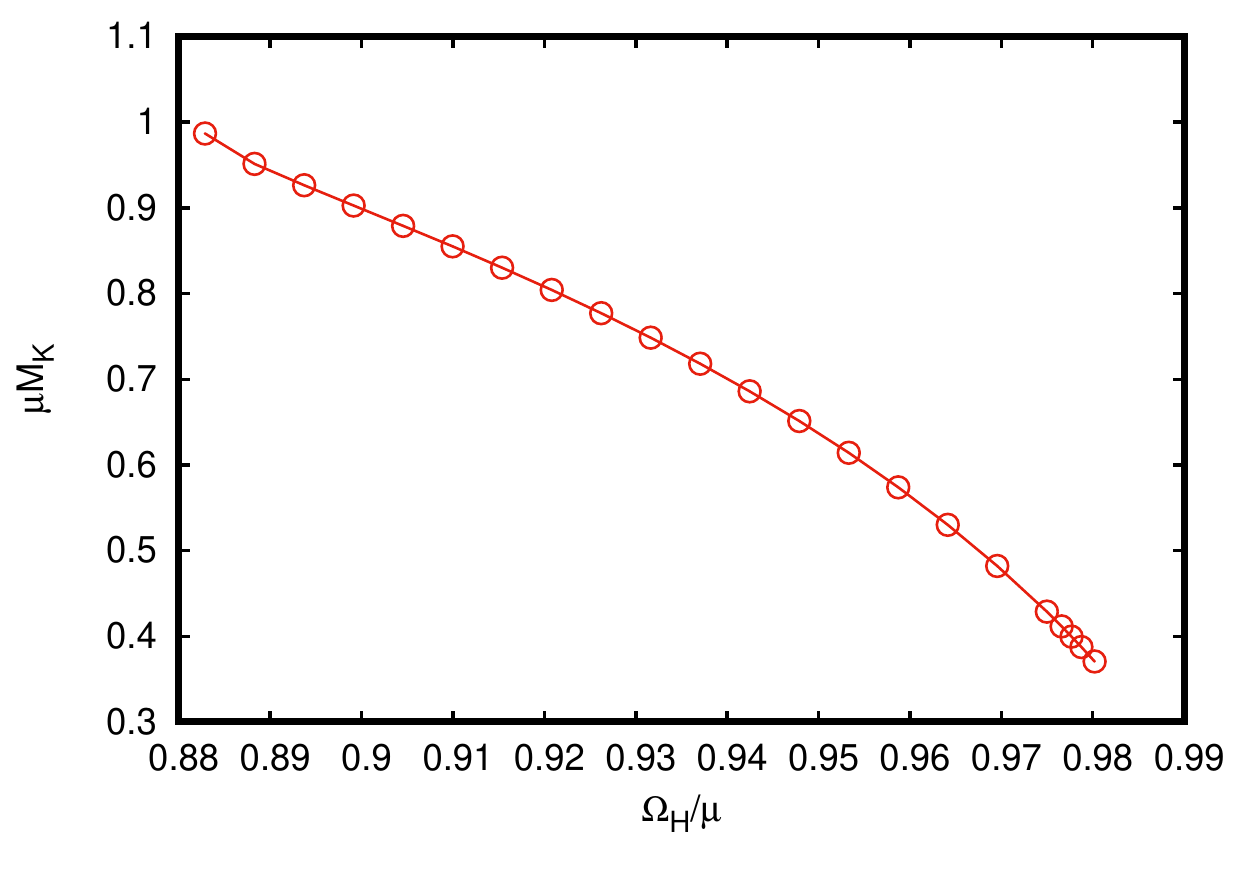}
    \includegraphics[width=0.45\textwidth]{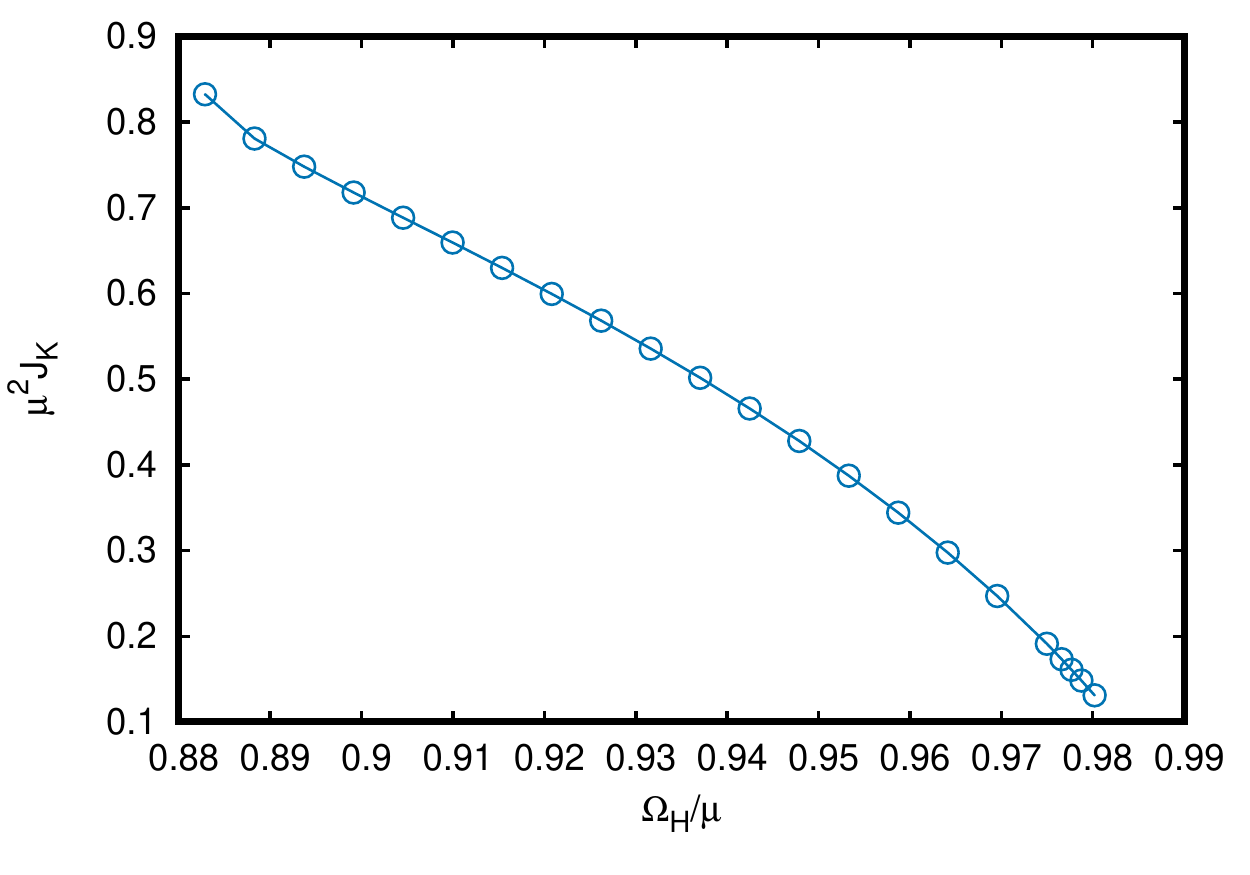}  
 \caption{Komar mass (top panel) and Komar angular momentum (bottom panel) as a function of $\Omega_H$ for hairy BH solutions with $\mu r_H = 0.057648$ fixed.}
 \label{fig:KomarMassAngular}
\end{figure}

Figure~\ref{fig:KomarMassAngular} depicts the Komar mass (\ref{Kmass1}) and the Komar angular momentum (\ref{JKomar}) for a sequence of hairy black hole solutions with different values 
of $\Omega_H$, but keeping fixed the horizon at $r_H = 0.057648/\mu$. 
Note that these global quantities are very sensitive to small variations of the angular velocity of the black hole. The larger the angular velocity, the lower the 
amplitude of the scalar field results (cf. Figure \ref{fig:ScalarField1}), and therefore the Komar quantities become smaller as the scalar field contribution decreases in the DOC.

As a measure of the contribution of the scalar field $\Psi$ to the black hole solution, it is convenient to define the dimensionless quantity $q$, in terms of the Noether charge $\mathcal{Q}$ and the total angular momentum $J_{\rm K}$, as follows
\be
q \equiv \frac{m\mathcal{Q}}{J_{\rm K}}\;.
\ee
From the expressions given by Eqs.~(\ref{KJH2}) and (\ref{JKomarS}) it is easy to see that for a Kerr black hole $q = 0$, since in this scenario the Noether charge vanishes identically, while for a boson star (no horizon is present) $q = 1$ because in this case $J_{\rm K}= J^{\Sigma_t} $ and $J^{\Sigma_t} = m\mathcal{Q}$
(i.e. $M^{{\mathcal H}_t}=0= J^{{\mathcal H}_t}$). By continuity we may conclude that the quantity $q$ takes values in the interval $(0, 1)$ when we consider a black hole solution with scalar hair.

\begin{figure}[h!]
 \centering
    \includegraphics[width=0.45\textwidth]{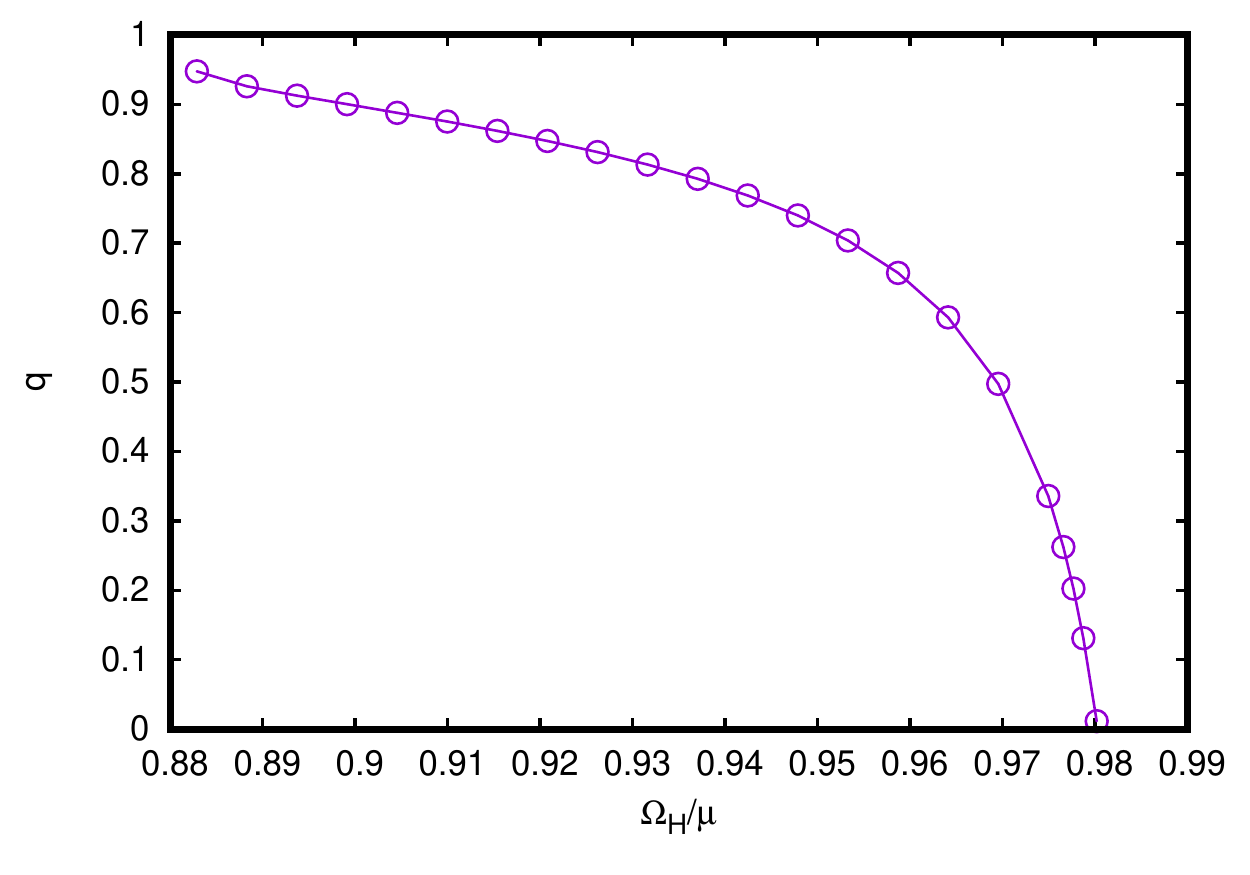}
    \includegraphics[width=0.45\textwidth]{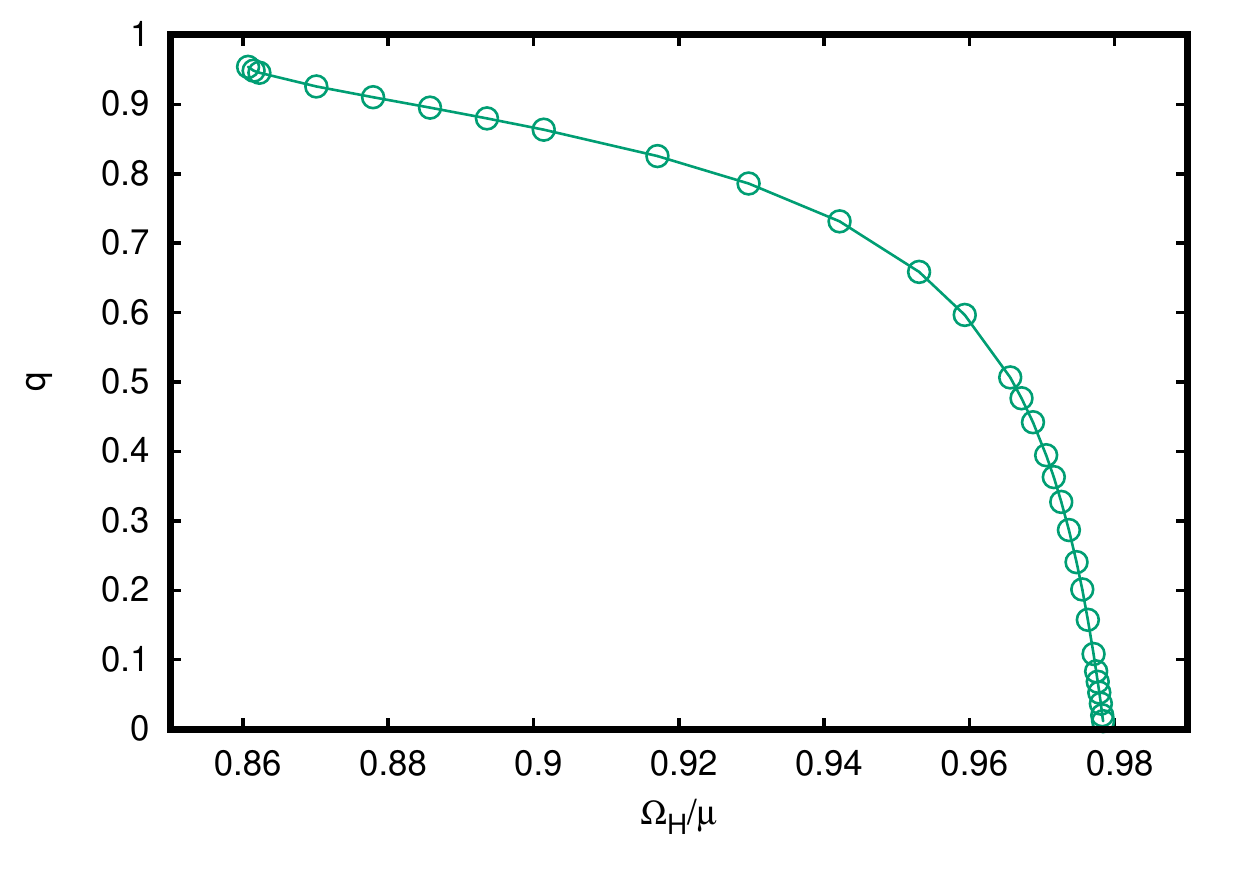}  
 \caption{The dimensionless quantity $q = m\mathcal{Q}/J_{\rm K}$ as a function of $\Omega_H$ for hairy black holes with event horizon at $\mu r_H = 0.057648$ (top panel) and $\mu r_H = 0.053651$ (bottom panel), respectively.}
 \label{fig:qNoether}
\end{figure}

Figure~\ref{fig:qNoether} depicts the quantity $q$ as a function of $\Omega _H$ for two different sequences of solutions of the Einstein-Klein-Gordon system corresponding to hairy black holes with event horizons fixed at $r_H = 0.057648/\mu$ (top panel) and $ r_H = 0.053651/\mu$ (bottom panel).
In both plots we observe how the quantity $q$ approaches 1 as the angular velocity of the black hole with scalar hair moves away from the value of $\Omega_H$ corresponding to a Kerr black hole from which the sequence of solutions were generated. This result is a consequence of the fact that the value of the amplitude of the scalar field $\phi$ increases as $\Omega_H$ decreases from the initial guess configuration.

\begin{figure}[h!]
 \centering
    \includegraphics[width=0.45\textwidth]{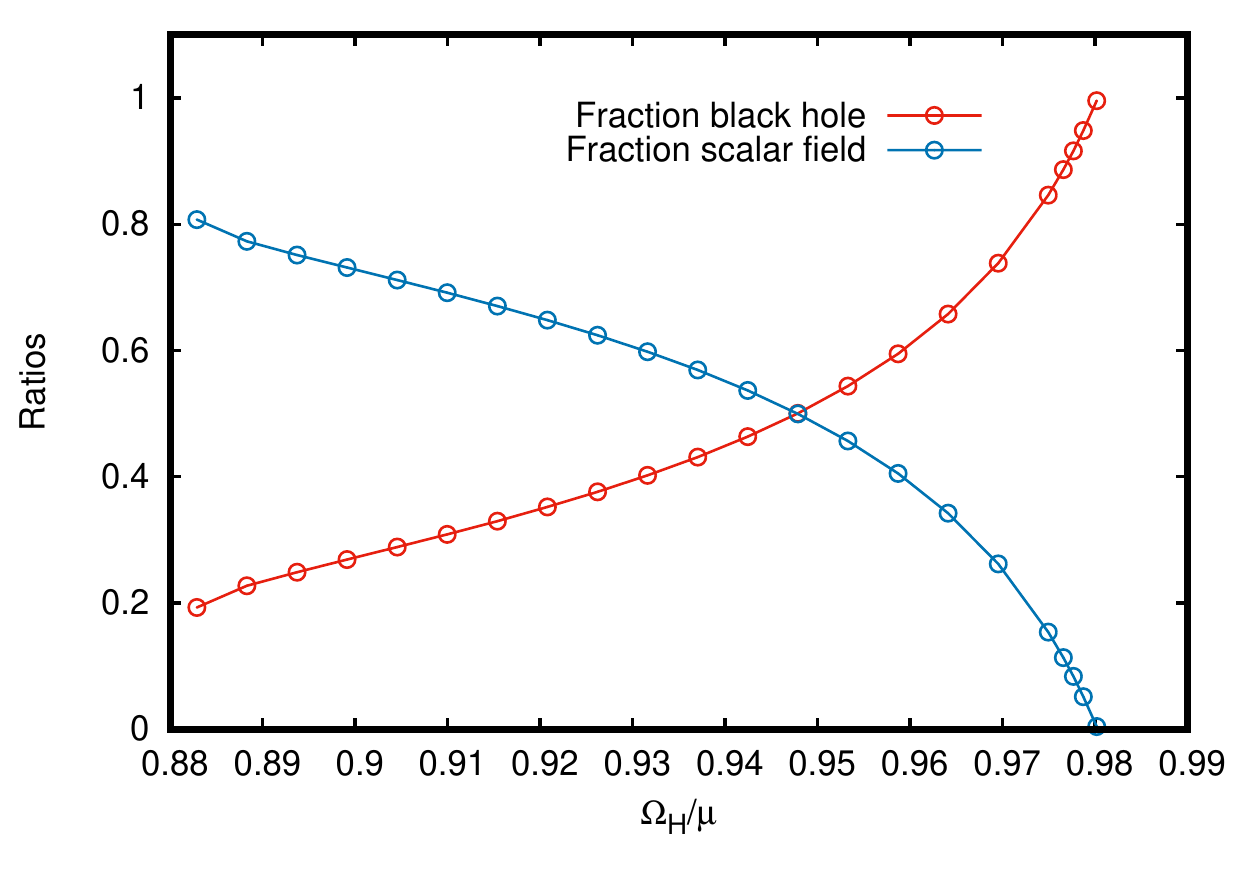}
    \includegraphics[width=0.45\textwidth]{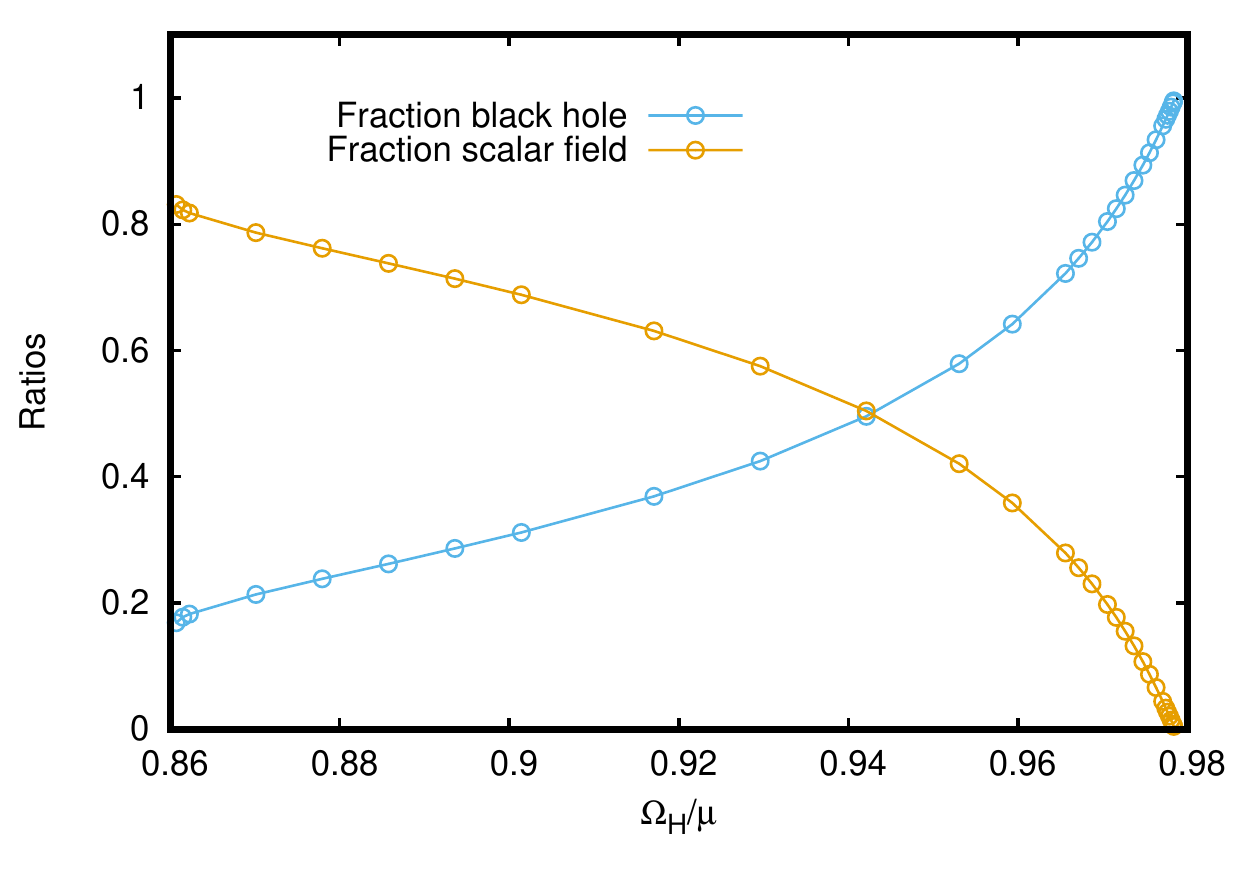}  
 \caption{Relative mass contributions $M^{\mathcal {H}_t}/M_{\rm ADM}$ and $M^{\Sigma_t}/M_{\rm ADM}$ as a function of $\Omega_H$. In these plots the event horizon is located at $\mu r_H = 0.057648$ (top panel) and $\mu r_H = 0.053651$ (bottom panel), respectively. }
 \label{fig:ratios}
\end{figure}

Figure~\ref{fig:ratios} shows the contribution of the 
black hole mass at the horizon, $M^{\mathcal {H}_t}/M_{\rm ADM}$, and the corresponding 
contribution of the scalar field mass
 $M^{\Sigma_t}/M_{\rm ADM}$, relative to the 
 total mass, for the family of hairy black holes associated with the values $\mu r_H = 0.057648$ and $\mu r_H = 0.053651$, respectively, as a function of $\Omega_H$. 
 When the former  contribution is lower, the latter is larger and vice versa because 
 both fractions add to one according to Eq.(\ref{Kmass2}), where $M_{\rm K}\equiv M_{\rm ADM}$. 

\begin{figure}[h!]
 \centering
    \includegraphics[width=0.45\textwidth]{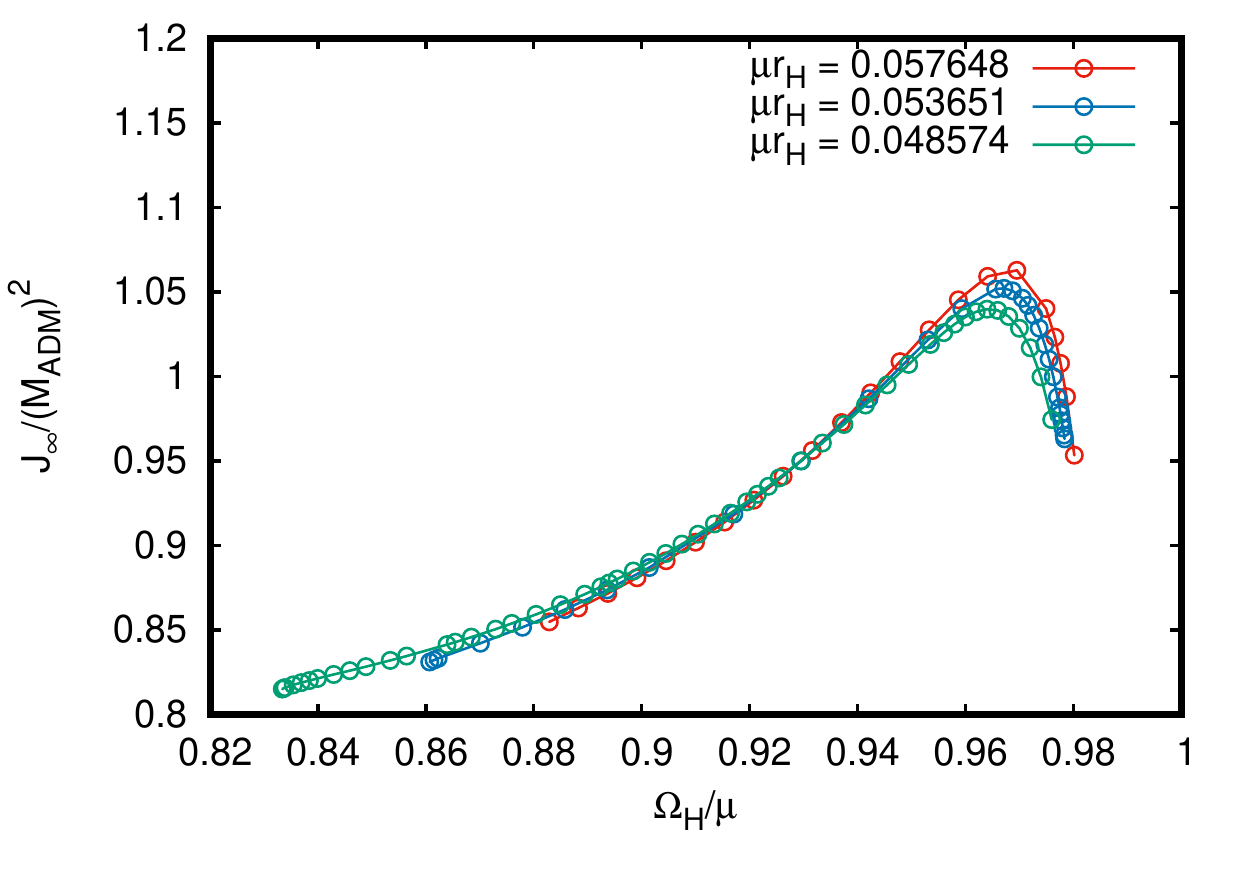}  
 \caption{The dimensionless parameter $J_{\infty}/M_{\rm ADM}^2$ as a function of $\Omega_H$ for hairy black holes with three different horizon values.}
 \label{fig:ParameterJM2}
\end{figure}

Figure~\ref{fig:ParameterJM2} depicts the dimensionless quantity $J_{\infty}/M^2_{\rm ADM}$ as a function of $\Omega_H$ for three different sequences of hairy black holes associated with the event horizons located at $\mu r_H = 0.057648$, $\mu r_H = 0.053651$ and $\mu r_H = 0.048574$, respectively. We appreciate 
that this quantity is not bounded from above by unity, unlike the Kerr black hole where $0 \leq |a/M| \leq 1$ ($a=J/M$ and $M_{\rm ADM}=M$ for Kerr)
 (cf. the Tables in  Appendix~\ref{app:HBHT}). This behavior
 is an indicator of how different the 
 hairy configurations can be relative to the vacuum Kerr solution.

Figure~\ref{fig:surfacegravityBHsSH} shows the surface gravity $\kappa$ (see Appendix~\ref{app:surgrav}) for the same sequence of solutions that appear in Figure~\ref {fig:ParameterJM2}.
\begin{figure}[h!]
 \centering
    \includegraphics[width=0.45\textwidth]{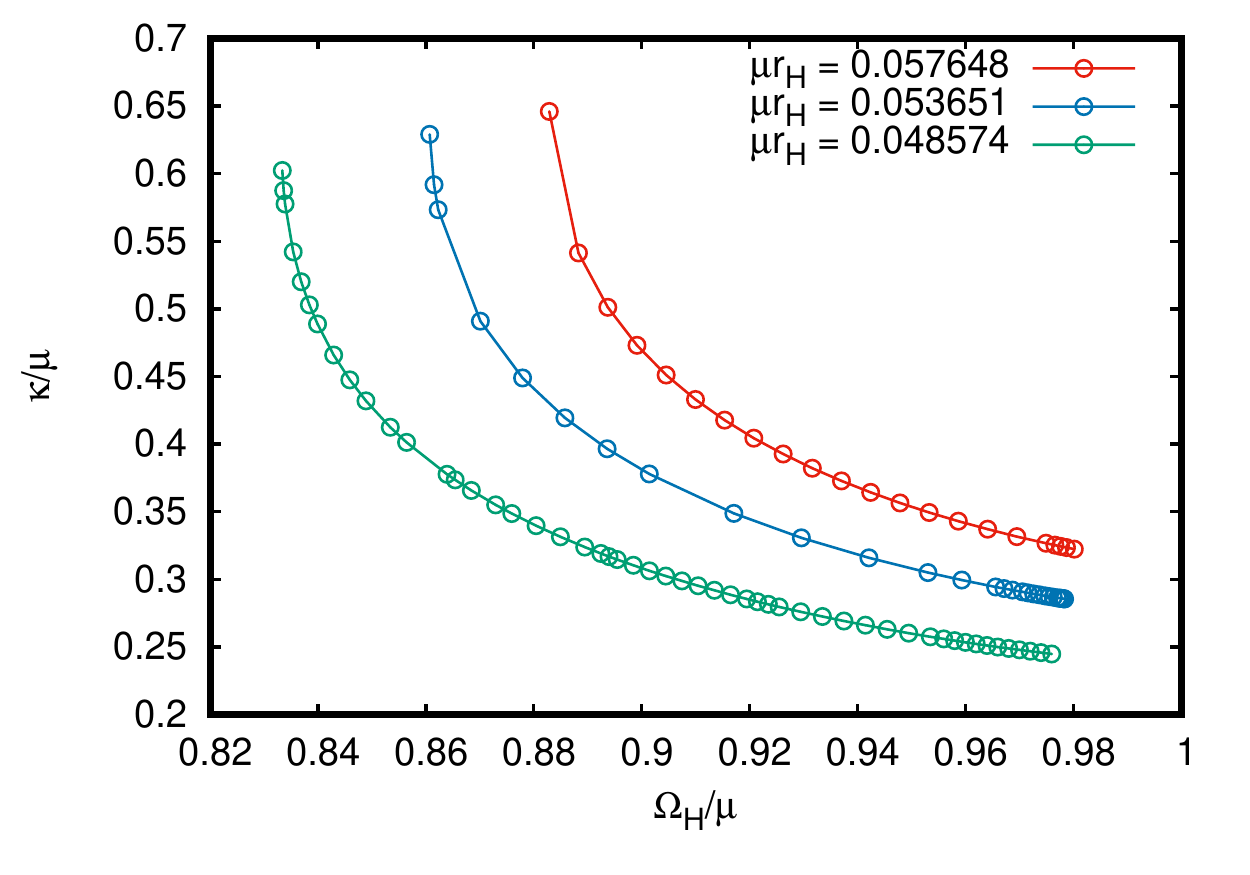}  
 \caption{Surface gravity $\kappa$ as a function of $\Omega_H$ for hairy BH's with three different horizon values.}
 \label{fig:surfacegravityBHsSH}
\end{figure} 

\begin{figure}[h!]
 \centering
    \includegraphics[width=0.45\textwidth]{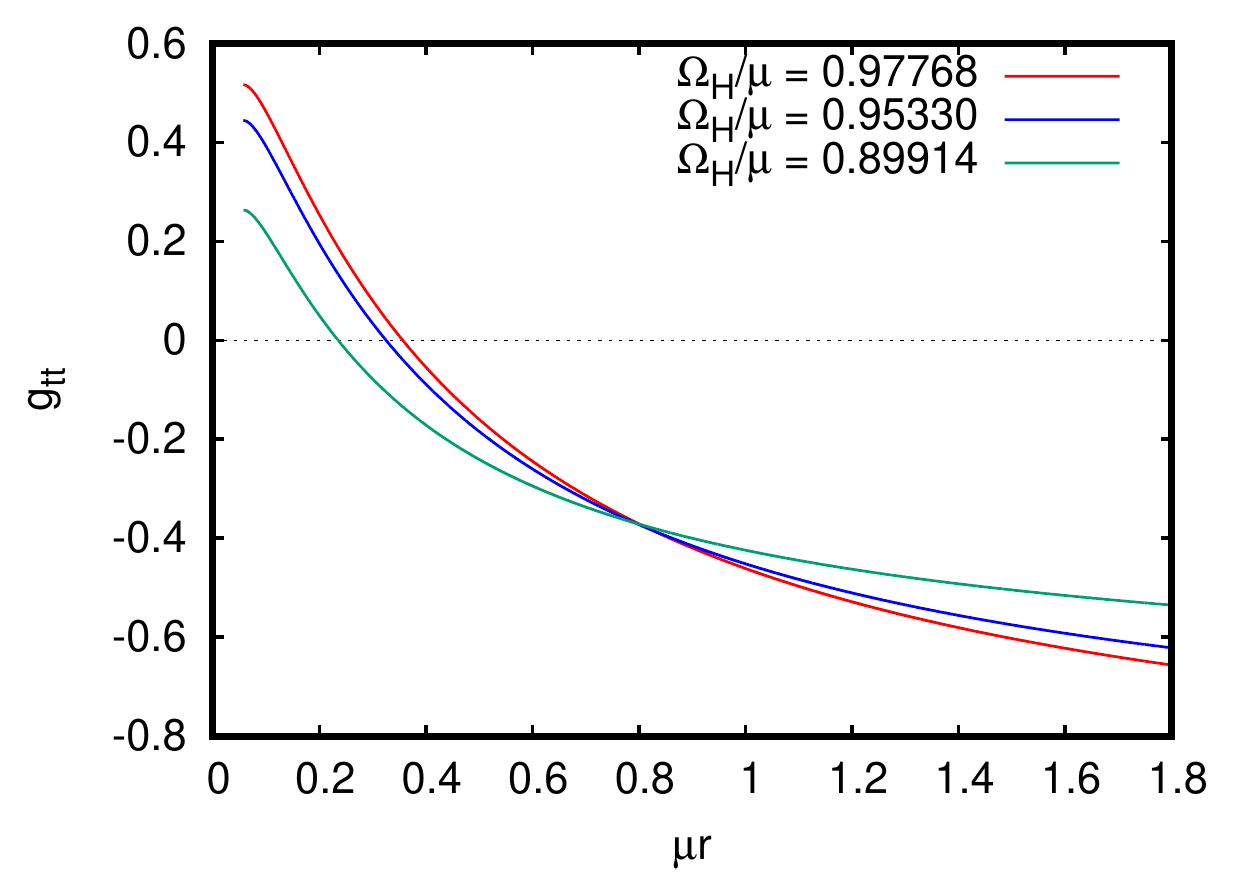}  
 \caption{Localization of ergosurfaces on the equatorial plane ($\theta = \pi/2$) for three hairy black hole with fixed $\mu r_H = 0.057648$.}
 \label{fig:Ergosphere}
\end{figure}

Figure~\ref{fig:Ergosphere} plots the metric component $g_{tt} = -N^2 + B^2(\beta^{\varphi})^2r^2\sin^2\theta$ as a function of the coordinate $r$ (on the equatorial plane), in order to locate the ergosurfaces associated with the regions where $g_{tt} =0$. Here we have considered three hairy black holes with fixed $\mu r_H = 0.057648$ but for three different values of the angular velocity $\Omega_H$, as shown in the figure.

\begin{figure}[h!]
 \centering
    \includegraphics[width=0.45\textwidth]{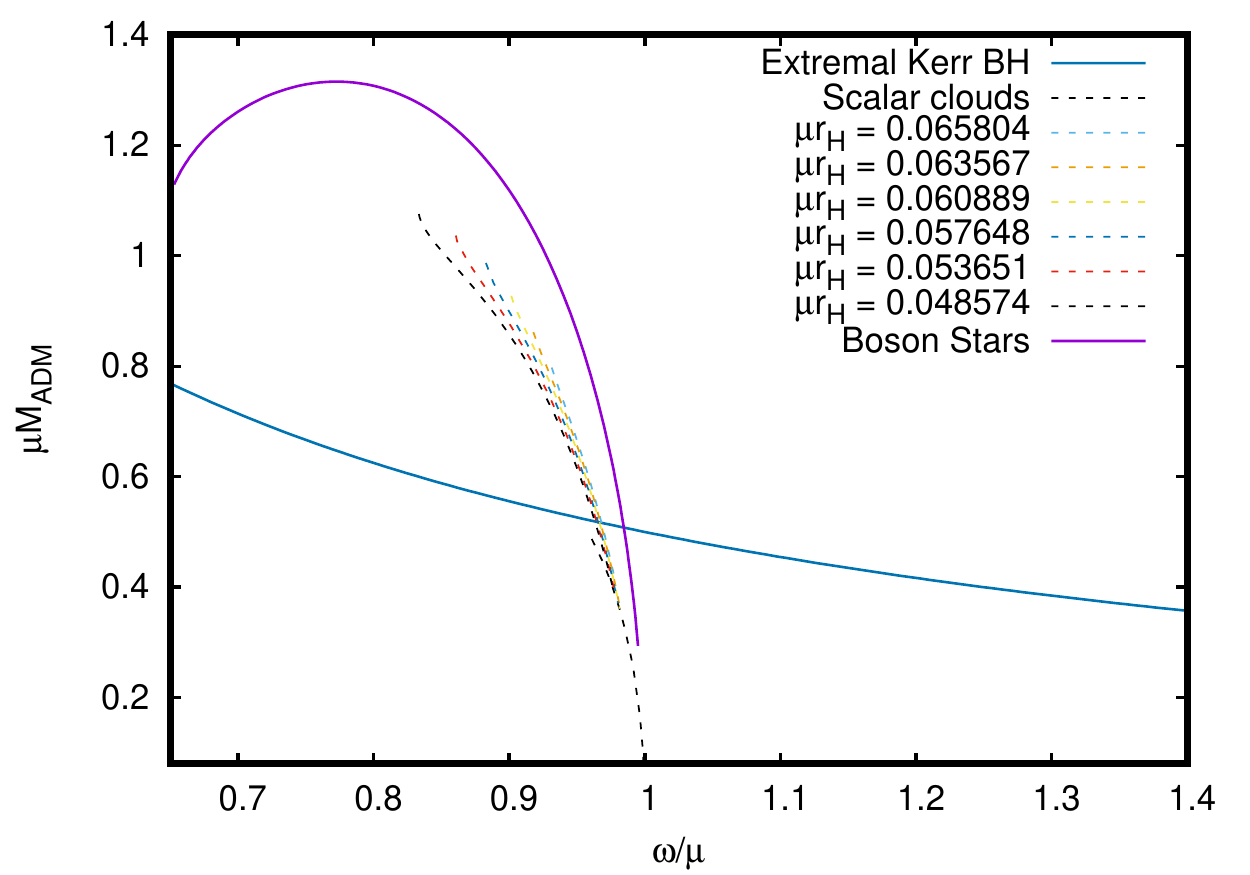}
 \caption{Existence lines for hairy black holes -- colored dashed lines --(diagram $M_{\rm ADM}$ vs $\omega$). 
 Each sequence has a fixed $r_H$, and six sequences are computed. The solid (magenta) line is associated with rotating boson stars, while the (blue) solid line corresponds to extremal Kerr BH's
 ($M=1/2\Omega_H$). The (black) dashed line that starts close to the values $\mu M=0.1$, $\omega/\mu=1$ and ends close to the extremal (blue) solid line is the existence line for {\it boson clouds} in the background of Kerr BH's. Here $\omega=\Omega_H$ ($m=1$) for the hairy solutions and the boson clouds.}
 \label{fig:ExistenceLines}
\end{figure}

Figure~\ref{fig:ExistenceLines} depicts the existence lines (diagram $M_{\rm ADM}$ vs $\omega$) associated with the hairy black holes solutions when solving the full EKG system for a fixed $r_H$ (we consider six different sequences)
and varying the value $\Omega_H$. The values of $r_H$ are shown in the figure for each sequence of solutions, and as stressed before, all of them correspond to the integers $n = 0$ and $m = 1$. The blue solid line corresponds to the (vacuum) extremal Kerr solution $|a| = M$, while the magenta solid line represents configurations associated with boson stars with $m = 1$. The latter were computed with the spectral code used in Ref.~\cite{Grandclement2014}. The black dashed line corresponds to the scalar cloud solutions in a Kerr spacetime in QI coordinates with $n = 0$ and $m =l= 1$ 
(see the Appendix \ref{app:ScalarClouds}).

\begin{figure}[h!]
 \centering
    \includegraphics[width=0.45\textwidth]{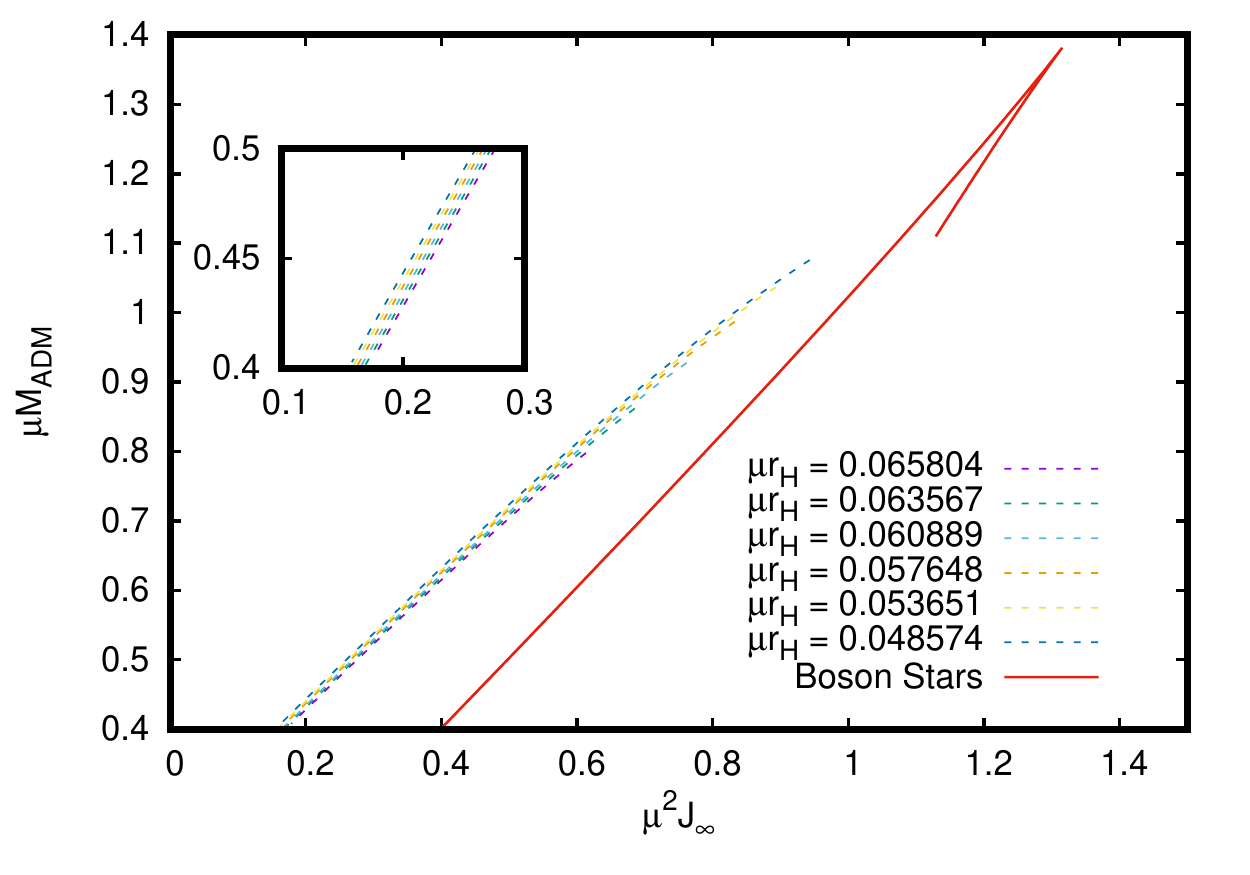}
 \caption{Existence lines for hairy black holes in a diagram $M_{\rm ADM}$ vs $J_{\infty}$.}
 \label{fig:MvsJ}
\end{figure}

Figure~\ref{fig:MvsJ} shows the existence lines in a diagram $M_{\rm ADM}$ vs $J_{\infty}$ for the same sequence of solutions that appear in Fig.~\ref{fig:ExistenceLines}.

\begin{figure}[h!]
 \centering
    \includegraphics[width=0.45\textwidth]{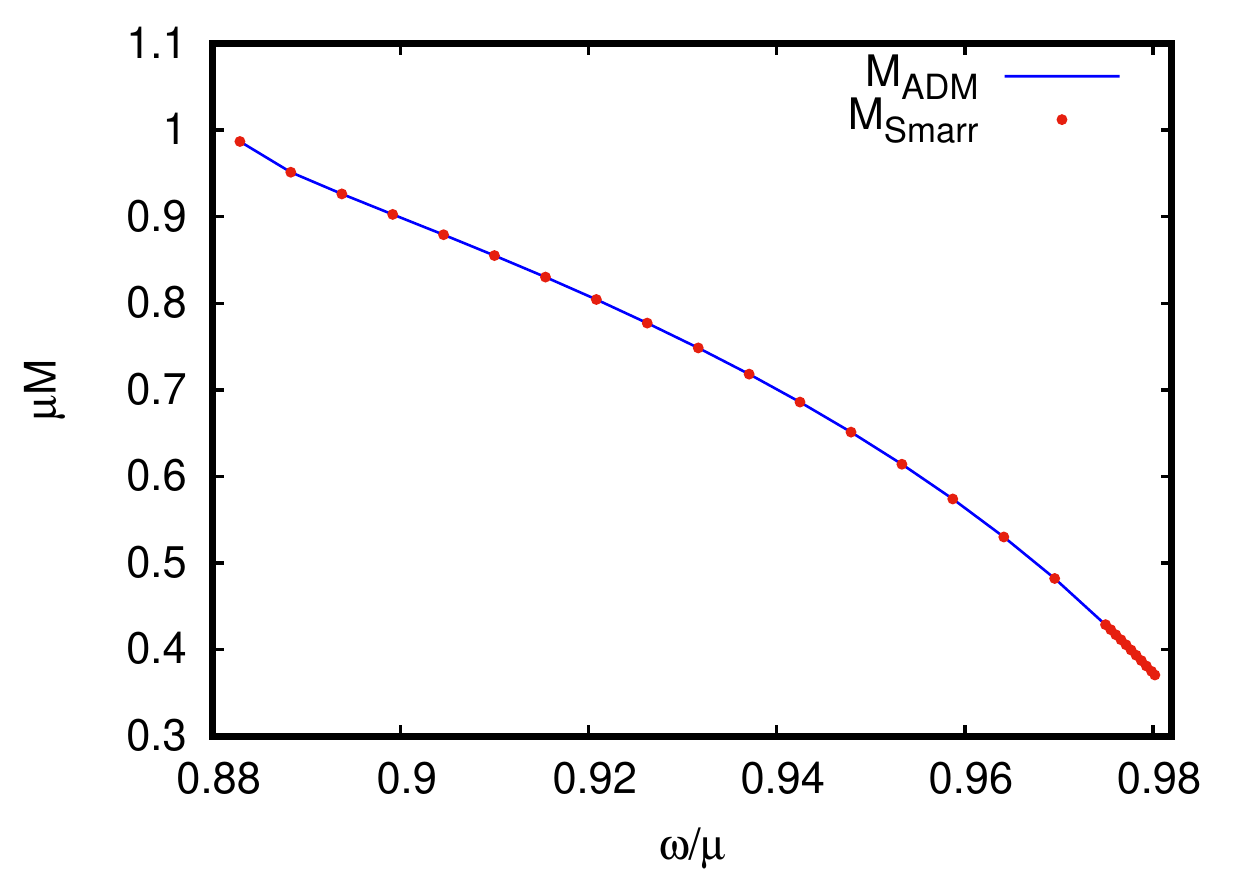}
 \caption{Comparison between the ADM mass and the (Smarr) mass obtained from Eq.~$(\ref{Massthermo0})$ for hair BH's with  fixed $\mu r_H = 0.057648$.}
 \label{fig:MADMvsMSmarr}
\end{figure}

In Figure~\ref{fig:MADMvsMSmarr} we compare the ADM mass (\ref{ADMmass}) with the value obtained using Eq.~(\ref{Massthermo0}), which involves thermodynamic quantities associated with the BH's event horizon. This is for a sequence of hairy black holes with $\mu r_H = 0.057648$ and different values of $\Omega_H$. The relative error between both quantities is $\sim 10^{-6}$. The Table~\ref{tab:Thermodynamics1} displays several values of the thermodynamic quantities $\{T_H, A_H, S_H\}$ associated with the hairy BH's.

\begin{figure}[h!]
 \centering
    \includegraphics[width=0.45\textwidth]{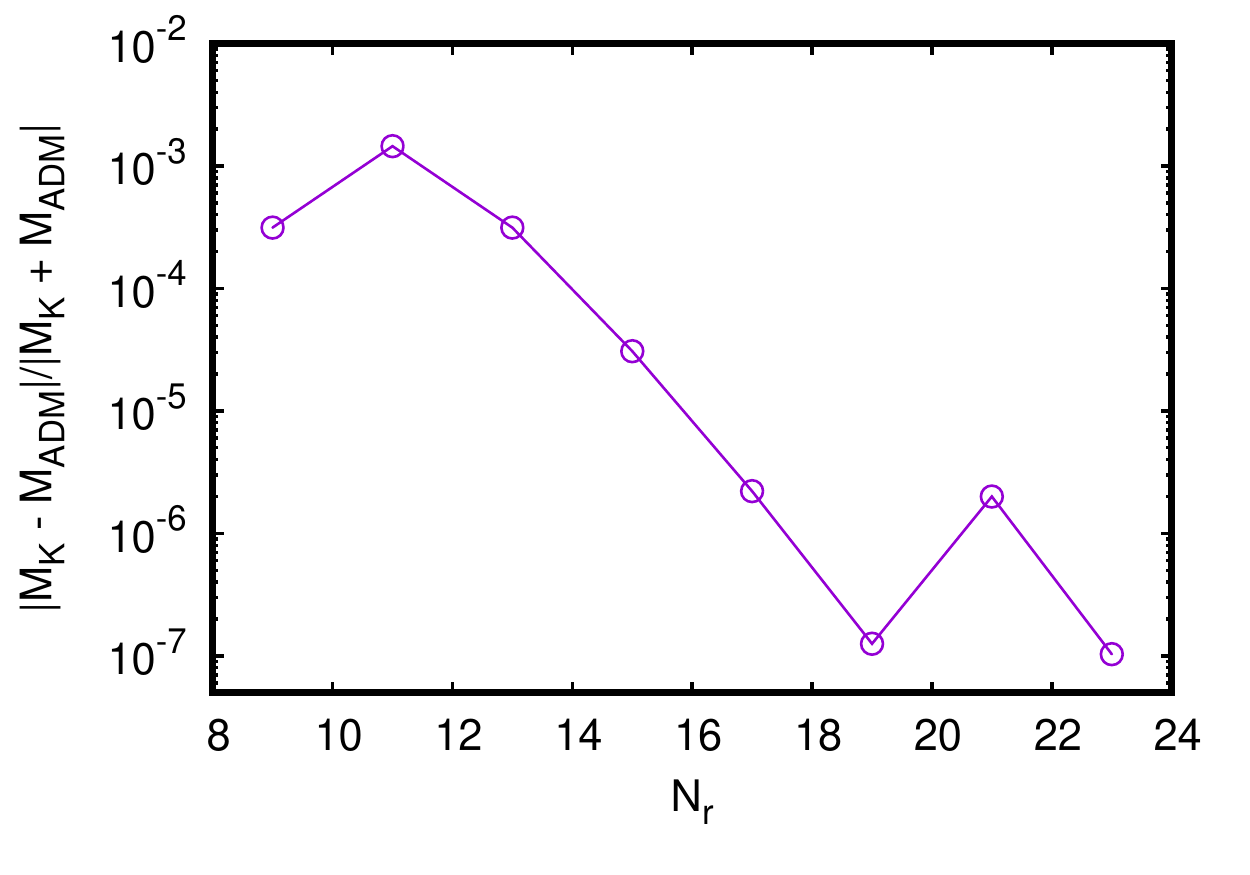}
    \includegraphics[width=0.45\textwidth]{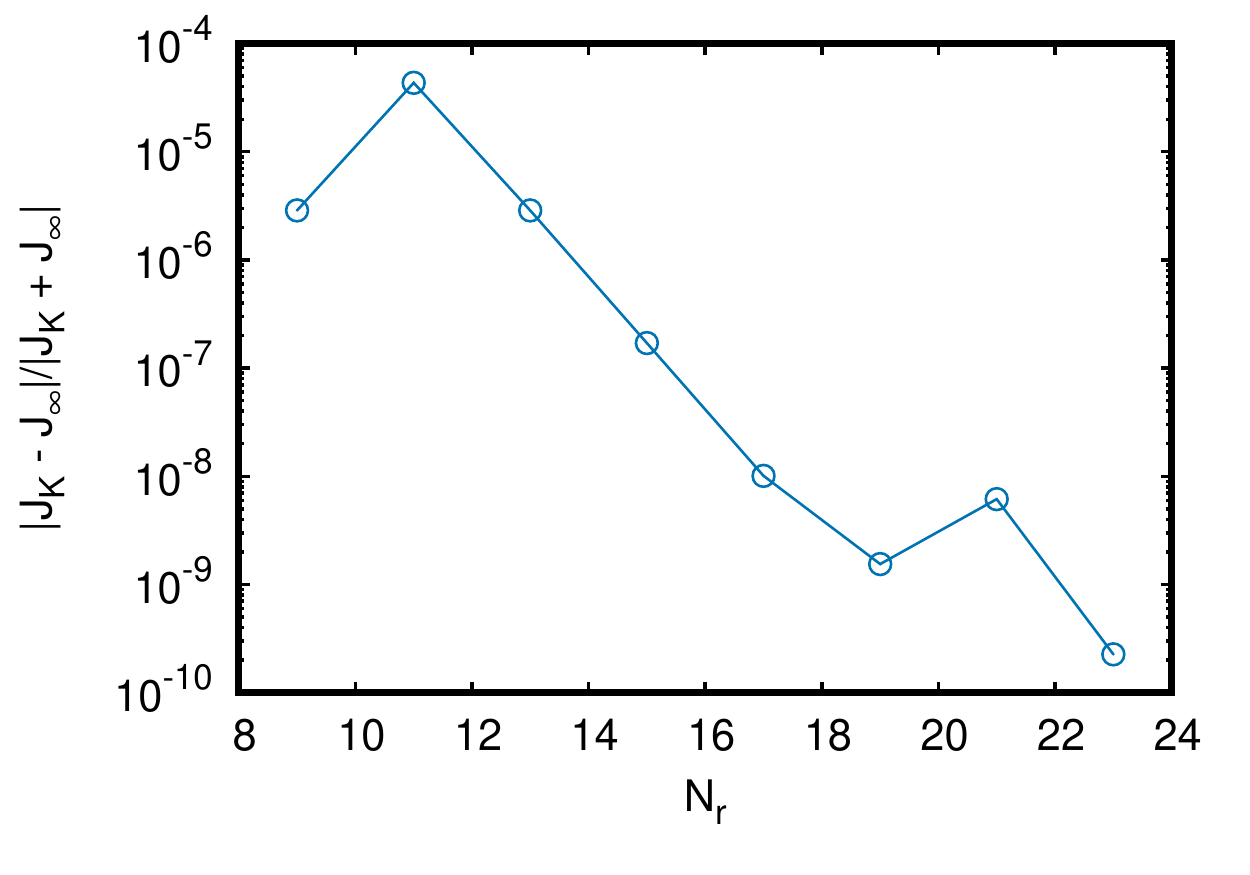}
 \caption{Error indicators between the Komar and ADM masses and
 between expressions (\ref{JKomar2}) ($J_{\rm K}$) and (\ref{JADM}) ($J_\infty$) of the total angular momentum associated with a hairy black hole with event horizon at $\mu r_H = 0{.}057648$ and angular velocity $\Omega_H/\mu = 0{.}88289$, as a function of the number of radial coefficients in the spectral expansion.}
 \label{fig:resolEKG1}
\end{figure}

\begin{figure}[h!]
 \centering
    \includegraphics[width=0.45\textwidth]{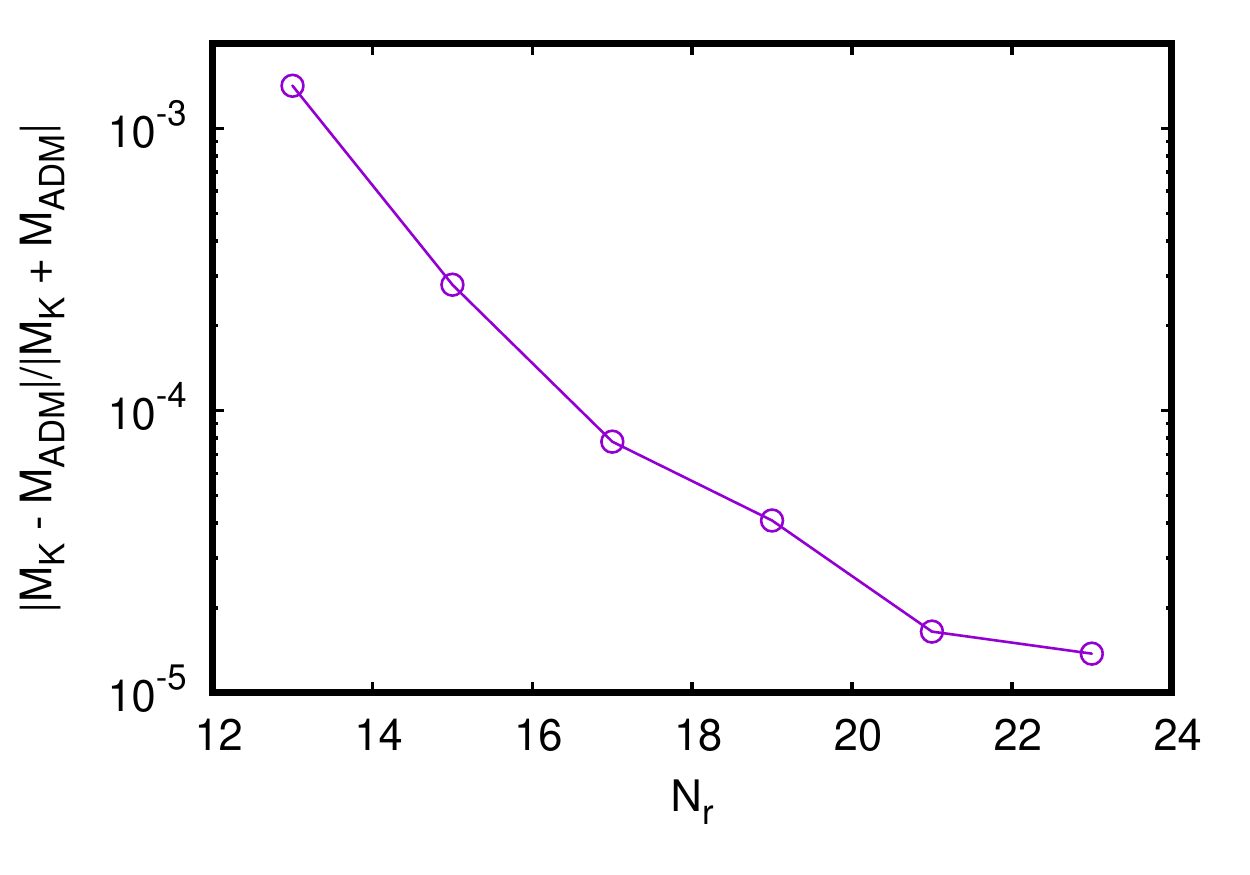}
    \includegraphics[width=0.45\textwidth]{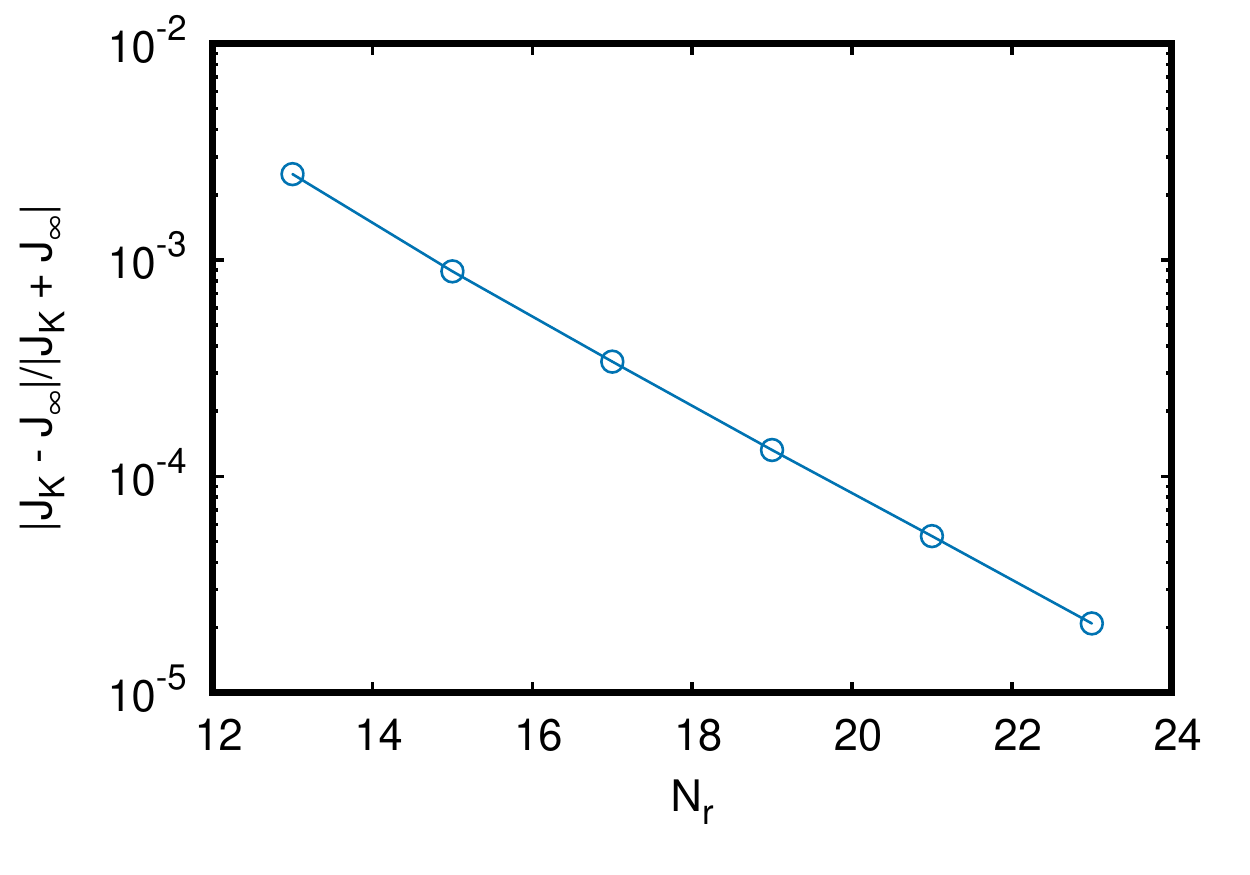}
 \caption{Similar to Fig.\ref{fig:resolEKG1}
 with $\mu r_H = 0{.}048574$ and angular velocity $\Omega_H/\mu = 0{.}97594$.}
 \label{fig:resolEKG2}
\end{figure}

\begin{figure}[h!]
 \centering
    \includegraphics[width=0.45\textwidth]{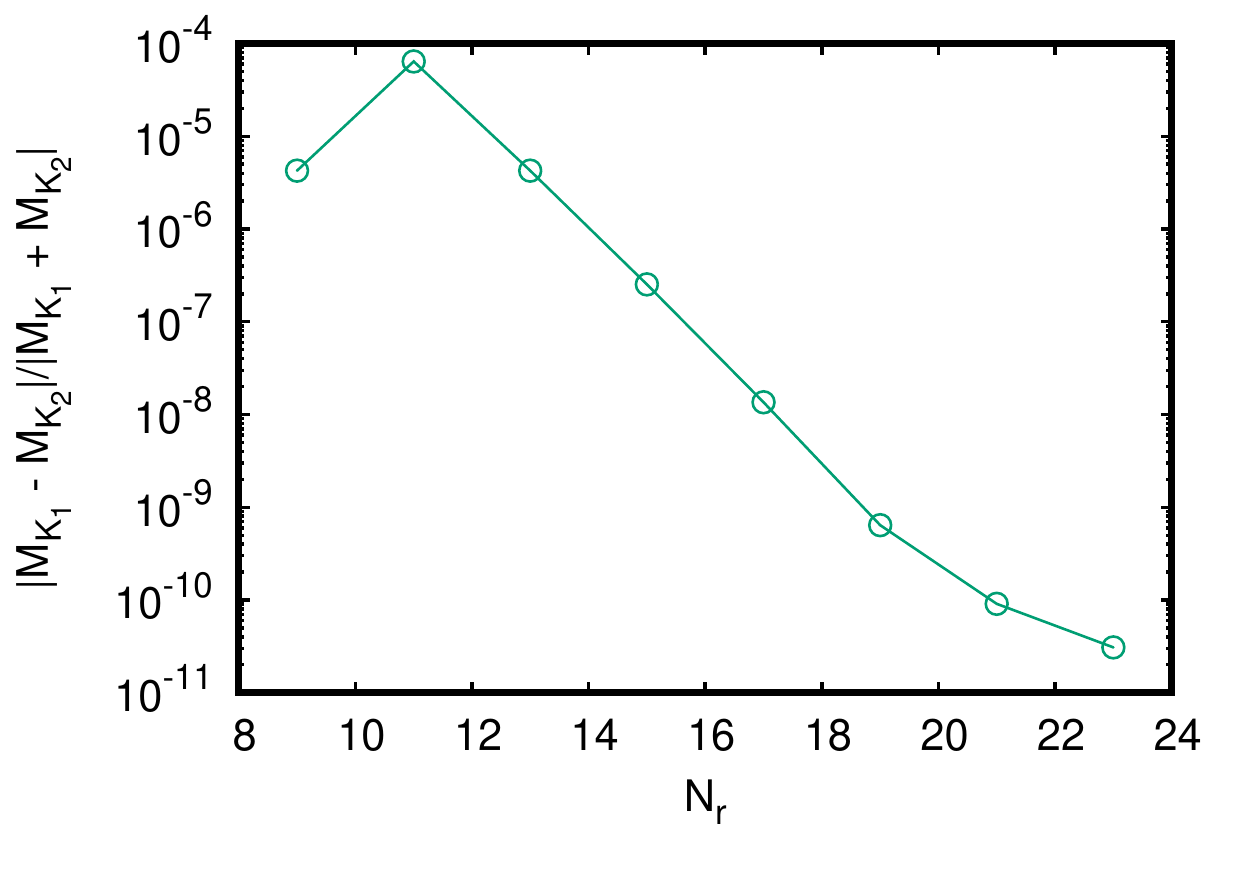}
    \includegraphics[width=0.45\textwidth]{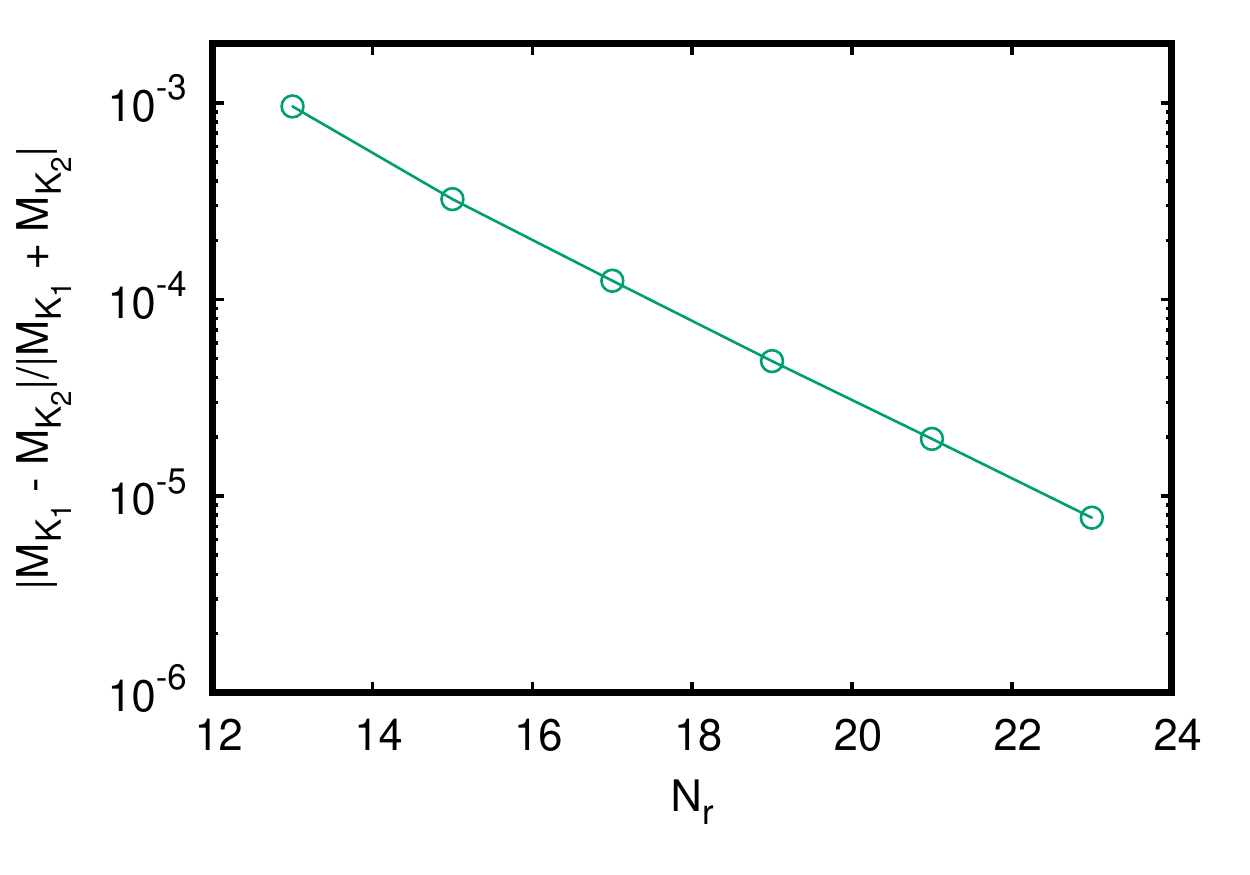}
\caption{Error indicators associated with the Komar mass $M_{\rm K}$ when using the two expressions given by (\ref{Kmass1}) and (\ref{Kmass2}), labeled as $M_{\rm K_1}$ and $M_{\rm K_2}$ in the figure, respectively. The top panel correspond to $\mu r_H = 0{.}057648$ and angular velocity $\Omega_H/\mu = 0{.}88289$, and the bottom panel to $\mu r_H = 0{.}048574$ and $\Omega_H/\mu = 0{.}97594$.}
 \label{fig:resolEKG3}
\end{figure}

Figures~\ref{fig:resolEKG1}-\ref{fig:resolEKG2} show the error indicators for different spectral resolutions of hairy BH solutions characterized by two different values of $r_H$ and $\Omega_H$. The top panel of each figure corresponds to the relative difference between the ADM mass and the Komar mass defined as $|M_{\rm K} - M_{\rm ADM}|/|M_{\rm K} + M_{\rm ADM}|$, while the bottom panel corresponds to the relative difference between expressions  (\ref{JADM}) and  (\ref{JKomar}) for the angular momentum defined as $|J_{\rm K} - J_\infty|/|J_{\rm K} + J_\infty|$. Another error indicator, depicted in 
Fig.~\ref{fig:resolEKG3}, is the relative difference between the two expressions for the Komar mass $M_{\rm K}$, given by Eqs.~(\ref{Kmass1}) and (\ref{Kmass2}), where 
in practice (\ref{Kmass1}) is computed using the expression (\ref{KmassH2}) but instead of evaluating at $r_H$ it is computed at $r\rightarrow \infty$. The error indicators display the characteristic exponential convergence with the expansion number of radial coefficients in the spectral decomposition \cite{Grandclement2009}.

\section{Comparison with previous studies}
\label{sec:compare}
The clouds solutions in a fixed Kerr background (in QIC) are used as initial guess to compute the full hairy solutions. Thus, we recovered first the same kind of clouds found earlier by Herdeiro \& Radu  \cite{Herdeiro2014} and also in Ref.\cite{Garcia2019}, where BL coordinates were employed instead. 
Some examples of these cloud configurations are depicted in Appendix \ref{app:ScalarClouds}. In contrast, Herdeiro \& Radu departed from a boson star configuration (i.e. a globally regular scalar field solution without a horizon) as ``initial guess" for the scalar field, and from that configuration they computed the rotating hairy BH's (RHBH) solutions.

Following the above strategy we obtained RHBH solutions similar to those reported in   Ref.~\cite{Herdeiro2014,Herdeiro2015}. The space of solutions can be summarized in  the {\it existence lines} depicted as plots $M_{\rm ADM}$ vs $\omega$ (cf. Fig.~\ref{fig:ExistenceLines}) which are similar to Figs. 4 and 6 of Ref.\cite{Herdeiro2015}.

A more direct and quantitative comparison between our results and those of Ref.\cite{Herdeiro2015} is difficult due to the use of different coordinates and parametrization of the metric. 
The main contrast between both analyses, ours and theirs, is that for the moment we cannot recover the globally regular boson star configurations directly from our spectral code since our parametrization for the metric and the implementation of the boundary conditions do not allow us to take the limit $r_H\rightarrow 0$ corresponding to a regular origin. In the present parametrization, this limit is (in principle) associated with an extremal RHBH. This can be better appreciated in the Kerr (vacuum) scenario from Eq.(\ref{rRrelat}) where the horizon for an extremal Kerr BH in BL coordinates is located at $R=M$, which corresponds to $r_H=0$ in the QI coordinates. We plan to overcome these limitations in the future, which in addition will allow us to study the exact extremal hairy configurations with exactly vanishing surface gravity $\kappa$, and not only in the limit of small values. Some of these values of $\kappa$ can be appreciated from the last rows of Tables \ref{tab:GlobalQuantities5} and \ref{tab:GlobalQuantities6} 
associated with a RHBH that is close to an extremal Kerr due to a small contribution of the scalar hair to the total mass (cf. columns 5 and 7 of those tables). As we discussed already, the solid (magenta) line of Fig.~\ref{fig:ExistenceLines} corresponding to boson stars was obtained from an independent and different, albeit similar, spectral code used in Ref.\cite{Grandclement2014}. There, rotating boson star models 
have been computed with regularity conditions imposed at the origin, as opposed to the regularity conditions at the horizon imposed here. Due to the limitations alluded before we appreciate from Fig.~\ref{fig:ExistenceLines} that the existence (dashed) lines of HRBH do not connect continuously to the boson-star (magenta) solid line, but finish in a configuration with $q< 1$. The existence lines do not have a constant $q$, but a constant $r_H$, therefore our lines ``cross along" the corresponding (dashed) curves of Fig.4 of Ref.\cite{Herdeiro2015} which are depicted for $q={\rm const.}$ 

Moreover, as in Ref.~\cite{Herdeiro2015}, we find that the dimensionless parameter $J_{\infty}/M_{\rm ADM}^2$ can be larger than unity in RHBH unlike the (hairless) Kerr BH's where $0\leq |J/M^2|\leq 1$ (see the last columns of Tables \ref{tab:GlobalQuantities1}--
\ref{tab:GlobalQuantities6}).

In the full non-linear treatment as well as in the cloud (linear) study --where the Kerr background is fixed-- the scalar field solution vanishes identically if the rotation is ``turned off" (e.g. taking $m=0$), which is consistent with the no-hair theorem in spherical symmetry~\cite{Sudarsky1995,Pena1997}\footnote{In this case, due to the synchronicity condition (\ref{fluxcond2}), when $m=0$ automatically $\omega=0$ and the scalar field becomes real valued.}. This is something that was also remarked in Ref.\cite{Herdeiro2015}.

\section{Conclusions and Outlook}
\label{sec:conclusion}
We have confirmed that rotating black holes can support hair under the form of a complex-valued scalar field using a completely independent code and a numerical method different from Ref.\cite{Herdeiro2015}, 
namely we used QI coordinates and spectral  methods. 
We have provided different numerical tests (convergence tests and comparison with the Kerr exact solution in vacuum) that indicate the robustness of our numerical analysis. The existence of a nontrivial configuration of this scalar field avoid the no-hair theorems due to the rotation of the black hole (see Refs.\cite{Garcia2019,Garcia2020} for a deeper discusion). We have computed plenty of hairy configurations, including the computation of several global quantities (see Appendix \ref{app:HBHT}), showing that the hair can contribute substantially to the total (ADM or Komar) mass of the BH. This indicates that the hairy solutions can differ drastically from the Kerr solution. However, for configurations having a ``diluted" hair, the solutions are similar to the cloud solutions 
(see Appendix \ref{app:ScalarClouds}) where the scalar field can coexist in the background of a Kerr BH. A particular feature of the hairy solutions is that the 
dimensionless parameter 
$J_\infty/M_{\rm ADM}^2$ can take values $|J_\infty/M_{\rm ADM}^2|>1$, even when solutions are relatively ``far" from {\it extremality} (e.g. the surface gravity $\kappa\sim 0.5\mu$) unlike the Kerr solution where $|a/M|\leq 1$, and the upper bound is reached in the extremal case $a=M$ where $\kappa=0$. As argued in Ref.~\cite{Herdeiro2015} this is because hairy configurations can connect continuously to the solitonic (rotating boson star) solutions where that dimensionless quantity can also exceeds unity.

From the observational point of view it is still premature to assess the impact of these hairy solutions for different reasons. So far, all direct detection of dark matter particles have failed. Thus, it is still unclear if a fundamental complex-valued scalar field could be associated with dark matter or not. On the other hand, it is also unclear if the dark matter phenomenology could be instead  explained by an alternative theory of gravity, since all the proposals in that direction have also failed to recover in addition all the success of GR. Notwithstanding, we have at our disposal several instruments that can validate, bound or rule out the existence of this kind of hairy solutions in a near future. For instance, since these solutions produce a spacetime different from the Kerr solution, their respective {\it shadows} can be different since the photon trajectories are affected differently (cf. Refs.\cite{shadows}). Moreover, the gravitational waves emitted from the collision of two RHBH can have a different pattern relative to the Kerr counterparts, and so the LIGO-VIRGO observatories, as well as the forthcoming detectors KAGRA, Einstein Telescope and LISA can put stringent bounds to such solutions or perhaps indicate deviations from the standard Kerr BH interpretation that might be explained by the RHBH. These are scenarios that are worth of study. Moreover, it has been argued \cite{Xrayconst} that RHBH could be constrained by the analysis of the expected X-ray spectra produced by accretion disks around BH's with the forthcoming X-ray missions, like Suzaku, eXTP and LOFT \cite{Xray}. Finally, recently an {\it unusual} gravitational wave detection GW200129 by the LIRGO-VIRGO collaboration triggered speculations about possible deviations from GR,
so that BH's from alternative theories could explain such ``anomalous" signals~(cf. \cite{Silva2022}). Nevertheless, a recent analysis~\cite{Hannam2022} presents strong evidence showing that such signals are due to the collision of two ``ordinary" Kerr BH's that, however,  precess violently and together with the tilt and the high spin of the primary BH with respect to the orbital angular momentum, can explain the unusual gravitational wave patterns. This example teaches us that the answer to an unexpected phenomena can be found within GR before jumping to much more exotic explanations. Thus, hairy BH's within GR itself can be a potential candidate to explain some unexpected features that might be revealed in the forthcoming observations related with BH astrophysics without the need to resort to much more drastic avenues like modified theories of gravity. Therefore, it is worth exploring different properties (e.g. stability~\cite{Ganchev2018,Degollado2018} and extremality) and variations around this kind of RHBH solutions.
\bigskip


\section*{Acknowledgments}
This work was supported partially by DGAPA--UNAM grants IN111719, IN105223, and
CONACYT (FORDECYT-PRONACES) grant 140630. G.G. acknowledges CONACYT scholarship 291036. G.G. and M.S. are indebted to E. Murrieta and L. Díaz for their help in installing and parallelizing the spectral code at the ICN cluster. E.G. and P.G. acknowledge funding by l'Agence Nationale de la Recherche, project StronG
ANR-22-CE31-0015-01.

\bigskip

\appendix
\section{Kerr metric in Quasi-Isotropic coordinates and numerical tests}
\label{app:KerrQI}

Before providing the Kerr solution in Quasi-Isotropic coordinates (QIC) we present the Kerr solution in Boyer-Lindquist
coordinates (BLC), $(t,R,\theta,\varphi)$, in order to contrast both solutions\footnote{We use
  the notation $R$ for the radial BLC, since $r$ is reserved here for the QIC.}:
\bea
&& ds^2 = - \frac{\left(\Delta -a^2\sin^2\theta\right)}{\Sigma}dt^2 -
  \frac{2a\sin^2\theta\left(R^2 + a^2 - \Delta\right)}{\Sigma}dtd\varphi \nonumber \\ 
&+& \frac{\left[\left(R^2 + a^2\right)^2 - \Delta a^2\sin^2\theta\right]}{\Sigma}\sin^2\theta d\varphi^2 + \frac{\Sigma}{\Delta}dr^2 + \Sigma d\theta^2 \, ,\label{KM1}
\eea
where
\bea
\label{KM2}
\Sigma &=& R^2 + a^2\cos^2\theta \;, \\
\label{KM3}
\Delta &=& R^2 + a^2 - 2MR = \left(R - R_+\right)\left(R - R_-\right) \;,
\eea
and
\bea
\label{KM4}
R_+ &=& M + \sqrt{M^2 - a^2} \;,\\ 
\label{KM5}
R_- &=& M - \sqrt{M^2 - a^2} \;.
\eea

\smallskip
The event horizon is located at $R_H = R_+$.

The lapse function is given by
\be
\label{KM6}
N^2 = \frac{\Delta\Sigma}{\left(R^2 + a^2\right)^2 - \Delta a^2\sin^2\theta},
\ee
\bigskip

The relationship between the BLC $(t,R,\theta,\varphi)$ and the QIC $(t,r,\theta,\varphi)$ is as follows~\cite{Gourgoulhon2022}:
\bea
R &=& r + M + \frac{M^2 -a^2}{4r} \;,\\
r &=& \frac{1}{2}\Big(R -M + \sqrt{\Delta}\Big)\nonumber \\
\label{rRrelat}
&=& \frac{1}{2}\Big(R -M + \sqrt{R^2- 2MR + a^2}\Big) \;.
\eea
\begin{widetext}
\bea
A^2 &=& \frac{-2 a^2 M (M+4 r)+8 a^2 r^2 \cos (2 \theta )+a^4+(M+2 r)^4}{16r^4}\;,\label{AQI} \\
\nonumber N^2 &=&  16r^4\left(1 - \frac{\sqrt{M^2-a^2}}{2r}\right)^2\left(1+\frac{\sqrt{M^2-a^2}}{2r}\right)^2\times\\
\nonumber && \left[-2 a^2 M (M+4 r)+8 a^2 r^2 \cos (2 \theta ) + a^4+(M+2 r)^4\right] \times \\
\nonumber &&\left[8a^2 r^2 \cos (2 \theta ) \left(a^2-M^2+4 r^2\right)^2 - 4a^6 \left(M^2+4 M r-2 r^2\right)  \right. \\
\nonumber &-& 4a^2 (M+2 r)^2 \left(18 M^2 r^2+8 M^3 r+M^4-8 M r^3-8 r^4\right)+a^8+(M+2 r)^8\\
&+& \left. a^4\left(96 M^2 r^2+48 M^3 r+6 M^4-64 M r^3+32 r^4\right)\right]^{-1}\;,\label{NQI}
\eea
\bea
\nonumber B^2 &=& \left[8 a^2 r^2 \cos (2 \theta ) \left(a^2-M^2+4 r^2\right)^2-4 a^6 \left(M^2+4 M r-2 r^2\right)\right.\\
\nonumber &+& a^4 \left(96 M^2 r^2+48 M^3 r+6 M^4-64 M r^3+32 r^4\right)\\
\nonumber &-& \left.4 a^2 (M+2 r)^2 \left(18 M^2 r^2+8 M^3 r+M^4-8 M r^3-8 r^4\right)+a^8+(M+2 r)^8\right]\times\\
&&\left[16r^4 \left(-2 a^2 M (M+4 r)+8 a^2 r^2 \cos (2 \theta )+a^4+(M+2 r)^4\right)\right]^{-1}\;,\label{BQI}   
\eea
and
\bea
\nonumber \beta^{\varphi} &=& \left[128 a M r^3 (a-M-2 r) (a+M+2 r)\right]\times\left[8 a^2 r^2 \cos (2 \theta ) \left(a^2-M^2+4 r^2\right)^2\right.\\
\nonumber &-& 4a^6 \left(M^2+4 M r-2 r^2\right)+a^4 \left(96 M^2 r^2+48 M^3 r+6 M^4-64 M r^3+32 r^4\right)\\
\nonumber &-& 4 a^2 (M+2 r)^2 \left(18 M^2 r^2+8 M^3 r+M^4-8 M r^3-8 r^4\right)\\
&+&\left. a^8+(M+2 r)^8\right]^{-1}\;.\label{ShiftQI}
\eea
\end{widetext}
We have checked using \textit{SageMath} that the above values of
$A$, $B$, $N$ and $\beta^{\varphi}$ define a metric that fulfils the vacuum 3+1-Einstein equations\footnote{cf. the notebook\\ {\small \url{https://nbviewer.org/github/sagemanifolds/SageManifolds/blob/master/Notebooks/SM_KerrQI_3p1.ipynb}}.}. We have also checked using \textit{Mathematica} that they satisfy the elliptic equations (\ref{eq:S11})--(\ref{eq:S44}) in the absence of matter. Moreover, in the non-rotating limit $a=0$, the
above variables reduce to the usual expressions for the \Sc metric in isotropic coordinates with $A=B$, $\beta^\varphi=0$, and 
\bea
N &=& \frac{1-\frac{M}{2r}}{1+\frac{M}{2r}} \;,\\
A &=& \left(1+\frac{M}{2r}\right)^2 \;.
\eea
\bigskip

\begin{figure}[h!]
 \begin{center}
   \label{fig:fN}
    \includegraphics[width=0.5\textwidth]{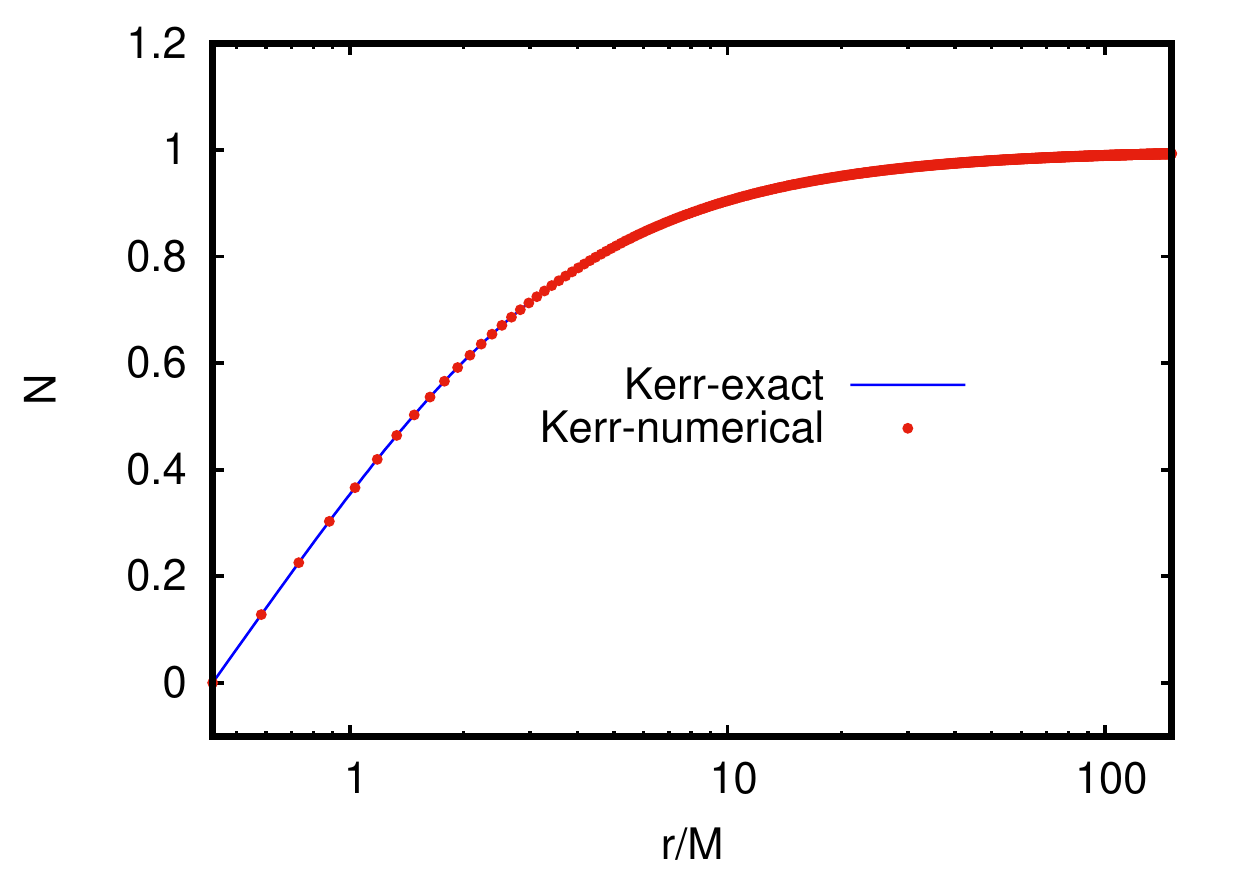}
   \label{fig:fA}
    \includegraphics[width=0.5\textwidth]{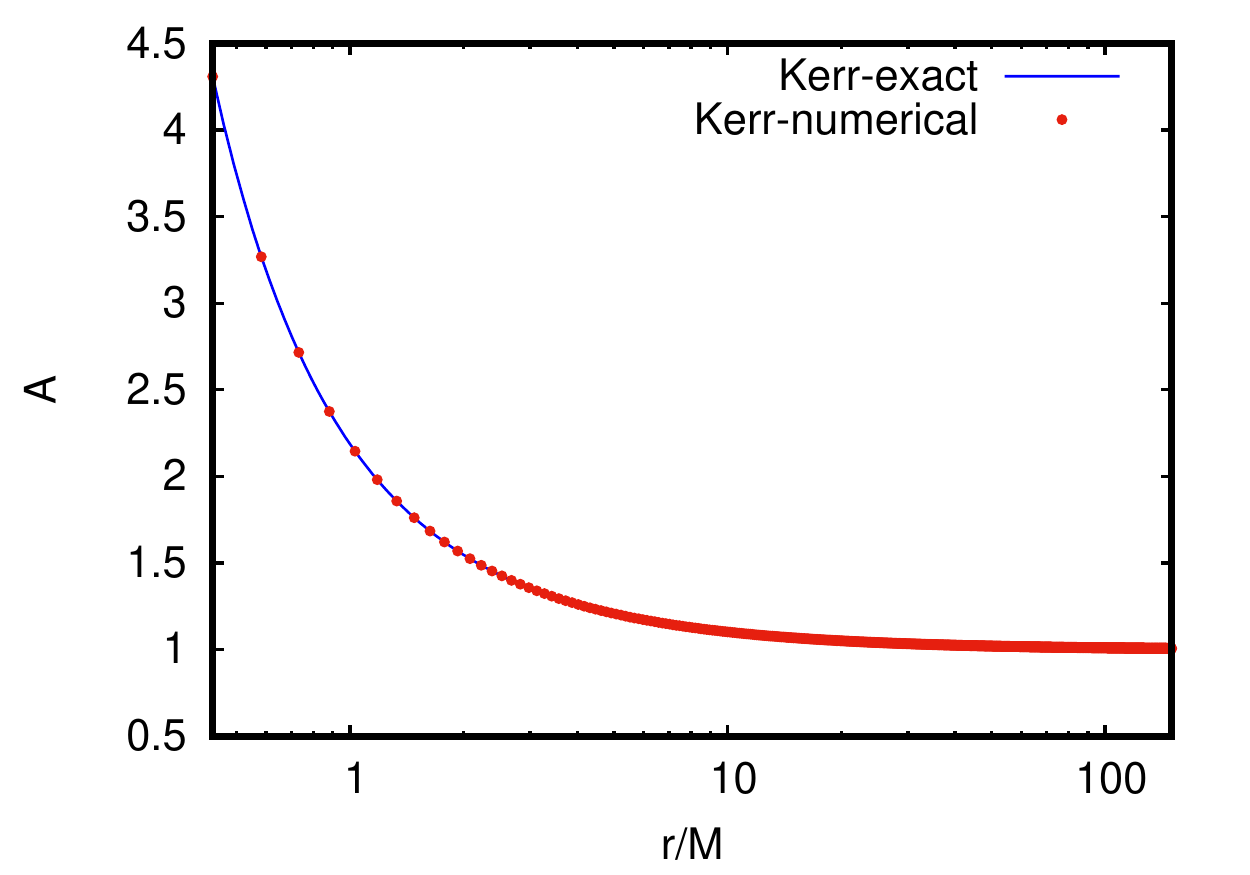}
 \caption{The red dots correspond to the numerical solution at the equatorial plane ($\theta=\pi/2$)
 for the lapse function $N$ (top panel) and the metric potential $A$ (bottom panel) obtained from solving the elliptical system using spectral methods (KADATH) and the blue solid lines corresponds to the exact Kerr solution taking $a/M = 0.5$.
 }\label{fig:fNA}
\end{center}
\end{figure}

\begin{figure}[h!]
 \begin{center}
   \label{fig:fB}
    \includegraphics[width=0.5\textwidth]{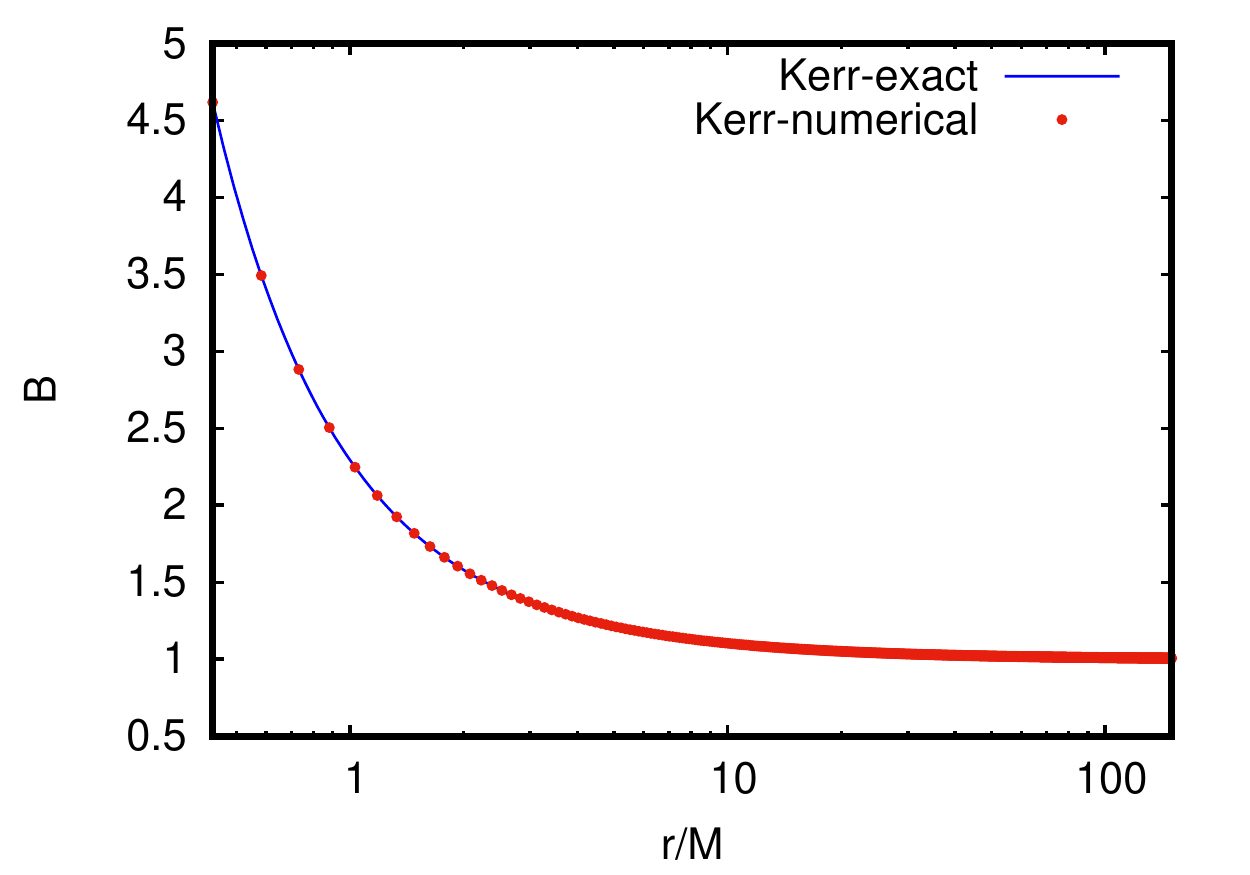}
   \label{fig:fNp}
    \includegraphics[width=0.5\textwidth]{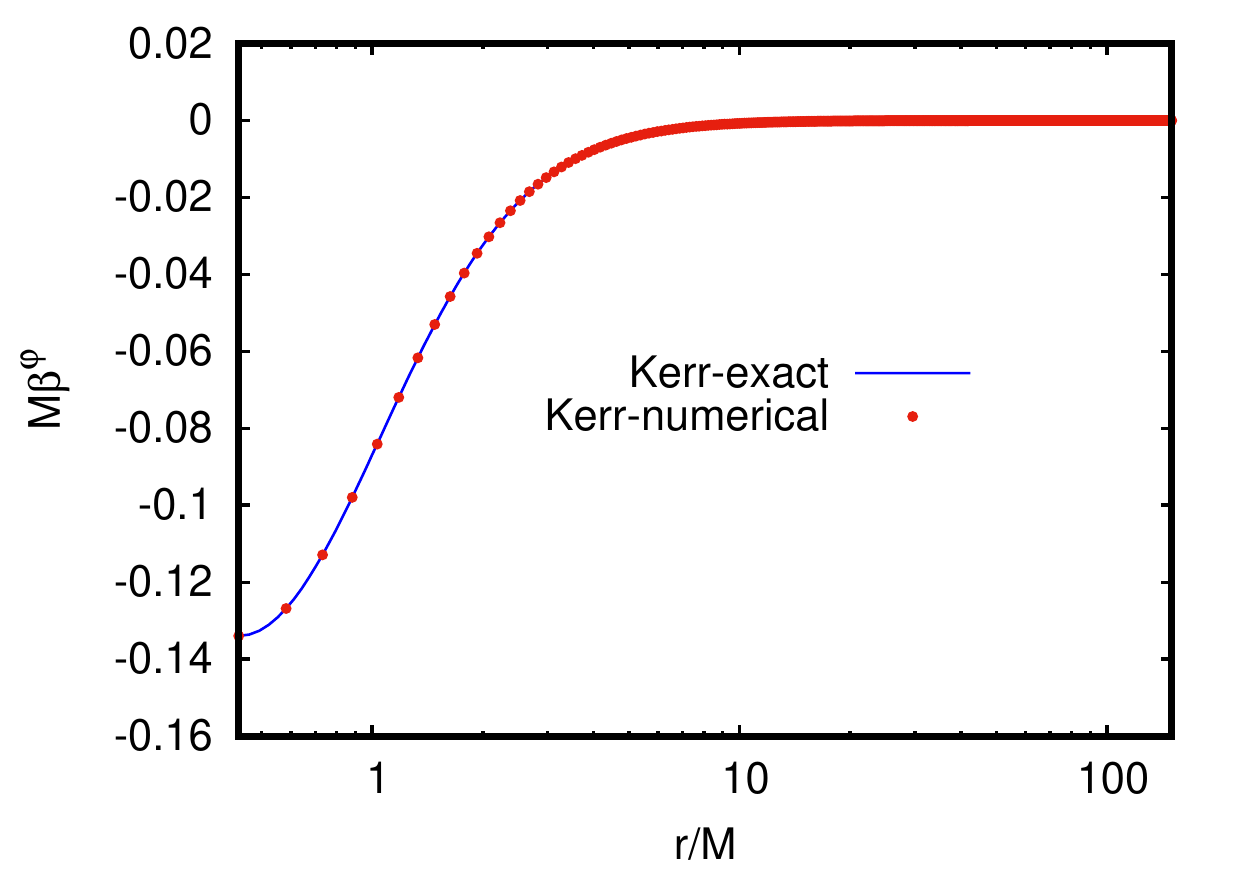}
 \caption{Similar to Fig.~\ref{fig:fNA}
  for the metric potential $B$ (top panel) and the shift $\beta^\varphi$ (bottom panel).}\label{fig:fBNp}
\end{center}
\end{figure}

\begin{figure}[h!]
 \centering
   \label{fig:ErrorN}
    \includegraphics[width=0.5\textwidth]{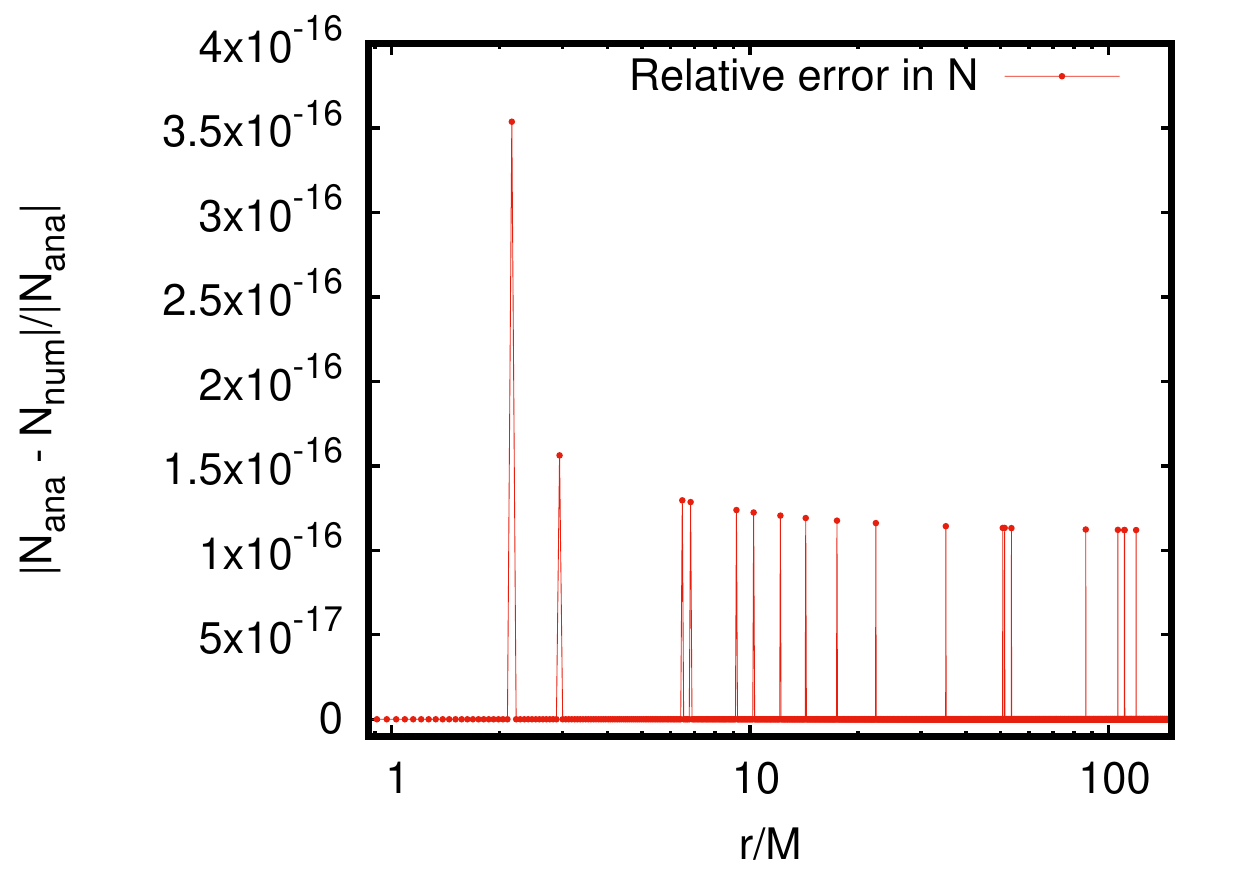}
   \label{fig:ErrorA}
    \includegraphics[width=0.5\textwidth]{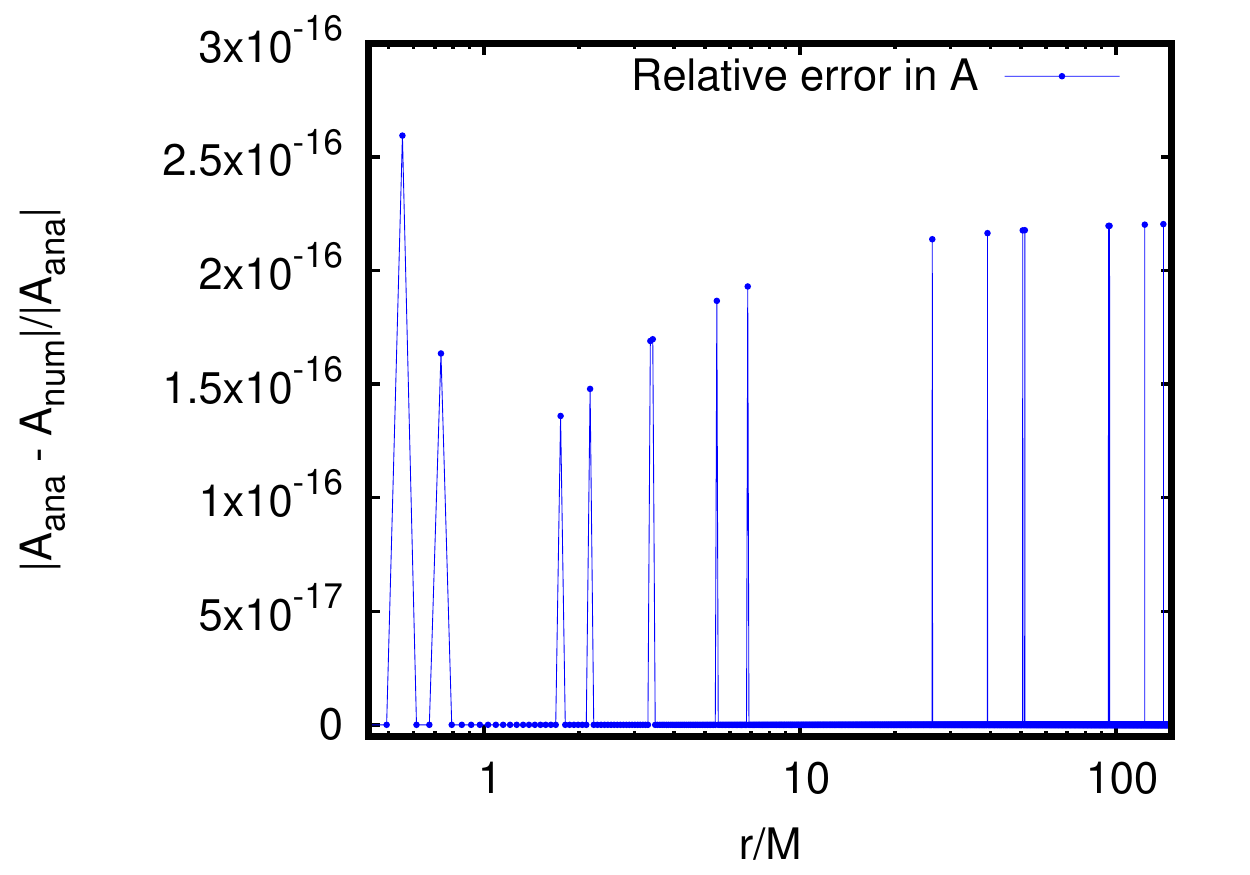}
 \caption{Relative errors associated with the solutions for $N$ (top panel) and $A$ (bottom panel) shown in Fig.~\ref{fig:fNA}.}
 \label{fig:ErrorNA}
\end{figure}

\begin{figure}[h!]
 \centering
   \label{fig:ErrorB}
    \includegraphics[width=0.5\textwidth]{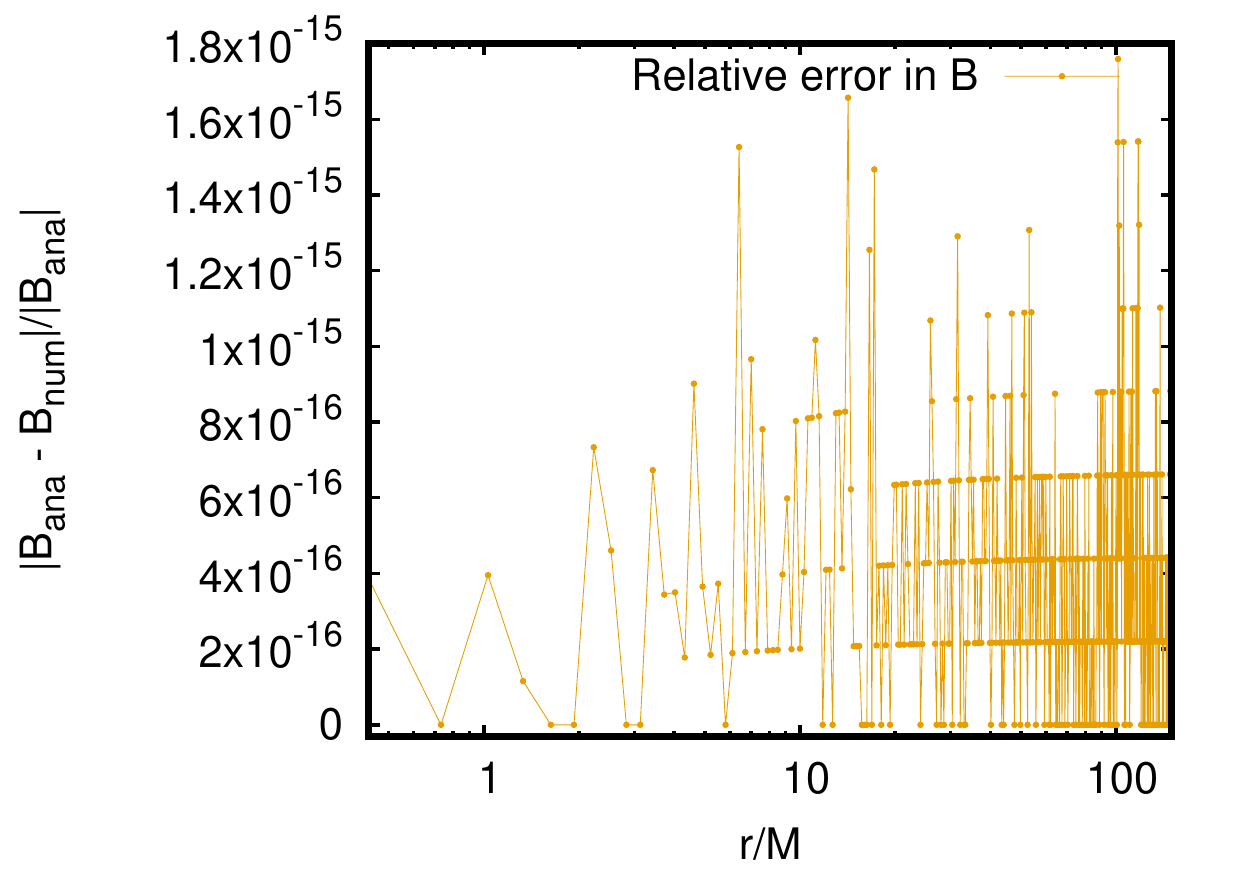}
   \label{fig:ErrorNp}
    \includegraphics[width=0.5\textwidth]{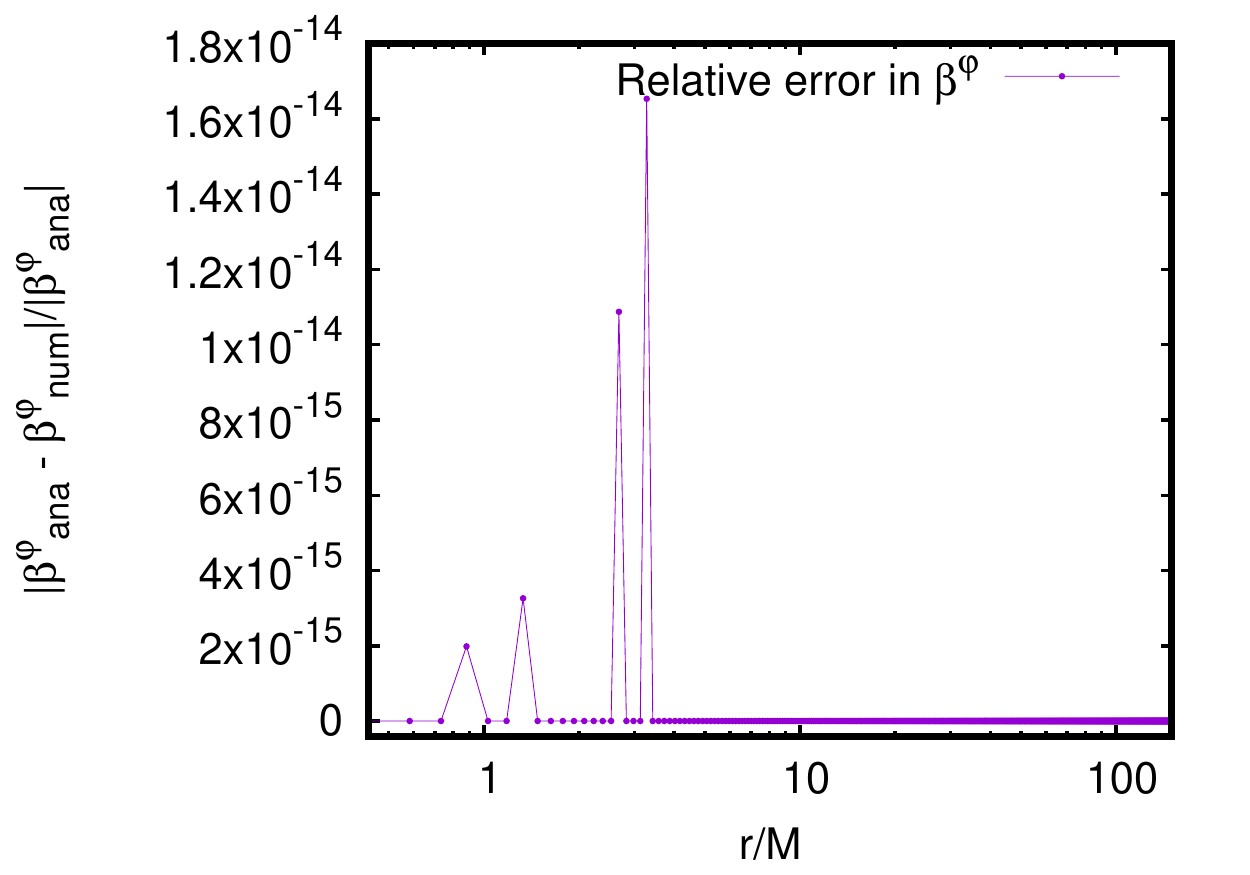}
 \caption{Relative errors associated with solutions for $B$ (top panel) and $\beta^{\varphi}$ (bottom panel) shown in Fig.~\ref{fig:fBNp}.}
 \label{fig:ErrorBNp}
 \end{figure}

Figures \ref{fig:fNA} and \ref{fig:fBNp} depict the numerical and the exact solutions for Kerr spacetime. Figures \ref{fig:ErrorNA} and \ref{fig:ErrorBNp} show the corresponding relative errors between those solutions. 

\section{Apparent Horizon}
\label{app:appHor}
We follow the approach of Refs.~\cite{Baumgarte2010,Alcubierre2008} and consider a smooth, closed, 2-dimensional (spatial) surface $\mathcal{S}$ embedded in $\Sigma_t$, and take $s^{a}$ as its outward pointing unit normal tangent to $\Sigma_t$ satisfying $s^{a}s_{a} = 1$ and $s^{a}n_{a} = 0$. The two-dimensional metric $m_{ab}$ 
on $\mathcal{S}$ induced by $\gamma_{ab}$ is
\be
\label{metricS}
m_{ab} = \gamma_{ab} - s_{a}s_{b} = g_{ab} + n_{a}n_{b} - s_{a}s_{b}\;.
\ee
Next, we define the following two null 
vectors 
\begin{equation}
\label{tangentskl}
k^{a} \equiv \frac{1}{\sqrt{2}}(n^{a} + s^{a})\qquad {\rm and}\qquad l^{a} \equiv \frac{1}{\sqrt{2}}(n^{a} - s^{a})\;. 
\end{equation}
satisfying $k^{a}k_{a} = 0$,  $m_{ab}k^{b} = 0$, $l^{a}l_{a} = 0$ and $m_{ab}l^{b} = 0$. 
These vectors at $\mathcal{S}$ are tangent 
and future pointing to the null geodesics 
that emanate from $\mathcal{S}$ and whose 
projection on $\Sigma_t$ are orthogonal to $\mathcal{S}$. The normalization is chosen so that $k^{a}l_{a} = -1$. 
The vectors $k^{a}$ and $l^{a}$ 
are tangent to the so called {\it outgoing} and {\it ingoing} null geodesics.
From Eqs.~(\ref{metricS}) and (\ref{tangentskl}) one obtains
\be
m_{ab} = g_{ab} + k_{a}l_{b} + l_{a}k_{b}\;,
\ee
The expansion associated with the outgoing null geodesics is defined by
\be
\label{expansion}
H \equiv m^{ab}\nabla_{a}k_{b}\;.
\ee
The outer-trapped surface is a surface $\mathcal{S}$ with $H<0$ and the trapped region is any region of $\Sigma_t$ containing outer-trapped surfaces. Moreover, 
when $H=0$, the surface $\mathcal{S}$ 
is called {\it marginally trapped}. The {\it apparent horizon} is defined as the outermost {\it marginally trapped surface}. 
 The expansion~(\ref{expansion}) can be written in terms of 
3+1 variables as follows~\cite{Alcubierre2008}:
\be
\label{expansionAl}
H = D_as^a - K + K_{ab}s^as^b\;.
\ee
We can apply this formula to the metric (\ref{metric}), and take as outward pointing unit (radial) normal 
$s^{a} = \frac{1}{A}\left(\frac{\partial}{\partial r}\right)^a$. Given $s^{a}$ and since $K_{rr} = 0$, the last term in (\ref{expansionAl}) vanishes. Moreover, since $K=K^a_{\;a} = 0$, the expansion is
\be
\label{Hstationary}
H = D_as^a = \frac{1}{\sqrt{\gamma}}\partial_r(\sqrt{\gamma}s^r)\;
\ee
where as before $\gamma = A^4B^2r^4\sin^2\theta$ is the determinant of the spatial metric $\gamma_{ab}$.
In this way, Eq.~(\ref{Hstationary}) reads
\bea
H &=& -\frac{1}{A^2}\frac{\partial A}{\partial r} + \frac{1}{A}\left(\frac{2}{A}\frac{\partial A}{\partial r} + \frac{1}{B}\frac{\partial B}{\partial r} + \frac{2}{r}\right) \nonumber \\
\label{Hstationary3}
&=& \frac{\partial}{\partial r}\left(\frac{1}{A}\right) + \frac{\Gamma}{A}\;,
\eea
where 
\be
\Gamma \equiv \left(\frac{2}{A}\frac{\partial A}{\partial r} + \frac{1}{B}\frac{\partial B}{\partial r} + \frac{2}{r}\right)\;.
\ee

In Sec.~\ref{sec: BC} Eq.~(\ref{Hstationary3}) was used in (\ref{Hstationary2}) to impose regularity conditions for the variables $A$ and $B$ at the horizon. For instance, in Kerr spacetime the apparent horizon coincides with the event horizon
at $r_H = \sqrt{M^2 - a^2}/2$, and thus
\be
H|_{r_H} = 0\;.
\ee
We checked using {\it Mathematica} that the expansion (\ref{Hstationary3}) vanishes at $r_H$
for the Kerr metric in QIC 
(cf. Appendix \ref{app:KerrQI}).

\section{Surface gravity formulae}
\label{app:surgrav}
In order to obtain a simple expression for the surface gravity in terms of derivatives for the lapse function
$N$, the starting point is Eq.(\ref{SurGra1}). That equation leads to \cite{Wald1984}
\be
\label{SurGra3}
\kappa = \lim \left(V {\mathtt a}\right) \equiv  \left(V {\mathtt a}\right)_{{\mathcal H}^+}.
\ee
where
\bea
\label{SurGra4}
V &=& \left(-\chi^a\chi_a\right)^{1/2} \;,\\
\label{SurGra5}
{\mathtt a} &=& \left(a^c a_c\right)^{1/2} \;,\\
a^c &=& \frac{\chi^b\nabla_b\chi^c}{\left( -\chi^a\chi_a\right)} = \frac{\chi^b\nabla_b\chi^c}{V^2} \;.
\eea
The vector $n^a$ (\ref{normaln}) which is normal  to $\Sigma_t$ is given in terms of the lapse function $N$ and the shift
vector $\beta^a$ as follows~\cite{Wald1984}:
\be
\label{SurGra7}
n^a = \frac{1}{N}\left(\frac{\partial}{\partial t}\right)^{a} - \frac{\beta^{i}}{N}\left(\frac{\partial}{\partial x^i}\right)^{a}.
\ee
At the horizon and for {\it axisymmetric} and {\it circular} spacetime, $n^a$ reduces to (\ref{normaln}),
and when comparing with (\ref{Killingchi}) we conclude 
\be
\label{SurGra8}
\chi^a = Nn^a\big|_{{\cal H}^+} \;.
\ee
Since $n^a n_a = -1$ the squared norm of the helical Killing field is related to the lapse function
by
\be
\label{SurGra9}
-\chi^a\chi_a\big|_{{\cal H}^+} = N^2\big|_{{\cal H}^+} = V^2\big|_{{\cal H}^+} =0\;.
\ee
Moreover, it is not difficult to prove that the acceleration (\ref{SurGra5}) of the Killing field $\chi^a$ coincides with the acceleration of the normal observers when $n^a$ is given by (\ref{SurGra8}):
\bea
\nonumber a^c_\perp &=& n^a \nabla_a n^c = D^c\ln N, \\
& = & \gamma^c_a\nabla^a\ln N,\label{SurGra10}
\eea
as mentioned below Eq.~(\ref{KmassH}), $\gamma^c_a$ is the {\it 3-metric} (or projector) onto $\Sigma_t$ and $D^c$ is the covariant derivative compatible with the 3-metric \cite{Wald1984,Gourgoulhon2012}.

From (\ref{SurGra9}) and (\ref{SurGra10}) the surface gravity given by (\ref{SurGra3}) reads
\be
\label{SurGra11}
\kappa = \left[Na_{\perp}\right]_{{\cal H}^+} \,,
\ee
where
\bea
\label{SurGra12}
\nonumber a_\perp = {\mathtt a} &=& \left[\left(D^c\ln N\right)\left(D_c\ln N\right)\right]^{1/2} \\
& = & \left[\frac{\gamma^{ab}}{N^2}\left(D_aN\right)\left(D_b N\right)\right]^{1/2}.
\eea
The desired formula for the surface gravity in terms of the spatial derivatives
of the lapse function and the 3-metric is:
\bea
\nonumber \kappa &=&  \left[\gamma^{ab}\left(D_a N\right)\left(D_b N\right)\right]^{1/2}_{{\cal H}^+} \\
\label{SurGra13}
& = & \frac{1}{2}\left[\frac{\gamma^{ab}}{N^2}\left(D_aN^2\right)\left(D_b N^2\right)\right]^{1/2}_{{\cal H}^+} \,.
\eea
When the formula (\ref{SurGra13}) is applied to the metric (\ref{metric}) one obtains 
\bea
\label{kappaapp}
\kappa &= & \left\{\frac{1}{A}\left[\left(\partial_rN\right)^2 + \frac{1}{r^2}\left(\partial_{\theta}N\right)^2\right]^{1/2}\right\}_{r_H}
\nonumber \\
&=& \left\{\frac{1}{2AN}\left[\left(\partial_rN^2\right)^2 + \frac{1}{r^2}\left(\partial_{\theta}N^2\right)^2\right]^{1/2}\right\}_{r_H}
\,,
\eea
where we used the following expressions for the relevant 3-metric components and its inverse
\bea
\gamma_{rr} &=& A^2 \,,\; \gamma_{\theta\theta} = r^2A^2 \,, \\
\gamma^{rr} &=& \frac{1}{A^2} \,,\; \gamma^{\theta\theta} = \frac{1}{r^2A^2} \,.
\eea

For instance, 
the surface gravity (\ref{kappa}) for \Sc and Kerr spacetimes in QIC leads to the well known expressions~\cite{Wald1984}:
\bea
\kappa_{\rm Kerr} &=&  \frac{\sqrt{M^2 - a^2}}{2M\left(M + \sqrt{M^2 - a^2}\right)} \;,\\ 
\kappa_{\rm Schw} &=&\frac{1}{4M} \;.
\eea
which can be found more straightforwardly in BL coordinates and its non-rotating limit $a=0$.

\section{Scalar clouds}
\label{app:ScalarClouds}

Using the KADATH library, a code was developed to obtain numerical solutions of  {\it scalar clouds}, which are solutions of Eq.~(\ref{KG}) in the background of a Kerr spacetime assuming Eqs.~(\ref{Psians}) and (\ref{fluxcond2}) with an asymptotically vanishing field. This kind of solutions were analyzed thoroughly in Refs.~\cite{Hod,Herdeiro2014,Herdeiro2015,Garcia2019} using BL coordinates, and due to the linearity of Eq.~(\ref{Psians}) one can use separation of variables that leads to Teukolsky equations for the angular $S_{lm}(\theta)$ and radial parts $R_{nlm}$
of the boson field $\Psi$, which are associated, repectively, with the spheroidal harmonics that depend on the integers $l$ and $m$ ($|m|\leq l$) and the radial function that depends also on the positive integer $n$ (number of nodes). In this context the product $S_{lm}(\theta) R_{nlm}$ (in BL coordinates) is the equivalent of 
the function $\phi(r,\theta)$ of Eq.~(\ref{Psians}) (in QI coordinates). Furthermore in Ref.~\cite{Garcia2019}
the radial Teukolsky equation is solved using the Kerr parameter $a$ as an {\it eigenvalue}, so below we report such eigenvalues for the cloud solutions computed here.

We computed scalar clouds using QIC 
within the aim of using them as input and as initial guess in the spectral code built to find solutions of the full EKG system and thus generate an initial hairy black hole solution. From the latter it was then possible to build a sequence of solutions with the value $r_H$ fixed by varying the frequency $\Omega_H$ gradually.

\begin{figure}[h!]
  \centering
    \includegraphics[width=0.45\textwidth]{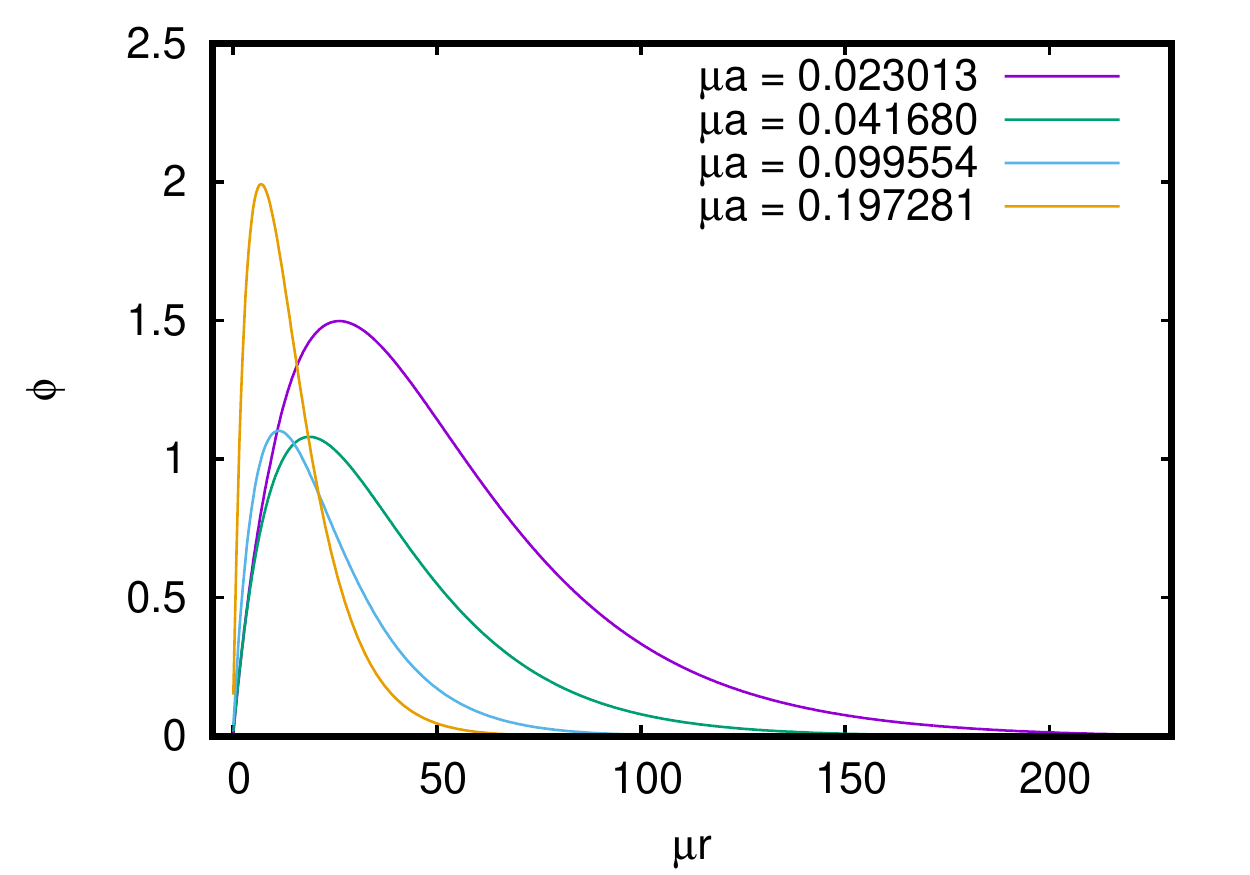}
  \caption{Scalar clouds around a Kerr black hole in quasi-isotropic coordinates with \textit{quantum numbers} $n = 0$ (nodeless) and $m =l=1$ (at the equatorial plane $\theta=\pi/2$).}
  \label{fig:TestClouds1}
\end{figure}

\begin{figure}[h!]
 \centering
   \label{fig:CompN}
    \includegraphics[width=0.45\textwidth]{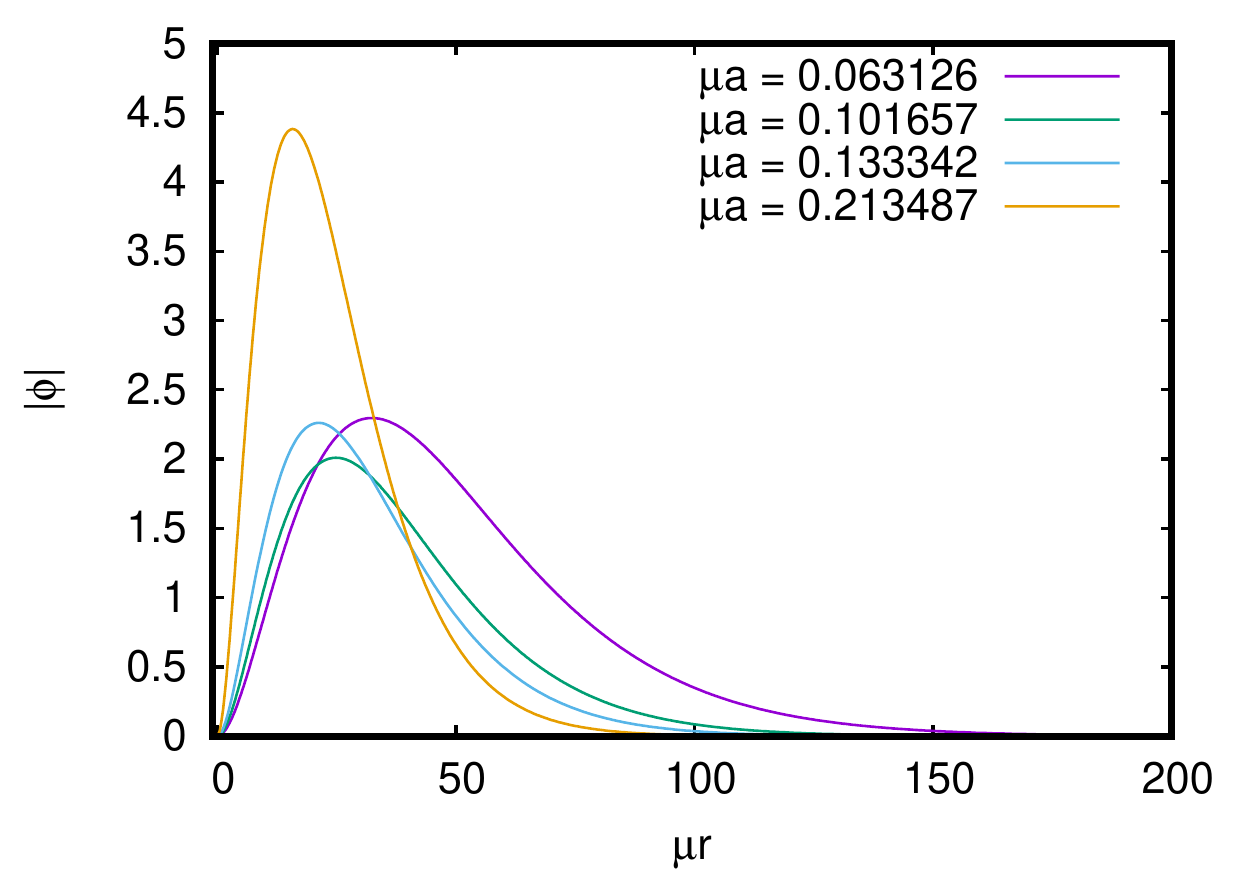}
   \label{fig:CompA}
    \includegraphics[width=0.45\textwidth]{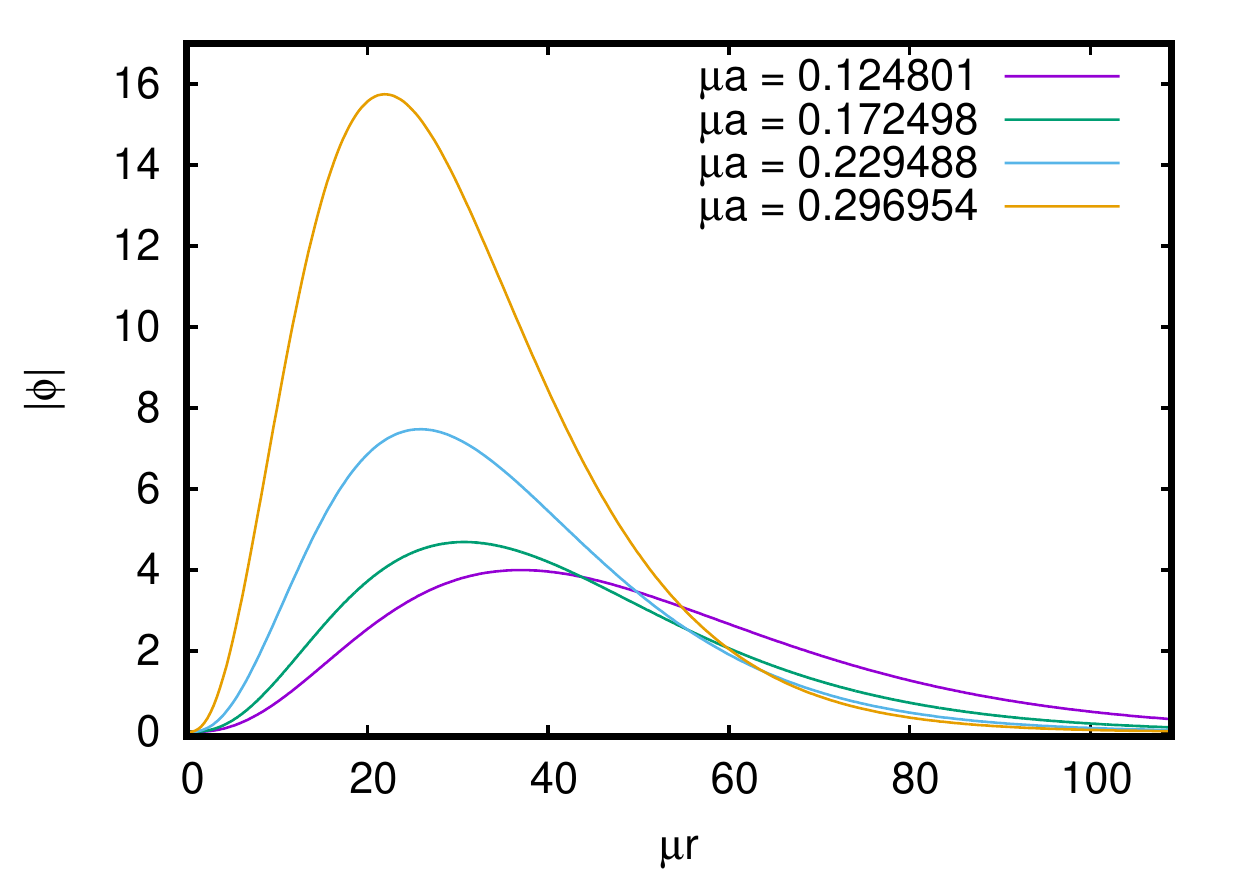}
    \caption{Scalar clouds around a Kerr black hole in quasi-isotropic coordinates with \textit{quantum numbers} $n = 0$ (nodeless) and integers $m =l= 2$ (top panel), and  $m =l= 3$ (bottom panel) respectively (at the equatorial plane $\theta=\pi/2$).}
 \label{fig:TestClouds23}
\end{figure}

Figures \ref{fig:TestClouds1} and ~\ref{fig:TestClouds23} depict some examples of cloud solutions at the equatorial plane. Although we do not compute hairy solutions for $m>1$ in this work, for completeness we present cloud solutions for $m=l=1,2,3$.

The Tables~\ref{tab:spectrum1QI}-\ref{tab:spectrum3QI} show the eigenvalues $\mu a$ that were found for different values of $r_{H}$ and $\Omega_H$ when we solve the Klein-Gordon equation in the 
background of Kerr spacetime in QI coordinates. The numerical data presented in these Tables correspond to configurations with $n = 0$ (nodeless). The results obtained using the spectral code for cloud solutions are consistent with those reported in Ref.~\cite{Garcia2019} for $l=m$ and $n=0$.
\begin{table}[h!]
\begin{center}
\resizebox{7.5cm}{!}{
\begin{tabular}{|c|c|c|c|c|c|}
\hline 
$\boldsymbol{\mu R_{H}}$ & $\boldsymbol{\mu r_{H}}$ & $\boldsymbol{\mu a}$ & $\boldsymbol{\mu M}$ & $\boldsymbol{\Omega_H/\mu}$ & $\boldsymbol{a/M}$ \\
\hline \hline 
0.15 & 0.036618	& 0.023013 & 0.076766 &	0.999244 & 0.299778\\ \hline
0.20 & 0.047829	& 0.041680 & 0.104344	& 0.998601 & 0.399449\\ \hline
0.30 & 0.066740	& 0.099554 & 0.166518	& 0.996433 & 0.597859\\ \hline
0.40 & 0.075676	& 0.197281 & 0.248650	& 0.991755 & 0.793407\\ \hline
0.42 & 0.074966	& 0.224626 & 0.270068	& 0.990165 & 0.831739\\ \hline
0.46 & 0.068260	& 0.293262 & 0.323482	& 0.985409 & 0.906580\\ \hline
0.48 & 0.060220	& 0.338793 & 0.359564	& 0.981483 & 0.942232\\ \hline
0.50 & 0.045465	& 0.398850 & 0.409084	& 0.974959 & 0.974984\\ \hline
0.51 & 0.033064	& 0.438945 & 0.443898	& 0.969403 & 0.988842\\ \hline
0.52 & 0.013913	& 0.491451 & 0.492238	& 0.959881 & 0.998401\\ \hline
\end{tabular}}
\caption{Eigenvalues for a scalar cloud with parameters $m =l= 1$. Some solutions 
associated with these values appear in Fig.~\ref{fig:TestClouds1}.}
\label{tab:spectrum1QI}
\end{center}
\end{table}

\begin{table}[h!]
\begin{center}
\resizebox{7.5cm}{!}{
\begin{tabular}{|c|c|c|c|c|c|}
\hline 
$\boldsymbol{\mu R_{H}}$ & $\boldsymbol{\mu r_{H}}$ & $\boldsymbol{\mu a}$ & $\boldsymbol{\mu M}$ & $\boldsymbol{\Omega_H/\mu}$ & $\boldsymbol{a/M}$ \\
\hline \hline 
0.20 & 0.049490	& 0.020192 & 0.101018	& 0.499733 & 0.199886\\ \hline
0.35 & 0.084653	& 0.063126 & 0.180692	& 0.499081 & 0.349357\\ \hline
0.44 & 0.104129	& 0.101657 & 0.231744	& 0.498472 & 0.438660\\ \hline
0.50 & 0.116110	& 0.133342 & 0.267780	& 0.497954 & 0.497954\\ \hline
0.62 & 0.136623	& 0.213487 & 0.346756	& 0.496505 & 0.615670\\ \hline
0.71 & 0.147570	& 0.291552 & 0.414862	& 0.494907 & 0.702769\\ \hline
0.80 & 0.152408	& 0.390252 & 0.495186	& 0.492558 & 0.788092\\ \hline
0.92 & 0.141434	& 0.570923 & 0.637156	& 0.486970 & 0.896049\\ \hline
1.01 & 0.104874	& 0.772441 & 0.800412	& 0.477674 & 0.965054\\ \hline
1.10 & 0.024718	& 1.051402 & 1.052564	& 0.453219 & 0.998896\\ \hline
\end{tabular}}
\caption{Eigenvalues for a scalar cloud with parameters $m =l= 2$. 
Some solutions 
associated with these values appear in Fig.~\ref{fig:TestClouds23} (top panel).}
\label{tab:spectrum2QI}
\end{center}
\end{table}

\begin{table}[h!]
\begin{center}
\resizebox{7.5cm}{!}{
\begin{tabular}{|c|c|c|c|c|c|}
\hline 
$\boldsymbol{\mu R_{H}}$ & $\boldsymbol{\mu r_{H}}$ & $\boldsymbol{\mu a}$ & $\boldsymbol{\mu M}$ & $\boldsymbol{\Omega_H/\mu}$ & $\boldsymbol{a/M}$ \\
\hline \hline 
0.50 & 0.121336	& 0.085596 & 0.257326	& 0.332636 & 0.332635\\ \hline
0.60 & 0.143511	& 0.124801 & 0.312980	& 0.332296 & 0.398752\\ \hline
0.70 & 0.164374	& 0.172498 & 0.371250	& 0.331863 & 0.464640\\ \hline
0.80 & 0.183543	& 0.229488 & 0.432916	& 0.331311 & 0.530098\\ \hline
0.90 & 0.200505	& 0.296954 & 0.498990	& 0.330616 & 0.595110\\ \hline
1.00 & 0.214570	& 0.376460 & 0.570862	& 0.329729 & 0.659459\\ \hline
1.10 & 0.224748	& 0.470229 & 0.650508	& 0.328574 & 0.722864\\ \hline
1.20 & 0.229554	& 0.581511 & 0.740902	& 0.327026 & 0.784869\\ \hline
1.30 & 0.226633	& 0.715231 & 0.846762	& 0.324863 & 0.844666\\ \hline
1.50 & 0.173267	& 1.048926 & 1.104686	& 0.327146 & 0.949524\\ \hline
\end{tabular}}
\caption{Eigenvalues for a scalar cloud with parameters $m = l=3$. Some solutions 
associated with these values appear in Fig.~\ref{fig:TestClouds23} (bottom panel).}
\label{tab:spectrum3QI}
\end{center}
\end{table}


\section{Hairy black holes properties}
\label{app:HBHT}

The Tables~\ref{tab:GlobalQuantities1}-\ref{tab:GlobalQuantities6} display some properties of hairy black holes
associated with the numerical solutions of the Einstein-Klein-Gordon system discussed in  Section~\ref{sec:numint}.


\begin{table*}[h!]
\begin{center}
\resizebox{15.1cm}{!}{
\begin{tabular}{|c|c|c|c|c|c|c|c|c|c|c|c|c|}
\hline 
$\boldsymbol{\Omega_{H}/\mu}$ & $\boldsymbol{\mu M_{\rm ADM}}$ & $\boldsymbol{\mu^2J_{\infty}}$ & $\boldsymbol{M_{\rm BH}}$(\textdiscount) & $\boldsymbol{M_{\Psi}}$(\textdiscount) & $\boldsymbol{\mu M_{\rm BH}}$ & $\boldsymbol{\mu M_{\Psi}}$ & $\boldsymbol{\mu^2J_{\rm BH}}$ & $\boldsymbol{\mu^2J_{\Psi}}$ & $\boldsymbol{\mu^2\mathcal{Q}}$ & $\boldsymbol{q}$ & $\boldsymbol{\kappa/\mu}$ & $\boldsymbol{J_{\infty}/{M^2_{\rm ADM}}}$ \\
\hline \hline
0.92988 &	0.81204	& 0.63310	& 27.4350	& 72.5650	& 0.22278	& 0.58926	& 0.04947	& 0.58363	& 0.58363	& 0.92186 &	0.67177 &	0.9601\\ \hline
0.93018	& 0.80851	& 0.62792	& 27.9207	& 72.0793	& 0.22574	& 0.58277	& 0.05104	& 0.57688	& 0.57688	& 0.91871	& 0.65875	& 0.9606\\ \hline
0.93078	& 0.80279	& 0.61994	& 28.6299	& 71.3702	& 0.22984	& 0.57295	& 0.05320	& 0.56674	& 0.56674	& 0.91418	& 0.64185	& 0.9619\\ \hline
0.93137	& 0.79774	& 0.61315	& 29.2102	& 70.7900	& 0.23302	& 0.56472	& 0.05487	& 0.55828	& 0.55828	& 0.91051	& 0.62954	& 0.9635\\ \hline
0.93316	& 0.78405	& 0.59545	& 30.6678	& 69.3326	& 0.24045	& 0.54360	& 0.05873	& 0.53673	& 0.53673	& 0.90138	& 0.60325	& 0.9686\\ \hline
0.93494	& 0.77126	& 0.57951	& 31.9509	& 68.0498	& 0.24643	& 0.52484	& 0.06179	& 0.51772	& 0.51772	& 0.89338	& 0.58424	& 0.9742\\ \hline
0.93673	& 0.75883	& 0.56435	& 33.1642	& 66.8369	& 0.25166	& 0.50718	& 0.06444	& 0.49991	& 0.49991	& 0.88581	& 0.56894	& 0.9801\\ \hline
0.93852	& 0.74655	& 0.54961	& 34.3487	& 65.6529	& 0.25643	& 0.49013	& 0.06684	& 0.48278	& 0.48278	& 0.87839	& 0.55596	& 0.9862\\ \hline
0.94030	& 0.73428	& 0.53510	& 35.5265	& 64.4756	& 0.26086	& 0.47343	& 0.06905	& 0.46606	& 0.46605	& 0.87096	& 0.54462	& 0.9925\\ \hline
0.94150	& 0.72608	& 0.52549	& 36.3152	& 63.6873	& 0.26368	& 0.46242	& 0.07044	& 0.45505	& 0.45505	& 0.86596	& 0.53777	& 0.9968\\ \hline
0.94269	& 0.71784	& 0.51590	& 37.1107	& 62.8922	& 0.26640	& 0.45146	& 0.07178	& 0.44413	& 0.44413	& 0.86087	& 0.53138	& 1.0012\\ \hline
0.94388	& 0.70955	& 0.50632	& 37.9157	& 62.0877	& 0.26903	& 0.44054	& 0.07307	& 0.43325	& 0.43325	& 0.85569	& 0.52539	& 1.0057\\ \hline
0.94507	& 0.70119	& 0.49671	& 38.7332	& 61.2715	& 0.27159	& 0.42963	& 0.07432	& 0.42240	& 0.42239	& 0.85038	& 0.51975	& 1.0103\\ \hline
0.94626	& 0.69277	& 0.48707	& 39.5647	& 60.4401	& 0.27409	& 0.41871	& 0.07553	& 0.41155	& 0.41155	& 0.84493	& 0.51442	& 1.0149\\ \hline
0.94745	& 0.68426	& 0.47740	& 40.4131	& 59.5920	& 0.27653	& 0.40776	& 0.07671	& 0.40069	& 0.40069	& 0.83932	& 0.50936	& 1.0196\\ \hline
0.94864	& 0.67567	& 0.46766	& 41.2803	& 58.7247	& 0.27892	& 0.39678	& 0.07786	& 0.38980	& 0.38980	& 0.83352	& 0.50455	& 1.0244\\ \hline
0.94983	& 0.66697	& 0.45786	& 42.1689	& 57.8360	& 0.28126	& 0.38575	& 0.07898	& 0.37888	& 0.37888	& 0.82750	& 0.49996	& 1.0292\\ \hline
0.95103	& 0.65818	& 0.44798	& 43.0808	& 56.9235	& 0.28355	& 0.37466	& 0.08008	& 0.36791	& 0.36791	& 0.82125	& 0.49558	& 1.0341\\ \hline
0.95222	& 0.64927	& 0.43802	& 44.0187	& 55.9849	& 0.28580	& 0.36349	& 0.08115	& 0.35687	& 0.35687	& 0.81474	& 0.49139	& 1.0391\\ \hline
0.95341	& 0.64025	& 0.42797	& 44.9849	& 55.0173	& 0.28801	& 0.35225	& 0.08220	& 0.34577	& 0.34577	& 0.80793	& 0.48736	& 1.0440\\ \hline
0.95460	& 0.63110	& 0.41781	& 45.9822	& 54.0182	& 0.29019	& 0.34091	& 0.08323	& 0.33458	& 0.33458	& 0.80079	& 0.48350	& 1.0490\\ \hline
0.95579	& 0.62182	& 0.40754	& 47.0133	& 52.9847	& 0.29234	& 0.32947	& 0.08425	& 0.32330	& 0.32330	& 0.79328	& 0.47978	& 1.0540\\ \hline
0.95698	& 0.61239	& 0.39716	& 48.0814	& 51.9135	& 0.29445	& 0.31792	& 0.08525	& 0.31191	& 0.31191	& 0.78536	& 0.47620	& 1.0590\\ \hline
0.95817	& 0.60283	& 0.38665	& 49.1897	& 50.8011	& 0.29653	& 0.30624	& 0.08623	& 0.30042	& 0.30042	& 0.77698	& 0.47275	& 1.0640\\ \hline
0.95936	& 0.59310	& 0.37600	& 50.3420	& 49.6441	& 0.29858	& 0.29444	& 0.08720	& 0.28880	& 0.28880	& 0.76808	& 0.46942	& 1.0689\\ \hline
0.96055	& 0.58322	& 0.36521	& 51.5422	& 48.4378	& 0.30060	& 0.28250	& 0.08816	& 0.27705	& 0.27705	& 0.75861	& 0.46620	& 1.0737\\ \hline
0.96175	& 0.57316	& 0.35426	& 52.7952	& 47.1786	& 0.30260	& 0.27041	& 0.08910	& 0.26516	& 0.26516	& 0.74849	& 0.46308	& 1.0784\\ \hline
0.96294	& 0.56292	& 0.34315	& 54.1058	& 45.8615	& 0.30457	& 0.25816	& 0.09003	& 0.25313	& 0.25312	& 0.73765	& 0.46007	& 1.0829\\ \hline
0.96413	& 0.55248	& 0.33187	& 55.4799	& 44.4812	& 0.30652	& 0.24575	& 0.09094	& 0.24093	& 0.24093	& 0.72598	& 0.45716	& 1.0873\\ \hline
0.96532	& 0.54185	& 0.32041	& 56.9237	& 43.0318	& 0.30844	& 0.23317	& 0.09184	& 0.22857	& 0.22857	& 0.71337	& 0.45433	& 1.0913\\ \hline
0.96651	& 0.53099	& 0.30876	& 58.4446	& 41.5069	& 0.31034	& 0.22040	& 0.09272	& 0.21604	& 0.21604	& 0.69971	& 0.45159	& 1.0951\\ \hline
0.96770	& 0.51991	& 0.29690	& 60.0511	& 39.8994	& 0.31221	& 0.20744	& 0.09358	& 0.20332	& 0.20332	& 0.68483	& 0.44893	& 1.0984\\ \hline
0.96889	& 0.50858	& 0.28482	& 61.7524	& 38.2011	& 0.31406	& 0.19429	& 0.09440	& 0.19042	& 0.19042	& 0.66855	& 0.44635	& 1.1012\\ \hline
0.97008	& 0.49700	& 0.27252	& 63.5594	& 36.4027	& 0.31589	& 0.18092	& 0.09520	& 0.17732	& 0.17732	& 0.65066	& 0.44385	& 1.1033\\ \hline
0.97128	& 0.48515	& 0.25997	& 65.4846	& 34.4931	& 0.31770	& 0.16734	& 0.09597	& 0.16401	& 0.16401	& 0.63086	& 0.44142	& 1.1045\\ \hline
0.97247	& 0.47302	& 0.24717	& 67.5419	& 32.4590	& 0.31949	& 0.15354	& 0.09669	& 0.15048	& 0.15048	& 0.60881	& 0.43906	& 1.1047\\ \hline
0.97366	& 0.46059	& 0.23409	& 69.7474	& 30.2842	& 0.32125	& 0.13949	& 0.09738	& 0.13671	& 0.13671	& 0.58402	& 0.43677	& 1.1035\\ \hline
0.97425	& 0.45426	& 0.22744	& 70.9117	& 29.1379	& 0.32212	& 0.13236	& 0.09771	& 0.12974	& 0.12973	& 0.57040	& 0.43565	& 1.1022\\ \hline
0.97485	& 0.44785	& 0.22072	& 72.1200	& 27.9484	& 0.32299	& 0.12517	& 0.09803	& 0.12269	& 0.12269	& 0.55586	& 0.43454	& 1.1005\\ \hline
0.97544	& 0.44136	& 0.21392	& 73.3756	& 26.7126	& 0.32385	& 0.11790	& 0.09835	& 0.11557	& 0.11557	& 0.54025	& 0.43345	& 1.0981\\ \hline
0.97604	& 0.43479	& 0.20704	& 74.6812	& 25.4264	& 0.32471	& 0.11055	& 0.09866	& 0.10838	& 0.10837	& 0.52345	& 0.43238	& 1.0952\\ \hline
0.97664	& 0.42814	& 0.20007	& 76.0405	& 24.0859	& 0.32556	& 0.10312	& 0.09898	& 0.10110	& 0.10110	& 0.50530	& 0.43132	& 1.0915\\ \hline
0.97723	& 0.42141	& 0.19302	& 77.4568	& 22.6862	& 0.32641	& 0.09560	& 0.09929	& 0.09373	& 0.09373	& 0.48559	& 0.43028	& 1.0869\\ \hline
0.97783	& 0.41458	& 0.18588	& 78.9343	& 21.2222	& 0.32725	& 0.08798	& 0.09962	& 0.08627	& 0.08627	& 0.46410	& 0.42925	& 1.0815\\ \hline
0.97842	& 0.40767	& 0.17865	& 80.4776	& 19.6881	& 0.32808	& 0.08026	& 0.09995	& 0.07870	& 0.07870	& 0.44054	& 0.42824	& 1.0749\\ \hline
\end{tabular}}
\caption{Global quantities and parameters associated with a sequence of rotating hairy black holes with an event horizon located at $\mu r_H = 0.065804$. The columns correspond, respectively, to the angular velocity  $\Omega_H$ (in units of $\mu$), the ADM mass $M_{\rm ADM}=M_{\rm K}$ (in units of $1/\mu$), the angular momentum $J_{\infty}$ (in units of $1/\mu^2$), the percentage of the BH mass (at the horizon) relative to the total mass, the percentage of the hair contribution to the total mass, the BH and hair masses $M_{\rm BH}$ and $M_\Psi$ (in units of $1/\mu$) defined as $M^{{\cal H}^+}$ and $M^{\Sigma_t}$ in Sec.~\ref{sec:globalq}, the BH and hair angular momenta $J_{\rm BH}$ and $J_\Psi$ (in units of $1/\mu^2$) defined as $J^{{\cal H}^+}$ and $J^{\Sigma_t}$ in Sec.~\ref{sec:globalq}, the total boson number $\cal Q$ (in units of $1/\mu^2$), the dimensionless ratio $q= m{\cal Q}/J_{\rm K}$, the surface gravity $\kappa$ (in units of $\mu$), and the dimensionless 
parameter $J_{\infty}/M_{\rm  ADM}^2$. Note that in these examples $J_\Psi= {\cal Q}$
(taking $\hbar=1$) since throughout this 
work we take the azimuthal number $m = 1$ [cf. Eq.~(\ref{JmQ})].}\label{tab:GlobalQuantities1}
\end{center}
\end{table*}

\begin{table*}[h!]
\begin{center}
\resizebox{15.1cm}{!}{
\begin{tabular}{|c|c|c|c|c|c|c|c|c|c|c|c|c|}
\hline 
$\boldsymbol{\Omega_{H}/\mu}$ & $\boldsymbol{\mu M_{\rm ADM}}$ & $\boldsymbol{\mu^2J_{\infty}}$ & $\boldsymbol{M_{\rm BH}}$(\textdiscount) & $\boldsymbol{M_{\Psi}}$(\textdiscount) & $\boldsymbol{\mu M_{\rm BH}}$ & $\boldsymbol{\mu M_{\Psi}}$ & $\boldsymbol{\mu^2J_{\rm BH}}$ & $\boldsymbol{\mu^2J_{\Psi}}$ & $\boldsymbol{\mu^2\mathcal{Q}}$ & $\boldsymbol{q}$ & $\boldsymbol{\kappa/\mu}$ & $\boldsymbol{J_{\infty}/{M^2_{\rm ADM}}}$ \\
\hline \hline
0.91781	& 0.86192	& 0.68500	& 25.7905	& 74.2090	& 0.22229	& 0.63962	& 0.05246	& 0.63254	& 0.63254	& 0.92342	& 0.62823	& 0.92206\\ \hline
0.92070	& 0.84161	& 0.65793	& 27.8515	& 72.1483	& 0.23440	& 0.60720	& 0.05883	& 0.59910	& 0.59910	& 0.91058	& 0.58459	& 0.92889\\ \hline
0.92359	& 0.82356	& 0.63511	& 29.5362	& 70.4637	& 0.24325	& 0.58031	& 0.06340	& 0.57171	& 0.57171	& 0.90018	& 0.55770	& 0.93638\\ \hline
0.92648	& 0.80613	& 0.61368	& 31.1157	& 68.8844	& 0.25083	& 0.55530	& 0.06725	& 0.54643	& 0.54642	& 0.89041	& 0.53729	& 0.94435\\ \hline
0.92938	& 0.78883	& 0.59285	& 32.6670	& 67.3333	& 0.25769	& 0.53114	& 0.07069	& 0.52216	& 0.52216	& 0.88076	& 0.52059	& 0.95275\\ \hline
0.93227	& 0.77142	& 0.57224	& 34.2286	& 65.7723	& 0.26405	& 0.50738	& 0.07385	& 0.49839	& 0.49839	& 0.87095	& 0.50636	& 0.96159\\ \hline
0.93516	& 0.75378	& 0.55163	& 35.8257	& 64.1759	& 0.27005	& 0.48375	& 0.07679	& 0.47484	& 0.47484	& 0.86080	& 0.49392	& 0.97085\\ \hline
0.93805	& 0.73579	& 0.53086	& 37.4786	& 62.5243	& 0.27576	& 0.46005	& 0.07956	& 0.45130	& 0.45130	& 0.85013	& 0.48286	& 0.98056\\ \hline
0.93921	& 0.72848	& 0.52249	& 38.1594	& 61.8439	& 0.27798	& 0.45052	& 0.08063	& 0.44186	& 0.44186	& 0.84568	& 0.47876	& 0.98455\\ \hline
0.94036	& 0.72110	& 0.51406	& 38.8532	& 61.1504	& 0.28017	& 0.44095	& 0.08168	& 0.43238	& 0.43238	& 0.84111	& 0.47482	& 0.98862\\ \hline
0.94152	& 0.71364	& 0.50558	& 39.5612	& 60.4427	& 0.28232	& 0.43134	& 0.08271	& 0.42287	& 0.42287	& 0.83641	& 0.47102	& 0.99274\\ \hline
0.94267	& 0.70609	& 0.49704	& 40.2846	& 59.7196	& 0.28445	& 0.42168	& 0.08372	& 0.41332	& 0.41332	& 0.83157	& 0.46737	& 0.99693\\ \hline
0.94383	& 0.69846	& 0.48842	& 41.0247	& 58.9797	& 0.28654	& 0.41195	& 0.08471	& 0.40372	& 0.40372	& 0.82656	& 0.46384	& 1.00118\\ \hline
0.94499	& 0.69074	& 0.47974	& 41.7827	& 58.2218	& 0.28861	& 0.40216	& 0.08569	& 0.39406	& 0.39405	& 0.82139	& 0.46043	& 1.00549\\ \hline
0.94614	& 0.68292	& 0.47098	& 42.5601	& 57.4443	& 0.29065	& 0.39230	& 0.08665	& 0.38433	& 0.38433	& 0.81602	& 0.45714	& 1.00986\\ \hline
0.94730	& 0.67501	& 0.46214	& 43.3583	& 56.6459	& 0.29267	& 0.38236	& 0.08760	& 0.37454	& 0.37454	& 0.81045	& 0.45395	& 1.01428\\ \hline
0.94846	& 0.66698	& 0.45320	& 44.1787	& 55.8249	& 0.29466	& 0.37234	& 0.08853	& 0.36467	& 0.36467	& 0.80465	& 0.45087	& 1.01874\\ \hline
0.94961	& 0.65885	& 0.44418	& 45.0233	& 54.9796	& 0.29664	& 0.36224	& 0.08946	& 0.35472	& 0.35472	& 0.79861	& 0.44788	& 1.02325\\ \hline
0.95077	& 0.65061	& 0.43505	& 45.8935	& 54.1081	& 0.29859	& 0.35203	& 0.09037	& 0.34469	& 0.34469	& 0.79229	& 0.44497	& 1.02777\\ \hline
0.95193	& 0.64225	& 0.42583	& 46.7916	& 53.2085	& 0.30052	& 0.34173	& 0.09126	& 0.33457	& 0.33457	& 0.78568	& 0.44216	& 1.03234\\ \hline
0.95308	& 0.63377	& 0.41649	& 47.7193	& 52.2786	& 0.30243	& 0.33133	& 0.09215	& 0.32434	& 0.32434	& 0.77875	& 0.43942	& 1.03691\\ \hline
0.95424	& 0.62516	& 0.40704	& 48.6790	& 51.3162	& 0.30432	& 0.32081	& 0.09303	& 0.31401	& 0.31401	& 0.77145	& 0.43676	& 1.04148\\ \hline
0.95540	& 0.61642	& 0.39746	& 49.6732	& 50.3187	& 0.30619	& 0.31017	& 0.09390	& 0.30357	& 0.30357	& 0.76376	& 0.43417	& 1.04605\\ \hline
0.95655	& 0.60753	& 0.38776	& 50.7045	& 49.2833	& 0.30805	& 0.29941	& 0.09476	& 0.29301	& 0.29301	& 0.75564	& 0.43166	& 1.05058\\ \hline
0.95771	& 0.59851	& 0.37793	& 51.7760	& 48.2072	& 0.30988	& 0.28852	& 0.09561	& 0.28233	& 0.28233	& 0.74703	& 0.42921	& 1.05506\\ \hline
0.95887	& 0.58933	& 0.36796	& 52.8908	& 47.0868	& 0.31170	& 0.27750	& 0.09645	& 0.27152	& 0.27152	& 0.73789	& 0.42682	& 1.05947\\ \hline
0.96002	& 0.57999	& 0.35785	& 54.0527	& 45.9192	& 0.31350	& 0.26633	& 0.09728	& 0.26057	& 0.26057	& 0.72815	& 0.42450	& 1.06380\\ \hline
0.96118	& 0.57049	& 0.34758	& 55.2659	& 44.7004	& 0.31528	& 0.25501	& 0.09810	& 0.24948	& 0.24948	& 0.71776	& 0.42224	& 1.06799\\ \hline
0.96233	& 0.56081	& 0.33716	& 56.5349	& 43.4262	& 0.31705	& 0.24354	& 0.09892	& 0.23824	& 0.23824	& 0.70662	& 0.42003	& 1.07204\\ \hline
0.96349	& 0.55094	& 0.32656	& 57.8647	& 42.0920	& 0.31880	& 0.23190	& 0.09972	& 0.22685	& 0.22685	& 0.69465	& 0.41788	& 1.07586\\ \hline
0.96465	& 0.54088	& 0.31580	& 59.2613	& 40.6928	& 0.32054	& 0.22010	& 0.10050	& 0.21530	& 0.21529	& 0.68175	& 0.41578	& 1.07944\\ \hline
0.96580	& 0.53062	& 0.30484	& 60.7309	& 39.2229	& 0.32225	& 0.20813	& 0.10127	& 0.20358	& 0.20358	& 0.66780	& 0.41374	& 1.08269\\ \hline
0.96696	& 0.52015	& 0.29370	& 62.2809	& 37.6761	& 0.32395	& 0.19597	& 0.10202	& 0.19168	& 0.19168	& 0.65265	& 0.41174	& 1.08555\\ \hline
0.96812	& 0.50945	& 0.28235	& 63.9195	& 36.0450	& 0.32564	& 0.18363	& 0.10274	& 0.17961	& 0.17961	& 0.63613	& 0.40979	& 1.08790\\ \hline
0.96927	& 0.49852	& 0.27079	& 65.6564	& 34.3214	& 0.32731	& 0.17110	& 0.10343	& 0.16736	& 0.16736	& 0.61803	& 0.40789	& 1.08962\\ \hline
0.97043	& 0.48733	& 0.25900	& 67.5017	& 32.4954	& 0.32896	& 0.15836	& 0.10410	& 0.15491	& 0.15490	& 0.59808	& 0.40604	& 1.09057\\ \hline
0.97159	& 0.47590	& 0.24698	& 69.4676	& 30.5550	& 0.33059	& 0.14541	& 0.10473	& 0.14225	& 0.14225	& 0.57594	& 0.40423	& 1.09052\\ \hline
0.97274	& 0.46420	& 0.23470	& 71.5675	& 28.4859	& 0.33221	& 0.13223	& 0.10534	& 0.12936	& 0.12936	& 0.55118	& 0.40246	& 1.08921\\ \hline
0.97390	& 0.45222	& 0.22216	& 73.8172	& 26.2704	& 0.33382	& 0.11880	& 0.10592	& 0.11624	& 0.11624	& 0.52321	& 0.40074	& 1.08633\\ \hline
0.97506	& 0.43997	& 0.20934	& 76.2343	& 23.8866	& 0.33540	& 0.10509	& 0.10650	& 0.10284	& 0.10284	& 0.49126	& 0.39906	& 1.08146\\ \hline
0.97621	& 0.42741	& 0.19621	& 78.8401	& 21.3080	& 0.33697	& 0.09107	& 0.10708	& 0.08913	& 0.08913	& 0.45427	& 0.39741	& 1.07406\\ \hline
0.97737	& 0.41456	& 0.18277	& 81.6599	& 18.5021	& 0.33853	& 0.07670	& 0.10769	& 0.07508	& 0.07508	& 0.41079	& 0.39581	& 1.06349\\ \hline
0.97852	& 0.40137	& 0.16898	& 84.7247	& 15.4311	& 0.34006	& 0.06194	& 0.10835	& 0.06063	& 0.06063	& 0.35881	& 0.39425	& 1.04893\\ \hline
0.97968	& 0.38783	& 0.15483	& 88.0747	& 12.0506	& 0.34158	& 0.04674	& 0.10907	& 0.04576	& 0.04576	& 0.29552	& 0.39273	& 1.02937\\ \hline
0.98084	& 0.37388	& 0.14028	& 91.7606	& 8.3133	& 0.34308	& 0.03108	& 0.10985	& 0.03043	& 0.03043	& 0.21692	& 0.39125	& 1.00351\\ \hline
0.98199	& 0.35947	& 0.12530	& 95.8506	& 4.1692	& 0.34456	& 0.01499	& 0.11063	& 0.01467	& 0.01467	& 0.11708	& 0.38980	& 0.96968\\ \hline
\end{tabular}}
\caption{Global quantities and parameters associated with a sequence of hairy black holes with horizon at $\mu r_H = 0.063567$. The description of columns are like in Table~\ref{tab:GlobalQuantities1}.
}\label{tab:GlobalQuantities2}
\end{center}
\end{table*}

\begin{table*}[h!]
\begin{center}
\resizebox{15.1cm}{!}{
\begin{tabular}{|c|c|c|c|c|c|c|c|c|c|c|c|c|}
\hline 
$\boldsymbol{\Omega_{H}/\mu}$ & $\boldsymbol{\mu M_{\rm ADM}}$ & $\boldsymbol{\mu^2J_{\infty}}$ & $\boldsymbol{M_{\rm BH}}$(\textdiscount) & $\boldsymbol{M_{\Psi}}$(\textdiscount) & $\boldsymbol{\mu M_{\rm BH}}$ & $\boldsymbol{\mu M_{\Psi}}$ & $\boldsymbol{\mu^2J_{\rm BH}}$ & $\boldsymbol{\mu^2J_{\Psi}}$ & $\boldsymbol{\mu^2\mathcal{Q}}$ & $\boldsymbol{q}$ & $\boldsymbol{\kappa/\mu}$ & $\boldsymbol{J_{\infty}/{M^2_{\rm ADM}}}$ \\
\hline \hline
0.90155	& 0.92761	& 0.76171	& 22.2587	& 77.7410	& 0.20647	& 0.72114	& 0.04785	& 0.71386	& 0.71386	& 0.93718	& 0.64054	& 0.88522\\ \hline
0.90659	& 0.89423	& 0.71549	& 25.5517	& 74.4479	& 0.22849	& 0.66574	& 0.05965	& 0.65584	& 0.65584	& 0.91663	& 0.55779	& 0.89476\\ \hline
0.91276	& 0.86166	& 0.67409	& 28.4042	& 71.5955	& 0.24475	& 0.61691	& 0.06806	& 0.60603	& 0.60603	& 0.89904	& 0.51331	& 0.90792\\ \hline
0.91836	& 0.83278	& 0.63880	& 30.8748	& 69.1252	& 0.25712	& 0.57566	& 0.07429	& 0.56451	& 0.56451	& 0.88370	& 0.48560	& 0.92111\\ \hline
0.92340	& 0.80636	& 0.60733	& 33.1416	& 66.8588	& 0.26724	& 0.53912	& 0.07929	& 0.52804	& 0.52804	& 0.86944	& 0.46589	& 0.93404\\ \hline
0.92845	& 0.77911	& 0.57548	& 35.5181	& 64.4831	& 0.27673	& 0.50239	& 0.08390	& 0.49158	& 0.49158	& 0.85421	& 0.44939	& 0.94806\\ \hline
0.93013	& 0.76979	& 0.56471	& 36.3443	& 63.6572	& 0.27977	& 0.49003	& 0.08537	& 0.47935	& 0.47935	& 0.84884	& 0.44443	& 0.95298\\ \hline
0.93293	& 0.75394	& 0.54653	& 37.7680	& 62.2342	& 0.28475	& 0.46921	& 0.08774	& 0.45879	& 0.45879	& 0.83947	& 0.43668	& 0.96146\\ \hline
0.93573	& 0.73768	& 0.52801	& 39.2588	& 60.7441	& 0.28960	& 0.44810	& 0.09003	& 0.43798	& 0.43798	& 0.82949	& 0.42948	& 0.97030\\ \hline
0.93853	& 0.72095	& 0.50911	& 40.8278	& 59.1755	& 0.29435	& 0.42663	& 0.09226	& 0.41686	& 0.41686	& 0.81880	& 0.42278	& 0.97948\\ \hline
0.94133	& 0.70372	& 0.48978	& 42.4878	& 57.5159	& 0.29899	& 0.40475	& 0.09441	& 0.39537	& 0.39537	& 0.80724	& 0.41650	& 0.98901\\ \hline
0.94414	& 0.68594 & 0.46996	& 44.2529	& 55.7501	& 0.30355	& 0.38241	& 0.09651	& 0.37345	& 0.37345	& 0.79464	& 0.41061	& 0.99884\\ \hline
0.94694	& 0.66755	& 0.44961	& 46.1402	& 53.8608	& 0.30801	& 0.35955	& 0.09856	& 0.35105	& 0.35105	& 0.78079	& 0.40506	& 1.00894\\ \hline
0.95030	& 0.64464	& 0.42441	& 48.5944	& 51.4011	& 0.31326	& 0.33135	& 0.10096	& 0.32345	& 0.32345	& 0.76212	& 0.39882	& 1.02130\\ \hline
0.95198	& 0.63280	& 0.41145	& 49.9114	& 50.0800	& 0.31584	& 0.31690	& 0.10213	& 0.30932	& 0.30932	& 0.75178	& 0.39586	& 1.02752\\ \hline
0.95366	& 0.62068	& 0.39825	& 51.2965	& 48.6895	& 0.31839	& 0.30221	& 0.10330	& 0.29495	& 0.29495	& 0.74063	& 0.39299	& 1.03374\\ \hline
0.95534	& 0.60828	& 0.38477	& 52.7571	& 47.2225	& 0.32091	& 0.28725	& 0.10444	& 0.28033	& 0.28033	& 0.72856	& 0.39021	& 1.03990\\ \hline
0.95702	& 0.59558	& 0.37102	& 54.3017	& 45.6709	& 0.32341	& 0.27201	& 0.10558	& 0.26544	& 0.26544	& 0.71545	& 0.38753	& 1.04595\\ \hline
0.95870	& 0.58255	& 0.35696	& 55.9400	& 44.0253	& 0.32588	& 0.25647	& 0.10669	& 0.25027	& 0.25027	& 0.70111	& 0.38492	& 1.05184\\ \hline
0.96039	& 0.56918	& 0.34259	& 57.6836	& 42.2757	& 0.32832	& 0.24062	& 0.10779	& 0.23480	& 0.23480	& 0.68537	& 0.38240	& 1.05749\\ \hline
0.96207	& 0.55543	& 0.32788	& 59.5459	& 40.4101	& 0.33074	& 0.22445	& 0.10886	& 0.21901	& 0.21901	& 0.66798	& 0.37995	& 1.06278\\ \hline
0.96375	& 0.54129	& 0.31281	& 61.5426	& 38.4149	& 0.33313	& 0.20794	& 0.10990	& 0.20290	& 0.20290	& 0.64865	& 0.37758	& 1.06760\\ \hline
0.96543	& 0.52673	& 0.29735	& 63.6923	& 36.2741	& 0.33549	& 0.19107	& 0.11091	& 0.18645	& 0.18645	& 0.62702	& 0.37529	& 1.07174\\ \hline
0.96711	& 0.51172	& 0.28149	& 66.0166	& 33.9681	& 0.33782	& 0.17382	& 0.11186	& 0.16963	& 0.16963	& 0.60262	& 0.37306	& 1.07496\\ \hline
0.96879	& 0.49624	& 0.26519	& 68.5414	& 31.4727	& 0.34013	& 0.15618	& 0.11276	& 0.15243	& 0.15243	& 0.57480	& 0.37090	& 1.07689\\ \hline
0.96935	& 0.49096	& 0.25965	& 69.4328	& 30.5937	& 0.34089	& 0.15020	& 0.11304	& 0.14660	& 0.14660	& 0.56463	& 0.37019	& 1.07717\\ \hline
0.96991	& 0.48564	& 0.25406	& 70.3509	& 29.6889	& 0.34165	& 0.14418	& 0.11333	& 0.14073	& 0.14073	& 0.55394	& 0.36950	& 1.07723\\ \hline
0.97047	& 0.48025	& 0.24841	& 71.2972	& 28.7568	& 0.34241	& 0.13810	& 0.11360	& 0.13481	& 0.13481	& 0.54269	& 0.36880	& 1.07705\\ \hline
0.97103	& 0.47481	& 0.24271	& 72.2731	& 27.7957	& 0.34316	& 0.13198	& 0.11387	& 0.12884	& 0.12883	& 0.53082	& 0.36812	& 1.07659\\ \hline
0.97159	& 0.46930	& 0.23695	& 73.2801	& 26.8040	& 0.34391	& 0.12579	& 0.11414	& 0.12281	& 0.12281	& 0.51829	& 0.36744	& 1.07583\\ \hline
0.97215	& 0.46374	& 0.23113	& 74.3202	& 25.7794	& 0.34465	& 0.11955	& 0.11441	& 0.11672	& 0.11672	& 0.50501	& 0.36678	& 1.07473\\ \hline
0.97271	& 0.45812	& 0.22525	& 75.3943	& 24.7200	& 0.34540	& 0.11325	& 0.11467	& 0.11058	& 0.11058	& 0.49091	& 0.36611	& 1.07325\\ \hline
0.97327	& 0.45244	& 0.21930	& 76.5050	& 23.6234	& 0.34614	& 0.10688	& 0.11494	& 0.10437	& 0.10437	& 0.47591	& 0.36546	& 1.07135\\ \hline
0.97383	& 0.44669	& 0.21329	& 77.6539	& 22.4869	& 0.34687	& 0.10045	& 0.11520	& 0.09809	& 0.09809	& 0.45989	& 0.36481	& 1.06897\\ \hline
0.97439	& 0.44088	& 0.20722	& 78.8436	& 21.3078	& 0.34761	& 0.09394	& 0.11547	& 0.09175	& 0.09175	& 0.44275	& 0.36417	& 1.06608\\ \hline
0.97495	& 0.43501	& 0.20108	& 80.0758	& 20.0828	& 0.34833	& 0.08736	& 0.11575	& 0.08533	& 0.08533	& 0.42435	& 0.36353	& 1.06260\\ \hline
0.97551	& 0.42907	& 0.19486	& 81.3541	& 18.8088	& 0.34906	& 0.08070	& 0.11604	& 0.07883	& 0.07883	& 0.40454	& 0.36291	& 1.05849\\ \hline
0.97607	& 0.42305	& 0.18858	& 82.6806	& 17.4822	& 0.34978	& 0.07396	& 0.11633	& 0.07225	& 0.07225	& 0.38312	& 0.36229	& 1.05366\\ \hline
0.97663	& 0.41697	& 0.18222	& 84.0589	& 16.0993	& 0.35050	& 0.06713	& 0.11664	& 0.06558	& 0.06558	& 0.35990	& 0.36167	& 1.04803\\ \hline
0.97720	& 0.41082	& 0.17578	& 85.4926	& 14.6560	& 0.35122	& 0.06021	& 0.11696	& 0.05882	& 0.05882	& 0.33464	& 0.36106	& 1.04154\\ \hline
0.97776	& 0.40458	& 0.16926	& 86.9859	& 13.1483	& 0.35193	& 0.05320	& 0.11729	& 0.05197	& 0.05197	& 0.30706	& 0.36046	& 1.03407\\ \hline
0.97832	& 0.39827	& 0.16266	& 88.5432	& 11.5717	& 0.35264	& 0.04609	& 0.11764	& 0.04503	& 0.04503	& 0.27682	& 0.35987	& 1.02552\\ \hline
0.97888	& 0.39186	& 0.15598	& 90.1698	& 9.9219	& 0.35334	& 0.03888	& 0.11799	& 0.03799	& 0.03799	& 0.24354	& 0.35928	& 1.01578\\ \hline
0.97944	& 0.38537	& 0.14921	& 91.8719	& 8.1945	& 0.35404	& 0.03158	& 0.11836	& 0.03085	& 0.03085	& 0.20678	& 0.35870	& 1.00472\\ \hline
0.98000	& 0.37877	& 0.14235	& 93.6555	& 6.3854	& 0.35474	& 0.02419	& 0.11872	& 0.02363	& 0.02363	& 0.16600	& 0.35813	& 0.99218\\ \hline
0.98145	& 0.36113	& 0.12408	& 98.7263	& 1.2725	& 0.35653	& 0.00460	& 0.11959	& 0.00449	& 0.00449	& 0.03617	& 0.35667	& 0.95142\\ \hline
\end{tabular}}
\caption{Global quantities and parameters associated with a sequence of hairy black holes with horizon at $\mu r_H = 0.060889$. The description of columns are like in Table~\ref{tab:GlobalQuantities1}.}\label{tab:GlobalQuantities3}
\end{center}
\end{table*}

\begin{table*}[h!]
\begin{center}
\resizebox{15.1cm}{!}{
\begin{tabular}{|c|c|c|c|c|c|c|c|c|c|c|c|c|}
\hline 
$\boldsymbol{\Omega_{H}/\mu}$ & $\boldsymbol{\mu M_{\rm ADM}}$ & $\boldsymbol{\mu^2J_{\infty}}$ & $\boldsymbol{M_{\rm BH}}$(\textdiscount) & $\boldsymbol{M_{\Psi}}$(\textdiscount) & $\boldsymbol{\mu M_{\rm BH}}$ & $\boldsymbol{\mu M_{\Psi}}$ & $\boldsymbol{\mu^2J_{\rm BH}}$ & $\boldsymbol{\mu^2J_{\Psi}}$ & $\boldsymbol{\mu^2\mathcal{Q}}$ & $\boldsymbol{q}$ & $\boldsymbol{\kappa/\mu}$ & $\boldsymbol{J_{\infty}/{M^2_{\rm ADM}}}$ \\
\hline \hline
0.88289	& 0.98692	& 0.83265	& 19.2715	& 80.7280	& 0.19019	& 0.79672	& 0.04364	& 0.78901	& 0.78901	& 0.94760	& 0.64591	& 0.85487\\ \hline
0.88831	& 0.95134	& 0.78110	& 22.7088	& 77.2908	& 0.21604	& 0.73530	& 0.05785	& 0.72325	& 0.72325	& 0.92590	& 0.54137	& 0.86304\\ \hline
0.89372	& 0.92642	& 0.74810	& 24.8664	& 75.1332	& 0.23037	& 0.69605	& 0.06541	& 0.68269	& 0.68269	& 0.91260	& 0.50110	& 0.87166\\ \hline
0.89914	& 0.90275	& 0.71795	& 26.8647	& 73.1351	& 0.24252	& 0.66023	& 0.07168	& 0.64627	& 0.64627	& 0.90020	& 0.47300	& 0.88096\\ \hline
0.90456	& 0.87915	& 0.68869	& 28.8414	& 71.1583	& 0.25356	& 0.62559	& 0.07725	& 0.61144	& 0.61144	& 0.88780	& 0.45108	& 0.89104\\ \hline
0.90997	& 0.85509	& 0.65952	& 30.8603	& 69.1395	& 0.26388	& 0.59121	& 0.08237	& 0.57715	& 0.57715	& 0.87510	& 0.43304	& 0.90199\\ \hline
0.91539	& 0.83024	& 0.62995	& 32.9655	& 67.0344	& 0.27369	& 0.55655	& 0.08716	& 0.54279	& 0.54279	& 0.86160	& 0.41771	& 0.91390\\ \hline
0.92081	& 0.80434	& 0.59964	& 35.1965	& 64.8036	& 0.28310	& 0.52124	& 0.09167	& 0.50797	& 0.50797	& 0.84710	& 0.40440	& 0.92687\\ \hline
0.92622	& 0.77717	& 0.56833	& 37.5945	& 62.4057	& 0.29217	& 0.48500	& 0.09596	& 0.47237	& 0.47237	& 0.83120	& 0.39267	& 0.94095\\ \hline
0.93164	& 0.74851	& 0.53575	& 40.2075	& 59.7926	& 0.30096	& 0.44756	& 0.10004	& 0.43570	& 0.43570	& 0.81330	& 0.38220	& 0.95622\\ \hline
0.93705	& 0.71815	& 0.50165	& 43.0954	& 56.9048	& 0.30949	& 0.40866	& 0.10395	& 0.39770	& 0.39770	& 0.79280	& 0.37279	& 0.97269\\ \hline
0.94247	& 0.68582	& 0.46578	& 46.3358	& 53.6644	& 0.31778	& 0.36804	& 0.10769	& 0.35809	& 0.35809	& 0.76880	& 0.36428	& 0.99029\\ \hline
0.94789	& 0.65125	& 0.42784	& 50.0344	& 49.9659	& 0.32585	& 0.32540	& 0.11128	& 0.31657	& 0.31656	& 0.73990	& 0.35653	& 1.00878\\ \hline
0.95330	& 0.61407	& 0.38748	& 54.3405	& 45.6600	& 0.33369	& 0.28039	& 0.11471	& 0.27277	& 0.27277	& 0.70400	& 0.34945	& 1.02756\\ \hline
0.95872	& 0.57388	& 0.34427	& 59.4744	& 40.5262	& 0.34131	& 0.23257	& 0.11799	& 0.22627	& 0.22627	& 0.65730	& 0.34297	& 1.04535\\ \hline
0.96414	& 0.53011	& 0.29767	& 65.7786	& 34.2222	& 0.34870	& 0.18142	& 0.12113	& 0.17654	& 0.17654	& 0.59310	& 0.33704	& 1.05924\\ \hline
0.96955	& 0.48205	& 0.24696	& 73.8205	& 26.1805	& 0.35585	& 0.12620	& 0.12410	& 0.12285	& 0.12285	& 0.49750	& 0.33161	& 1.06278\\ \hline
0.97497	& 0.42865	& 0.19112	& 84.6222	& 15.3793	& 0.36274	& 0.06592	& 0.12692	& 0.06421	& 0.06421	& 0.33590	& 0.32665	& 1.04017\\ \hline
0.97551	& 0.42297	& 0.18521	& 85.9187	& 14.0827	& 0.36341	& 0.05957	& 0.12719	& 0.05802	& 0.05802	& 0.31330	& 0.32618	& 1.03524\\ \hline
0.97605	& 0.41721	& 0.17922	& 87.2645	& 12.7370	& 0.36408	& 0.05314	& 0.12746	& 0.05176	& 0.05176	& 0.28880	& 0.32572	& 1.02962\\ \hline
0.97660	& 0.41139	& 0.17317	& 88.6627	& 11.3388	& 0.36475	& 0.04665	& 0.12773	& 0.04544	& 0.04544	& 0.26240	& 0.32526	& 1.02322\\ \hline
0.97714	& 0.40548	& 0.16704	& 90.1166	& 9.8847	& 0.36541	& 0.04008	& 0.12800	& 0.03905	& 0.03905	& 0.23380	& 0.32480	& 1.01596\\ \hline
0.97768	& 0.39951	& 0.16084	& 91.6304	& 8.3711	& 0.36607	& 0.03344	& 0.12826	& 0.03258	& 0.03258	& 0.20260	& 0.32435	& 1.00775\\ \hline
0.97822	& 0.39345	& 0.15457	& 93.2080	& 6.7937	& 0.36673	& 0.02673	& 0.12852	& 0.02604	& 0.02604	& 0.16850	& 0.32390	& 0.99846\\ \hline
0.97876	& 0.38731	& 0.14821	& 94.8538	& 5.1478	& 0.36738	& 0.01994	& 0.12878	& 0.01943	& 0.01943	& 0.13110	& 0.32346	& 0.98799\\ \hline
0.97930	& 0.38109	& 0.14178	& 96.5726	& 3.4286	& 0.36803	& 0.01307	& 0.12904	& 0.01273	& 0.01273	& 0.08980	& 0.32302	& 0.97619\\ \hline
0.97985	& 0.37474	& 0.13520	& 98.3856	& 1.6161	& 0.36869	& 0.00606	& 0.12930	& 0.00590	& 0.00590	& 0.04370	& 0.32258	& 0.96281\\ \hline
0.98020	& 0.37063	& 0.13096	& 99.5876	& 0.4142	& 0.36910	& 0.00154	& 0.12947	& 0.00150	& 0.00150	& 0.01140	& 0.32231	& 0.95338\\ \hline
\end{tabular}}
\caption{Global quantities and parameters associated with a sequence of hairy black holes with horizon at $\mu r_H = 0.057648$. The description of columns are like in Table~\ref{tab:GlobalQuantities1}.}\label{tab:GlobalQuantities4}
\end{center}
\end{table*}

\begin{table*}[h!]
\begin{center}
\resizebox{15.1cm}{!}{
\begin{tabular}{|c|c|c|c|c|c|c|c|c|c|c|c|c|}
\hline 
$\boldsymbol{\Omega_{H}/\mu}$ & $\boldsymbol{\mu M_{\rm ADM}}$ & $\boldsymbol{\mu^2J_{\infty}}$ & $\boldsymbol{M_{\rm BH}}$(\textdiscount) & $\boldsymbol{M_{\Psi}}$(\textdiscount) & $\boldsymbol{\mu M_{\rm BH}}$ & $\boldsymbol{\mu M_{\Psi}}$ & $\boldsymbol{\mu^2J_{\rm BH}}$ & $\boldsymbol{\mu^2J_{\Psi}}$ & $\boldsymbol{\mu^2\mathcal{Q}}$ & $\boldsymbol{q}$ & $\boldsymbol{\kappa/\mu}$ & $\boldsymbol{J_{\infty}/{M^2_{\rm ADM}}}$ \\
\hline \hline
0.86072	& 1.03651	& 0.89289	& 16.8732	& 83.1265	& 0.17489	& 0.86162	& 0.04098	& 0.85191	& 0.85191	& 0.95411	& 0.62906	& 0.83109\\ \hline
0.86150	& 1.02755	& 0.87878	& 17.7308	& 82.2690	& 0.18219	& 0.84536	& 0.04518	& 0.83361	& 0.83361	& 0.94859	& 0.59177	& 0.83229\\ \hline
0.86228	& 1.02214	& 0.87057	& 18.2294	& 81.7704	& 0.18633	& 0.83581	& 0.04752	& 0.82305	& 0.82305	& 0.94541	& 0.57330	& 0.83327\\ \hline
0.87011	& 0.98697	& 0.82044	& 21.3387	& 78.6612	& 0.21061	& 0.77636	& 0.06089	& 0.75954	& 0.75954	& 0.92578	& 0.49086	& 0.84224\\ \hline
0.87794	& 0.95828	& 0.78209	& 23.8120	& 76.1879	& 0.22819	& 0.73010	& 0.07020	& 0.71190	& 0.71190	& 0.91025	& 0.44892	& 0.85166\\ \hline
0.88577	& 0.93038	& 0.74623	& 26.1956	& 73.8043	& 0.24372	& 0.68666	& 0.07818	& 0.66806	& 0.66805	& 0.89524	& 0.41942	& 0.86209\\ \hline
0.89360	& 0.90188	& 0.71072	& 28.6221	& 71.3779	& 0.25814	& 0.64374	& 0.08540	& 0.62532	& 0.62532	& 0.87984	& 0.39655	& 0.87378\\ \hline
0.90143	& 0.87205	& 0.67451	& 31.1686	& 68.8316	& 0.27181	& 0.60025	& 0.09210	& 0.58241	& 0.58241	& 0.86346	& 0.37790	& 0.88696\\ \hline
0.91708	& 0.80641	& 0.59740	& 36.8987	& 63.1033	& 0.29755	& 0.50887	& 0.10430	& 0.49310	& 0.49310	& 0.82541	& 0.34877	& 0.91866\\ \hline
0.92961	& 0.74612	& 0.52888	& 42.4793	& 57.5232	& 0.31695	& 0.42919	& 0.11315	& 0.41573	& 0.41573	& 0.78606	& 0.33071	& 0.95003\\ \hline
0.94213	& 0.67661	& 0.45179	& 49.5690	& 50.4171	& 0.33539	& 0.34113	& 0.12134	& 0.33044	& 0.33044	& 0.73142	& 0.31587	& 0.98685\\ \hline
0.95309	& 0.60581	& 0.37495	& 57.8986	& 42.0604	& 0.35076	& 0.25481	& 0.12803	& 0.24692	& 0.24692	& 0.65855	& 0.30495	& 1.02162\\ \hline
0.95936	& 0.55981	& 0.32594	& 64.1638	& 35.8367	& 0.35920	& 0.20062	& 0.13144	& 0.19450	& 0.19450	& 0.59674	& 0.29943	& 1.04006\\ \hline
0.96562	& 0.50880	& 0.27226	& 72.2019	& 27.9330	& 0.36736	& 0.14212	& 0.13435	& 0.13792	& 0.13792	& 0.50655	& 0.29439	& 1.05170\\ \hline
0.96718	& 0.49519	& 0.25798	& 74.5892	& 25.5800	& 0.36936	& 0.12667	& 0.13503	& 0.12295	& 0.12295	& 0.47660	& 0.29320	& 1.05207\\ \hline
0.96875	& 0.48122	& 0.24332	& 77.1652	& 23.0263	& 0.37134	& 0.11081	& 0.13573	& 0.10759	& 0.10759	& 0.44216	& 0.29204	& 1.05072\\ \hline
0.97058	& 0.46445	& 0.22569	& 80.4438	& 19.7474	& 0.37362	& 0.09172	& 0.13662	& 0.08908	& 0.08908	& 0.39468	& 0.29072	& 1.04628\\ \hline
0.97162	& 0.45461	& 0.21535	& 82.4684	& 17.7059	& 0.37491	& 0.08049	& 0.13717	& 0.07819	& 0.07819	& 0.36306	& 0.28998	& 1.04200\\ \hline
0.97266	& 0.44458	& 0.20480	& 84.6180	& 15.5260	& 0.37619	& 0.06902	& 0.13775	& 0.06705	& 0.06705	& 0.32740	& 0.28926	& 1.03621\\ \hline
0.97371	& 0.43432	& 0.19404	& 86.9081	& 13.1940	& 0.37746	& 0.05730	& 0.13837	& 0.05567	& 0.05567	& 0.28689	& 0.28855	& 1.02864\\ \hline
0.97475	& 0.42383	& 0.18305	& 89.3570	& 10.6957	& 0.37873	& 0.04533	& 0.13901	& 0.04404	& 0.04404	& 0.24057	& 0.28785	& 1.01899\\ \hline
0.97553	& 0.41579	& 0.17465	& 91.3115	& 8.7046	& 0.37966	& 0.03619	& 0.13949	& 0.03516	& 0.03516	& 0.20129	& 0.28733	& 1.01021\\ \hline
0.97632	& 0.40758	& 0.16610	& 93.3795	& 6.6060	& 0.38060	& 0.02692	& 0.13996	& 0.02615	& 0.02615	& 0.15743	& 0.28682	& 0.99989\\ \hline
0.97710	& 0.39920	& 0.15742	& 95.5735	& 4.3948	& 0.38153	& 0.01754	& 0.14038	& 0.01704	& 0.01704	& 0.10822	& 0.28632	& 0.98782\\ \hline
0.97746	& 0.39522	& 0.15331	& 96.6445	& 3.3227	& 0.38196	& 0.01313	& 0.14056	& 0.01275	& 0.01275	& 0.08316	& 0.28609	& 0.98151\\ \hline
0.97757	& 0.39407	& 0.15213	& 96.9562	& 3.0116	& 0.38208	& 0.01187	& 0.14061	& 0.01152	& 0.01152	& 0.07574	& 0.28602	& 0.97963\\ \hline
0.97767	& 0.39293	& 0.15095	& 97.2706	& 2.6984	& 0.38220	& 0.01060	& 0.14066	& 0.01029	& 0.01029	& 0.06819	& 0.28595	& 0.97770\\ \hline
0.97778	& 0.39178	& 0.14976	& 97.5877	& 2.3830	& 0.38232	& 0.00934	& 0.14070	& 0.00906	& 0.00906	& 0.06052	& 0.28589	& 0.97574\\ \hline
0.97788	& 0.39062	& 0.14858	& 97.9075	& 2.0656	& 0.38245	& 0.00807	& 0.14074	& 0.00783	& 0.00783	& 0.05272	& 0.28582	& 0.97373\\ \hline
0.97799	& 0.38946	& 0.14739	& 98.2297	& 1.7464	& 0.38257	& 0.00680	& 0.14079	& 0.00660	& 0.00660	& 0.04480	& 0.28576	& 0.97168\\ \hline
0.97809	& 0.38830	& 0.14619	& 98.5563	& 1.4226	& 0.38269	& 0.00552	& 0.14083	& 0.00536	& 0.00536	& 0.03668	& 0.28569	& 0.96959\\ \hline
0.97820	& 0.38713	& 0.14499	& 98.8849	& 1.0982	& 0.38282	& 0.00425	& 0.14087	& 0.00413	& 0.00413	& 0.02846	& 0.28563	& 0.96746\\ \hline
0.97830	& 0.38596	& 0.14380	& 99.2155	& 0.7726	& 0.38294	& 0.00298	& 0.14090	& 0.00290	& 0.00289	& 0.02013	& 0.28556	& 0.96529\\ \hline
0.97835	& 0.38538	& 0.14320	& 99.3810	& 0.6107	& 0.38300	& 0.00235	& 0.14092	& 0.00228	& 0.00228	& 0.01595	& 0.28553	& 0.96419\\ \hline
0.97840	& 0.38479	& 0.14259	& 99.5504	& 0.4424	& 0.38306	& 0.00170	& 0.14094	& 0.00165	& 0.00165	& 0.01159	& 0.28550	& 0.96305\\ \hline
\end{tabular}}
\caption{Global quantities and parameters associated with a sequence of hairy black holes with horizon at $\mu r_H = 0.053651$. The description of columns are like in Table~\ref{tab:GlobalQuantities1}.}\label{tab:GlobalQuantities5}
\end{center}
\end{table*}

\begin{table*}[h!]
\begin{center}
\resizebox{15.1cm}{!}{
\begin{tabular}{|c|c|c|c|c|c|c|c|c|c|c|c|c|}
\hline 
$\boldsymbol{\Omega_{H}/\mu}$ & $\boldsymbol{\mu M_{\rm ADM}}$ & $\boldsymbol{\mu^2J_{\infty}}$ & $\boldsymbol{M_{\rm BH}}$(\textdiscount) & $\boldsymbol{M_{\Psi}}$(\textdiscount) & $\boldsymbol{\mu M_{\rm BH}}$ & $\boldsymbol{\mu M_{\Psi}}$ & $\boldsymbol{\mu^2J_{\rm BH}}$ & $\boldsymbol{\mu^2J_{\Psi}}$ & $\boldsymbol{\mu^2\mathcal{Q}}$ & $\boldsymbol{q}$ & $\boldsymbol{\kappa/\mu}$ & $\boldsymbol{J_{\infty}/{M^2_{\rm ADM}}}$ \\
\hline \hline
0.83330	& 1.07819	& 0.94723	& 14.3146	& 85.6853	& 0.15434	& 0.92385	& 0.03672	& 0.91051	& 0.91051	& 0.96123	& 0.61389	& 0.81482\\ \hline
0.83365	& 1.07215	& 0.93753	& 14.8114	& 85.1885	& 0.15880	& 0.91335	& 0.03940	& 0.89814	& 0.89814	& 0.95798	& 0.58745	& 0.81560\\ \hline
0.83540	& 1.05970	& 0.91812	& 15.8508	& 84.1491	& 0.16797	& 0.89173	& 0.04479	& 0.87334	& 0.87334	& 0.95122	& 0.54214	& 0.81759\\ \hline
0.83991	& 1.04067	& 0.88954	& 17.4979	& 82.5019	& 0.18210	& 0.85857	& 0.05286	& 0.83668	& 0.83668	& 0.94057	& 0.48880	& 0.82137\\ \hline
0.84441	& 1.02576	& 0.86791	& 18.8317	& 81.1682	& 0.19317	& 0.83259	& 0.05903	& 0.80889	& 0.80888	& 0.93199	& 0.45626	& 0.82486\\ \hline
0.84891	& 1.01229	& 0.84887	& 20.0589	& 79.9410	& 0.20305	& 0.80924	& 0.06442	& 0.78444	& 0.78444	& 0.92411	& 0.43200	& 0.82838\\ \hline
0.85342	& 0.99947	& 0.83114	& 21.2382	& 78.7617	& 0.21227	& 0.78720	& 0.06937	& 0.76177	& 0.76177	& 0.91654	& 0.41247	& 0.83202\\ \hline
0.85792	& 0.98691	& 0.81412	& 22.3971	& 77.6028	& 0.22104	& 0.76587	& 0.07400	& 0.74012	& 0.74012	& 0.90910	& 0.39608	& 0.83586\\ \hline
0.86243	& 0.97440	& 0.79748	& 23.5519	& 76.4481	& 0.22949	& 0.74491	& 0.07840	& 0.71908	& 0.71908	& 0.90169	& 0.38196	& 0.83993\\ \hline
0.86843	& 0.95753	& 0.77548	& 25.1039	& 74.8960	& 0.24038	& 0.71715	& 0.08399	& 0.69149	& 0.69149	& 0.89170	& 0.36576	& 0.84580\\ \hline
0.87444	& 0.94021	& 0.75336	& 26.6885	& 73.3116	& 0.25093	& 0.68928	& 0.08930	& 0.66405	& 0.66405	& 0.88146	& 0.35182	& 0.85222\\ \hline
0.88044	& 0.92225	& 0.73085	& 28.3226	& 71.6775	& 0.26121	& 0.66105	& 0.09440	& 0.63645	& 0.63645	& 0.87084	& 0.33962	& 0.85927\\ \hline
0.88645	& 0.90349	& 0.70774	& 30.0227	& 69.9778	& 0.27125	& 0.63224	& 0.09930	& 0.60844	& 0.60844	& 0.85970	& 0.32881	& 0.86703\\ \hline
0.89246	& 0.88377	& 0.68386	& 31.8058	& 68.1949	& 0.28109	& 0.60268	& 0.10402	& 0.57984	& 0.57984	& 0.84789	& 0.31914	& 0.87557\\ \hline
0.89846	& 0.86296	& 0.65903	& 33.6912	& 66.3101	& 0.29074	& 0.57223	& 0.10859	& 0.55045	& 0.55045	& 0.83524	& 0.31041	& 0.88497\\ \hline
0.90447	& 0.84091	& 0.63311	& 35.7010	& 64.3009	& 0.30021	& 0.54072	& 0.11300	& 0.52011	& 0.52011	& 0.82152	& 0.30248	& 0.89531\\ \hline
0.91047	& 0.81749	& 0.60592	& 37.8620	& 62.1404	& 0.30952	& 0.50799	& 0.11726	& 0.48866	& 0.48866	& 0.80648	& 0.29524	& 0.90667\\ \hline
0.91648	& 0.79254	& 0.57732	& 40.2074	& 59.7955	& 0.31866	& 0.47390	& 0.12139	& 0.45594	& 0.45593	& 0.78974	& 0.28861	& 0.91913\\ \hline
0.92148	& 0.77046	& 0.55227	& 42.3316	& 57.6685	& 0.32615	& 0.44431	& 0.12473	& 0.42754	& 0.42754	& 0.77416	& 0.28348	& 0.93037\\ \hline
0.92549	& 0.75187	& 0.53136	& 44.1639	& 55.8324	& 0.33206	& 0.41979	& 0.12734	& 0.40402	& 0.40402	& 0.76035	& 0.27962	& 0.93993\\ \hline
0.92949	& 0.73240	& 0.50960	& 46.1347	& 53.8553	& 0.33789	& 0.39444	& 0.12990	& 0.37970	& 0.37970	& 0.74509	& 0.27597	& 0.95001\\ \hline
0.93350	& 0.71197	& 0.48691	& 48.2675	& 51.7133	& 0.34365	& 0.36818	& 0.13241	& 0.35451	& 0.35451	& 0.72806	& 0.27249	& 0.96058\\ \hline
0.93750	& 0.69048	& 0.46324	& 50.5920	& 49.3782	& 0.34933	& 0.34095	& 0.13487	& 0.32837	& 0.32837	& 0.70885	& 0.26919	& 0.97163\\ \hline
0.94150	& 0.66783	& 0.43848	& 53.1460	& 46.8158	& 0.35493	& 0.31265	& 0.13727	& 0.30121	& 0.30121	& 0.68694	& 0.26606	& 0.98313\\ \hline
0.94551	& 0.64390	& 0.41253	& 55.9783	& 43.9848	& 0.36044	& 0.28322	& 0.13959	& 0.27295	& 0.27295	& 0.66164	& 0.26307	& 0.99500\\ \hline
0.94951	& 0.61852	& 0.38528	& 59.1528	& 40.8329	& 0.36587	& 0.25256	& 0.14178	& 0.24351	& 0.24351	& 0.63203	& 0.26024	& 1.00709\\ \hline
0.95352	& 0.59156	& 0.35659	& 62.7514	& 37.2895	& 0.37122	& 0.22059	& 0.14379	& 0.21280	& 0.21280	& 0.59677	& 0.25754	& 1.01898\\ \hline
0.95602	& 0.57385	& 0.33786	& 65.2614	& 34.8317	& 0.37450	& 0.19988	& 0.14495	& 0.19291	& 0.19291	& 0.57097	& 0.25593	& 1.02596\\ \hline
0.95802	& 0.55919	& 0.32238	& 67.4383	& 32.7025	& 0.37711	& 0.18287	& 0.14583	& 0.17655	& 0.17655	& 0.54765	& 0.25468	& 1.03098\\ \hline
0.96002	& 0.54408	& 0.30645	& 69.7850	& 30.4025	& 0.37969	& 0.16541	& 0.14669	& 0.15977	& 0.15976	& 0.52134	& 0.25345	& 1.03522\\ \hline
0.96202	& 0.52852	& 0.29003	& 72.3226	& 27.9006	& 0.38224	& 0.14746	& 0.14755	& 0.14248	& 0.14248	& 0.49126	& 0.25226	& 1.03830\\ \hline
0.96403	& 0.51249	& 0.27309	& 75.0770	& 25.1586	& 0.38477	& 0.12894	& 0.14846	& 0.12463	& 0.12463	& 0.45636	& 0.25111	& 1.03976\\ \hline
0.96603	& 0.49598	& 0.25560	& 78.0811	& 22.1306	& 0.38726	& 0.10976	& 0.14947	& 0.10613	& 0.10613	& 0.41522	& 0.24998	& 1.03904\\ \hline
0.96803	& 0.47891	& 0.23750	& 81.3795	& 18.7640	& 0.38973	& 0.08986	& 0.15060	& 0.08690	& 0.08690	& 0.36590	& 0.24889	& 1.03551\\ \hline
0.97003	& 0.46119	& 0.21875	& 85.0339	& 15.0031	& 0.39217	& 0.06919	& 0.15184	& 0.06691	& 0.06691	& 0.30587	& 0.24782	& 1.02845\\ \hline
0.97203	& 0.44269	& 0.19930	& 89.1310	& 10.7926	& 0.39457	& 0.04778	& 0.15311	& 0.04619	& 0.04619	& 0.23174	& 0.24679	& 1.01695\\ \hline
0.97403	& 0.42323	& 0.17907	& 93.7895	& 6.08030	& 0.39694	& 0.02573	& 0.15421	& 0.02486	& 0.02486	& 0.13883	& 0.24579	& 0.99972\\ \hline
0.97604	& 0.40266	& 0.15800	& 99.1603	& 0.80870	& 0.39928	& 0.00326	& 0.15486	& 0.00314	& 0.00314	& 0.01990	& 0.24483	& 0.97451\\ \hline
\end{tabular}}
\caption{Global quantities and parameters associated with a sequence of hairy black holes with horizon at $\mu r_H = 0.048574$. The description of columns are like in Table~\ref{tab:GlobalQuantities1}.}\label{tab:GlobalQuantities6}
\end{center}
\end{table*}

\begin{table*}[h!]
\begin{center}
\resizebox{11cm}{!}{
\begin{tabular}{|c|c|c|c|c|c|c|c|c|c|c|c|c|}
\hline 
$\boldsymbol{\Omega_{H}/\mu}$ & $\boldsymbol{\mu M_{\rm ADM}}$ & $\boldsymbol{\mu^2J_{\infty}}$ & $\boldsymbol{\kappa/\mu}$ & $\boldsymbol{T_H/\mu}$ & $\boldsymbol{\mu^2A_H}$ & $\boldsymbol{\mu^2S_H}$ & $\boldsymbol{\mu M_{\rm Smarr}}$\\
\hline \hline
0.88289 & 0.986917 & 0.832652 & 0.645915 & 0.102801 & 2.20098 & 0.55024 & 0.986913 \\ \hline
0.88831 & 0.951343 & 0.781093 & 0.541370 & 0.086162 & 2.62930 & 0.65732 & 0.951340 \\ \hline
0.89373 & 0.926418 & 0.748099 & 0.501102 & 0.079753 & 2.84506 & 0.71126 & 0.926414 \\ \hline
0.89914 & 0.902750 & 0.717944 & 0.472997 & 0.075280 & 3.01884 & 0.75471 & 0.902749 \\ \hline
0.90456 & 0.879152 & 0.688689 & 0.451077 & 0.071791 & 3.17039 & 0.79260 & 0.879150 \\ \hline
0.90997 & 0.855091 & 0.659515 & 0.433038 & 0.068920 & 3.30730 & 0.82683 & 0.855089 \\ \hline
0.91539 & 0.830236 & 0.629945 & 0.417709 & 0.066481 & 3.43346 & 0.85837 & 0.830235 \\ \hline
0.92081 & 0.804335 & 0.599639 & 0.404400 & 0.064362 & 3.55113 & 0.88778 & 0.804335 \\ \hline
0.92622 & 0.777166 & 0.568322 & 0.392667 & 0.062495 & 3.66175 & 0.91544 & 0.777167 \\ \hline
0.93164 & 0.748513 & 0.535743 & 0.382205 & 0.060830 & 3.76631 & 0.94158 & 0.748514 \\ \hline
0.93706 & 0.718148 & 0.501651 & 0.372796 & 0.059332 & 3.86549 & 0.96632 & 0.718150 \\ \hline
0.94247 & 0.685822 & 0.465783 & 0.364277 & 0.057977 & 3.95975 & 0.98994 & 0.685823 \\ \hline
0.94789 & 0.651244 & 0.427841 & 0.356527 & 0.056743 & 4.04944 & 1.01236 & 0.651246 \\ \hline
0.95331 & 0.614070 & 0.387477 & 0.349450 & 0.055617 & 4.13478 & 1.03369 & 0.614073 \\ \hline
0.95872 & 0.573875 & 0.344266 & 0.342973 & 0.054586 & 4.21589 & 1.05397 & 0.573878 \\ \hline
0.96414 & 0.530110 & 0.297665 & 0.337040 & 0.053642 & 4.29282 & 1.07320 & 0.530115 \\ \hline
0.96956 & 0.482045 & 0.246955 & 0.331609 & 0.052777 & 4.36551 & 1.09138 & 0.482050 \\ \hline
0.97497 & 0.428652 & 0.191123 & 0.326653 & 0.051989 & 4.43378 & 1.10845 & 0.428658 \\ \hline
0.97551 & 0.422967 & 0.185206 & 0.326183 & 0.051914 & 4.44036 & 1.11009 & 0.422973 \\ \hline
0.97606 & 0.417212 & 0.179221 & 0.325718 & 0.051840 & 4.44688 & 1.11172 & 0.417218 \\ \hline
0.97660 & 0.411385 & 0.173167 & 0.325257 & 0.051766 & 4.45336 & 1.11334 & 0.411391 \\ \hline 
0.97714 & 0.405485 & 0.167041 & 0.324800 & 0.051694 & 4.45979 & 1.11495 & 0.405490 \\ \hline 
0.97768 & 0.399507 & 0.160841 & 0.324348 & 0.051622 & 4.46617 & 1.11654 & 0.399513 \\ \hline
0.97822 & 0.393451 & 0.154566 & 0.323901 & 0.051551 & 4.47251 & 1.11813 & 0.393457 \\ \hline
0.97876 & 0.387314 & 0.148211 & 0.323459 & 0.051480 & 4.47879 & 1.11970 & 0.387320 \\ \hline
0.97931 & 0.381095 & 0.141775 & 0.323021 & 0.051410 & 4.48502 & 1.12125 & 0.381099 \\ \hline
0.97985 & 0.374785 & 0.135256 & 0.322588 & 0.051341 & 4.49120 & 1.12280 & 0.374793 \\ \hline
0.98020 & 0.370630 & 0.130963 & 0.322308 & 0.051297 & 4.49519 & 1.12380 & 0.370636\\ \hline 
\end{tabular}}
\caption{Thermodynamic properties associated with a hairy black hole with horizon at $\mu r_H = 0.057648$.}\label{tab:Thermodynamics1}
\end{center}
\end{table*}

\end{document}